\RequirePackage{ifpdf}
 \documentclass[letterpaper]{JHEP3}
 \pdfoutput=1

\usepackage{epsf,epsfig}
\usepackage{amssymb}
\usepackage{graphicx}  

\usepackage{amsmath}
\usepackage{amsfonts}
\usepackage{enumerate}%
\usepackage{enumitem}  

\usepackage{float}
\usepackage{subfigure}
 \usepackage{epsfig}

\usepackage{graphicx}

\newcounter{fig}

\newcommand{\beq}{\begin{equation}}
\newcommand{\eeq}{\end{equation}}
\newcommand{\beqs}{\begin{eqnarray}}
\newcommand{\eeqs}{\end{eqnarray}}

\newcommand{\be}{\begin{equation}}
\newcommand{\ee}{\end{equation}}
\newcommand{\bea}{\begin{eqnarray}}
\newcommand{\eea}{\end{eqnarray}}

\newcommand{\diff}{\mathrm{d}}

\numberwithin{equation}{section}

\abstract{ 
We construct a new class of black hole solutions in  
 five-dimensional
Einstein-Maxwell-Chern-Simons theory with a negative cosmological constant. 
These configurations are cohomogeneity-1, with  two equal-magnitude angular momenta. 
In the generic case, 
they possess a non-vanishing magnetic potential at infinity
with
a boundary metric which is the product of time and a 
squashed three-dimensional sphere. 
Both extremal and non-extremal black holes are studied.
The non-extremal black holes 
satisfying a certain relation between electric charge, angular momenta
and magnitude of the magnetic potential at infinity
do not trivialize in the limit of
vanishing event horizon size,
becoming particle-like 
(non-topological)
solitonic configurations.
Among the extremal black holes, we show the existence of a
new one-parameter family of supersymmetric solutions,
which bifurcate from a critical Gutowski-Reall  configuration.

}

\preprint{ }

\title{    
Squashed, magnetized black holes
in $D=5$ minimal gauged supergravity   
} 
  
 \author{
 {\large Jose Luis Bl\'azquez-Salcedo}$^{a}$, 
{\large Jutta Kunz}$^{a}$,
{\large Francisco Navarro-L\'erida}$^{b}$  
  and} 
  
  \author{
 {\large Eugen Radu}$^{c}$
\\ 
\\
$^{a}$
{\small Institut f\"ur Physik, Universit\"at Oldenburg, \\ Postfach 2503
D-26111 Oldenburg, Germany} 
\\
$^{b}$   
{\small Dept. de F\'{\i}sica Te\'orica I and UPARCOS, Ciencias F\'{\i}sicas,
 Universidad Complutense de Madrid, Av. Complutense s/n E-28040 Madrid, Spain}
\\
$^{c}$   
{\small  Departamento de F\'isica da Universidade de Aveiro and CIDMA, \\
 Campus de Santiago, 3810-183 Aveiro, Portugal}
 \\ 
{\small  E-mail: \emph{jose.blazquez.salcedo@uni-oldenburg.de,\\ jutta.kunz@uni-oldenburg.de, fnavarro@fis.ucm.es, eugen.radu@ua.pt}}
 }
 
\begin{document}

\section{Introduction and motivation}
 
The study of black objects in gravity models with a negative cosmological constant
 has attracted
recently considerable interest, 
being fueled by studies of the  Anti-de Sitter/Conformal Field Theory  (AdS/CFT) correspondence
\cite{Witten:1998qj},
\cite{Maldacena:1997re}.
This conjecture
basically
proposes a 'dictionary' between classical AdS bulk
gravitational solutions (in $D-$dimensions) and field theory states at strong coupling (in $(D-1)-$dimensions).

Of particular interest in this context are the solutions of five dimensional
${\cal N} = 4$ $SO(6)$ gauged supergravity, which is thought to be a consistent truncation of
type
IIB supergravity on $AdS_5\times S^5$ 
\cite{Cvetic:1999xp},
\cite{Gauntlett:2006ai}.
In its minimal version, the bosonic sector of this model
is just Einstein-Maxwell (EM) theory with a negative cosmological constant 
and a Chern-Simons (CS) U(1) term (with a fixed value of
the coupling constant).
Despite its (apparent) simplicity, this theory possesses
a variety of 
interesting 
solutions which have been investigated extensively 
over the last two decades.

Restricting to configurations possessing an event horizon, 
one remarks that 
most of the studies in the literature concentrate on two different classes of solutions.
First, there are the black holes (BHs) with a spherical horizon topology\footnote{Black rings 
with an $S^2\times S^1$ event horizon topology exist as well, 
approaching at infinity a globally AdS$_5$ background.
Such solutions have been constructed in \cite{Caldarelli:2008pz} using approximate methods, 
and fully nonperturbatively in \cite{Figueras:2014dta}.} 
in a globally AdS$_5$ spacetime background,
in which case the dual theory is formulated in a $D=4$ Einstein universe.
The Schwarzschild-AdS BH is the simplest example,
while the  most general such EMCS solutions rotate  in two
planes and possess four global
charges: the mass, the electric charge, and two angular momenta \cite{Chong:2005hr}.
A considerable  simplification is obtained for an Ansatz with
two equal-magnitude angular
momenta, an assumption  which factorizes the angular dependence of the problem.
These  BH solutions have been found in closed form 
by Cveti\v c, L\"u and Pope (CLP)
 in  
	\cite{Cvetic:2004hs}
	(see also \cite{Chong:2006zx}).
Remarkably, their extremal limit contains a subset of solutions that
 preserves some amount of supersymmetry 
\cite{Gutowski:2004ez}.
An extension of the CLP BHs
which
possesses
 an extra parameter $\Phi_m$ associated with a non-zero magnitude
of the magnetic potential at infinity
has been reported in 
the recent work \cite{Blazquez-Salcedo:2017cqm}.

Second, there are the black branes, which  
approach at infinity the Poincar\'e patch of the AdS spacetime.
These solutions have a Ricci flat horizon,
 while their dual field theory states reside
in a $D=4$ Minkowski spacetime.
The most general such configurations
appear to be those reported in 
\cite{D'Hoker:2009bc},
\cite{D'Hoker:2010ij};
in addition to the mass and electric charge, 
they possess an extra parameter corresponding to the magnitude of the magnetic field at infinity. 

However, it is worth remarking that the AdS/CFT correspondence 
does not  constrain  the way of approaching
the boundary of spacetime, 
asymptotically $locally$ AdS  (AlAdS) solutions  
being also relevant.
An interesting class of such configurations are the $D=5$ AdS black strings\footnote{These solutions have been
 generalized
to higher dimensions and a more general topology of the event horizon in \cite{Mann:2006yi}.}
originally found by Copsey and Horowitz in \cite{Copsey:2006br}.
These are natural AdS counterparts of the (better known) uniform black strings in a $D=5$
Kaluza-Klein theory, the horizon topology being $S^{2}\times S^1$.
Also, the conformal boundary, where the dual theory lives, is the product of time and $S^{2}\times S^1$.

 \medskip
 
{The main purpose of this work is to report the existence
of a new
 class of solutions of the $D=5$ minimal gauged supergravity model.
These solutions possess a squashed sphere in the boundary metric and can be viewed as interpolating 
between 
(some versions of) 
the three classes of black objects mentioned above.}

{Moreover, we find that a particular set of these configurations 
has special properties,
forming a new one-parameter 
family of supersymmetric BHs.} 

{
A discussion of the basic properties of these solutions 
was given in the recent work 
\cite{Blazquez-Salcedo:2017kig},
in a slightly different context. }


\begin{figure}[t]
\centering
\includegraphics[scale=0.4,angle=-90]{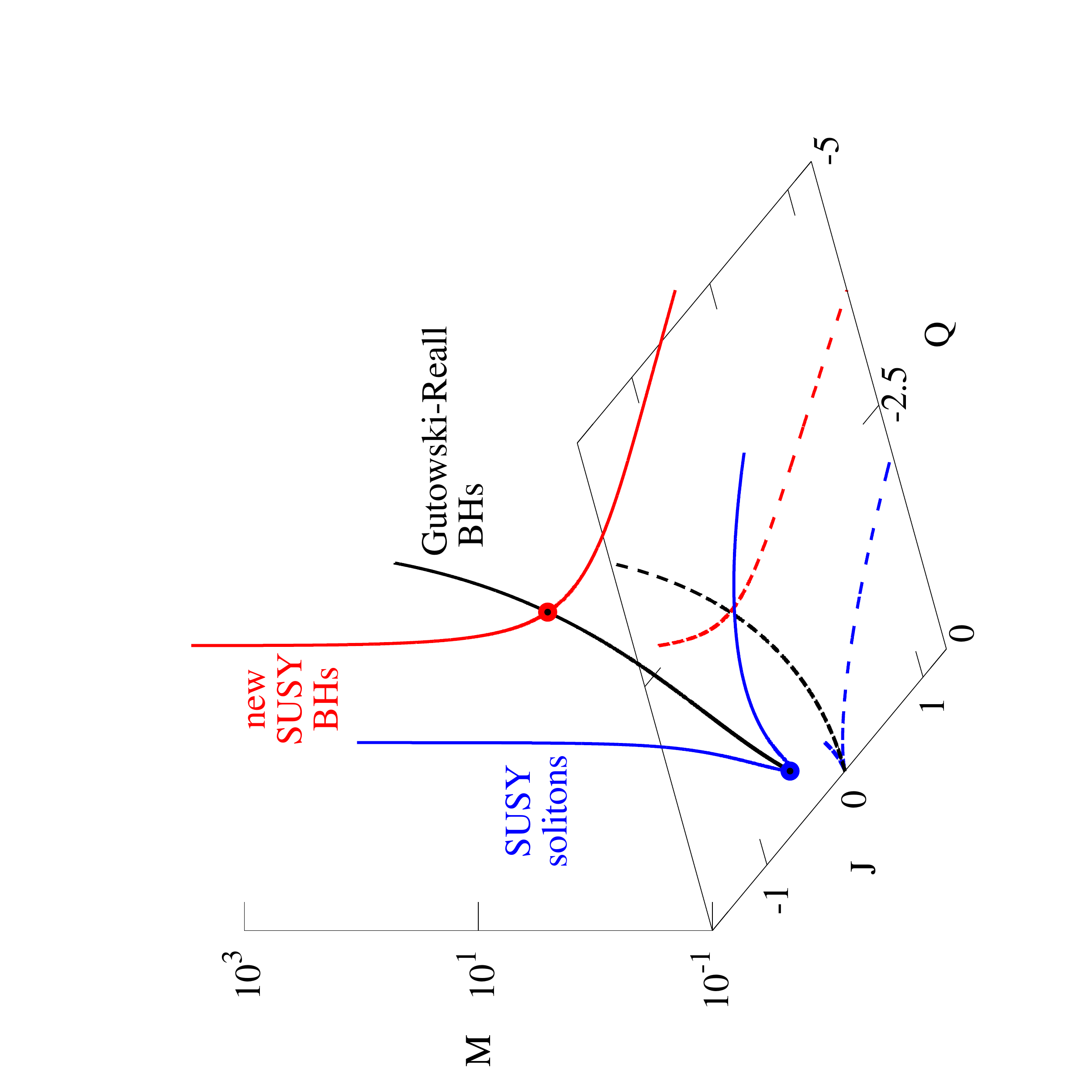}
\caption{The  {\it mass-angular momentum-electric charge}  diagram 
				is shown for 
				the supersymmetric solitons in \cite{Cassani:2014zwa},
				the Gutowski-Reall BHs \cite{Gutowski:2004ez}
				and the new  supersymmetric BHs in this work.}
\label{MJQ} 
\end{figure}


\begin{figure}[t]
\centering
\subfigure[]{\includegraphics[scale=0.29,angle=-90]{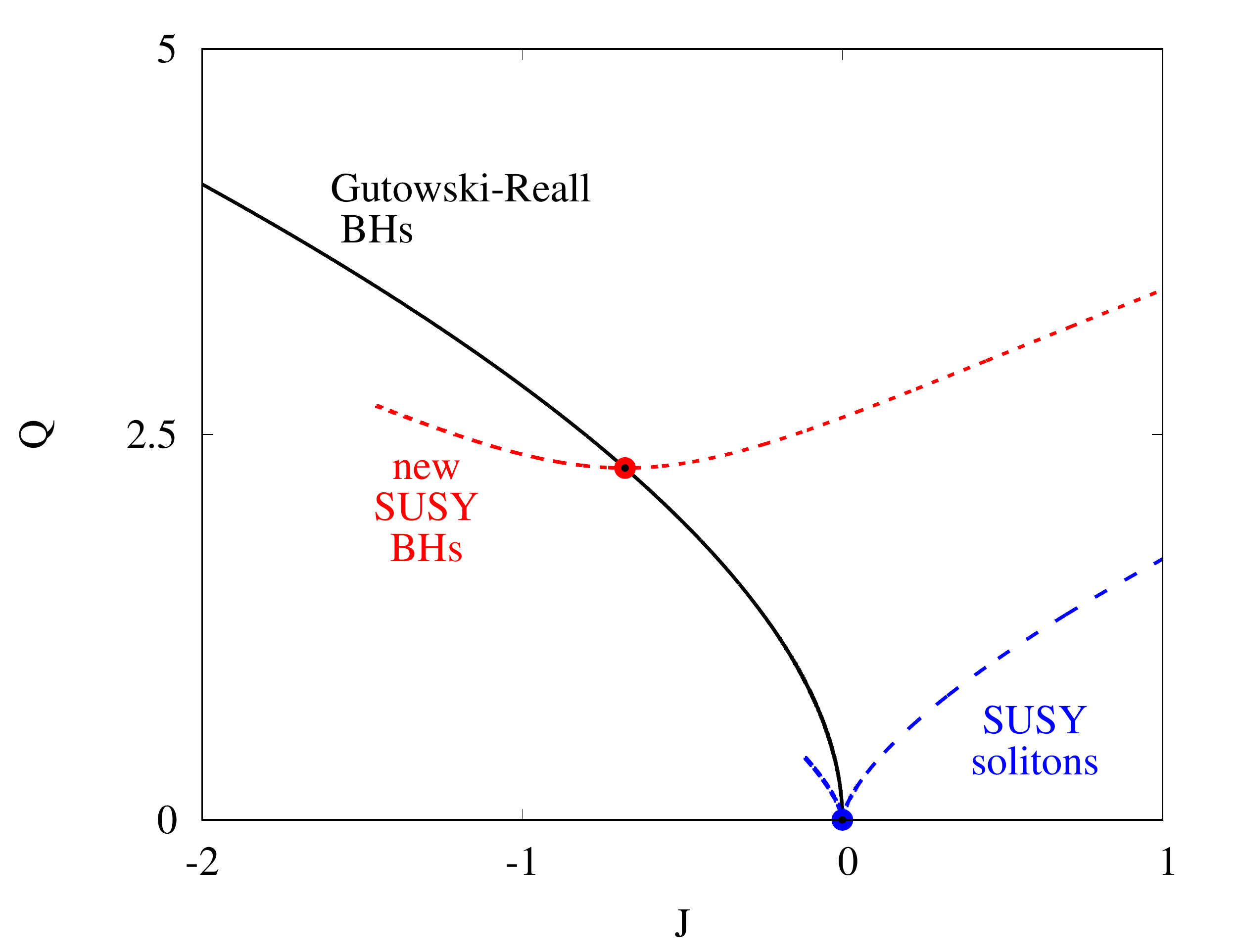}
\label{fig_2d_susy_JQ}}
\subfigure[]{\includegraphics[scale=0.29,angle=-90]{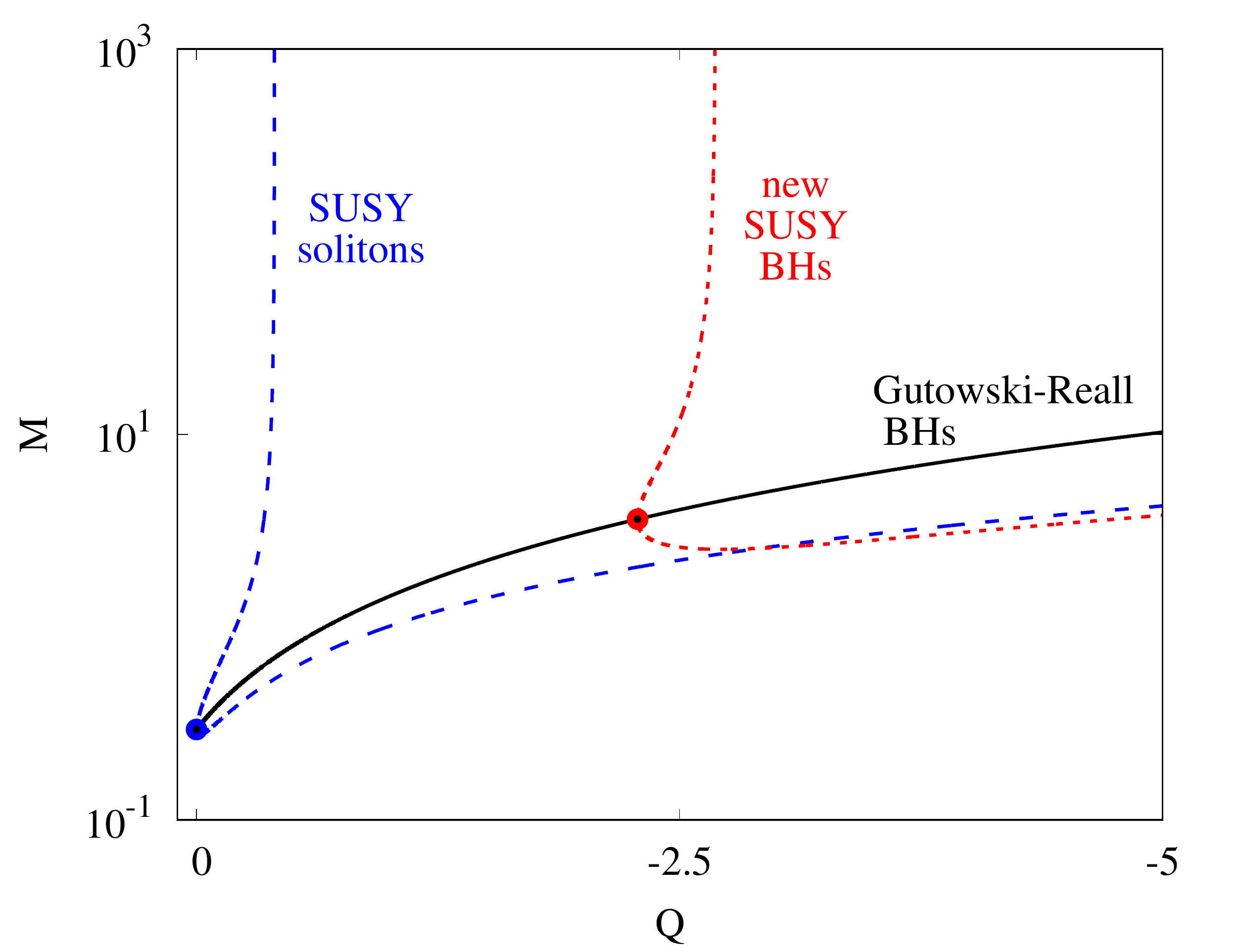}
\label{fig_2d_susy_MQ}}
\caption{(a) The  {\it electric charge-angular momentum}  diagram 
				is shown for 
				the supersymmetric solitons in \cite{Cassani:2014zwa},
				the Gutowski-Reall BHs \cite{Gutowski:2004ez}
				and the new  supersymmetric BHs in this work. (b) A similar figure with the  {\it mass-electric charge}  diagram 
				for the three families of supersymmetric solutions.}
\label{fig_2d_susy}
\end{figure}

\subsection{Summary of results}
 
In a convenient set of coordinates, 
the conformal boundary metric of the 
solutions in this work
reads 
\begin{eqnarray}
\label{g0}
 ds_{(bdry)}^2= 
 L^2d\Omega_{(v)}^2- dt^2,~~{\rm where}~~d\Omega_{(v)}^2=\frac{1}{4}\left( d\theta^2+\sin^2\theta d\phi^2 + (d  \bar {\psi} + v \cos\theta d \phi)^2 \right) ,
\end{eqnarray} 
with 
$v $ a control parameter,
$d\Omega_{(v)}^2$ the metric on a squashed $S^3$ sphere
and $0\leq \theta<\pi$, $0\leq \phi< 2\pi$, $0\leq  \bar \psi <4\pi v$.
The presence of a squashed  sphere in the boundary geometry of some asymptotically $locally$
AdS configurations has been found before in the literature,
the   $D=4$ nutty  instantons 
(reviewed in Appendix A)
being 
perhaps the best known case. 

Clearly
the sphere in (\ref{g0})
becomes a round one for $v=1$,
 in which case the solutions approach a globally AdS background.
Another case of interest is $v=0$, the bulk solutions becoming AdS black strings and vortices,
with a boundary which is the product of time and $S^{2}\times S^1$ (with $S^1$ parametrized by $\bar {\psi}$,
whose periodicity is arbitrary in this limit). 
Finally, for large values of $v$, 
one can show that,  after a proper rescaling, the boundary geometry (\ref{g0}) 
is the product of time and a twisted $R^3$ part.

\medskip

In this work we provide evidence for the existence  of a 
family of black objects 
with a conformal boundary given by (\ref{g0}).
The solutions are constructed  numerically
 within the framework of the $D=5$
minimal gauged supergravity model, and possess a 
gauge potential with both a magnetic and an electric part;
they also rotate in the bulk,
 with equal-magnitude angular momenta.

\medskip

The main properties of the generic nonextremal solutions 
can be summarized as follows:
\begin{enumerate}[label=(\roman*)]
\item
They possess an event horizon of spherical topology and are regular on and outside the horizon.
Also, they do not present other pathologies (such as closed timelike curves (CTCs)).
\item
In addition to the mass $M$, the electric charge $Q$
and the angular momenta $J$, 
the new solutions possess an extra parameter 
$c_m$
associated with a non-zero magnitude
of the magnetic potential at infinity.
\item
A particular set of  BHs with $c_m\neq 0$ does not trivialize as the horizon size 
shrinks to zero, a limit which describes a one-parameter family of {\it squashed spinning charged solitons}.
The angular momentum  $J$ and the electric charge $ Q$ of these solutions are determined by 
the
magnetic flux at infinity $\Phi_m$
through the base space $S^2$ of the $S^1$-fibration,
with 
$
 J=  \Phi_m Q.
$
%
%
\item
The generic BH solutions possess an extremal limit, 
with a nonzero event horizon area.
Moreover, supersymmetric BHs exist as well,
forming a one-parameter family.
These BHs  
 bifurcate from a critical Gutowski-Reall \cite{Gutowski:2004ez}
configuration,
their mass, angular momenta and electric charge having relatively
simple expressions in terms of the squashing parameter $v$
only.
 \end{enumerate}
We note that 
supersymmetric solitons exist as well  within the same framework,
being investigated in 
 the interesting work \cite{Cassani:2014zwa}.
However, they do not correspond to a limit of the supersymmetric BHs. 

The  $mass-angular~momentum-electric~charge$ 
diagrams summarizing the picture for these three different types of supersymmetric 
solutions are shown in Figure \ref{MJQ} and in Figure \ref{fig_2d_susy}.

\section{The general framework} 
 
\subsection{The model and Ansatz} 

The action for $D=5$ minimal
gauged supergravity  is given by 
\begin{equation} 
\label{EMCSac}
I= -\frac{1}{16\pi } \int_{{\cal M}} d^5\mathrm{x}
\sqrt{-g}
\biggl[ 
R +\frac{12}{L^2} 
-F_{\mu \nu} F^{\mu \nu}  
-
\frac{2 \lambda }{3\sqrt{3}}  \varepsilon^{\mu\nu\alpha\beta\gamma}A_{\mu}F_{\nu\alpha}F_{\beta\gamma} 
 \biggr ]  
-\frac{1}{8 \pi }\int_{\partial\mathcal{M}} d^4 \mathrm{x}\sqrt{-h}K,
\end{equation}
where $R$ is the curvature scalar, $L$ is the AdS length scale   
(which is fixed by the cosmological constant $\Lambda=-6/L^2$)
and 
$A_\mu $ is the gauge potential with the field strength tensor 
$ F_{\mu \nu} = \partial_\mu A_\nu -\partial_\nu A_\mu $.
 Also $\lambda$ is the CS coupling constant, with $\lambda=1$ in the minimal gauged supergravity case.
Since a number of basic results do not depend on the precise value of $\lambda$,
we shall  keep it general 
in all relations below,
such that (\ref{EMCSac})
will describe a generic Einstein--Maxwell--Chern-Simons (EMCS) model.
However, the numerical results will cover the SUGRA case only. 
Finally,  $K$ is the trace 
of the extrinsic curvature for the boundary $\partial\mathcal{M}$ and $h$ is
the induced  metric of the boundary.  
 
The field equations of this model consist of the Einstein equations
\begin{equation}
\label{Einstein_equation}
R_{\mu\nu} -\frac{1}{2}R g_{\mu \nu}
-\frac{6}{L^2}g_{\mu\nu}=2\left(F_{\mu\rho}{F^{\rho}}_{\nu}-\frac{1}{4}F^2 \right) ,
\end{equation}
together with the Maxwell--Chern-Simons   (MCS)  equations
\begin{equation}
\label{Maxwell_equation}
\nabla_{\nu} F^{\mu\nu} 
+ \frac{\lambda}{2\sqrt{3}}\varepsilon^{\mu\nu\alpha\beta\gamma}F_{\nu\alpha}F_{\beta\gamma}=0.
\end{equation}

 
A general parametrization of the metric Ansatz 
which covers both the generic and the supersymmetric configurations
possesses a local $SU(2)\times U(1)\times U(1)$ symmetry and
reads
\begin{eqnarray}
\label{metric}
&&ds^2 =
-F_0(r) dt^2+
 F_1(r) dr^2
  + \frac{1}{4} F_2(r)  ( \sigma_1^2+ \sigma_2^2)+ \frac{1}{4}F_3(r) \big(  \sigma_3- 2W(r) dt \big)^2
,
\end{eqnarray} 
with  
 $\sigma_i$ the left-invariant one-forms on $S^3$,
 \begin{eqnarray}
 \nonumber
\sigma_1=\cos \psi d \theta+\sin\psi \sin \theta d \phi,
~~
\label{sigma}
\sigma_2=-\sin \psi d \theta+\cos\psi \sin  \theta d \phi,
~~
\sigma_3= d\psi  + \cos d \theta d \phi,
\end{eqnarray}
 the coordinates $\theta$, $\phi$, $\psi$ being the Euler angles on $S^3$,
with the usual range (in particular, a periodicity $4\pi$ for $\psi$).
Also, note the existence of gauge freedom degree in 
the line element (\ref{metric})), which will be fixed  by convenience.

A gauge field Ansatz compatible with the symmetries of (\ref{metric})
contains an electric potential  and a magnetic one, with 
\begin{eqnarray}
\label{U1}
A=a_0(r)dt + \frac{1}{2} a_k(r)  \sigma_3.
\end{eqnarray}

The general configurations satisfy 
the following set of equations which follow from (\ref{Einstein_equation}), (\ref{Maxwell_equation}): 
\begin{eqnarray}
\label{eqF0}
\nonumber
&&
\frac{2F_2''}{F_2}
+\frac{F_3''}{F_3}
-\frac{F_1'F_2'}{F_1F_2}
-\frac{F_1'F_3'}{2F_1F_3}
-\frac{1}{2} \left(\frac{F_2'}{F_2}-\frac{F_3'}{F_3}\right)^2 
-\frac{8F_1}{F_2}
+\frac{2F_1F_3}{F_2^2}
+\frac{F_3W'^2}{2F_0}
\\
\nonumber 
&&
{~~~~~~~~~~~~~~~~}
-\frac{12 F_1} {L^2}
+2\left(\frac{a_k'^2}{F_3}+\frac{4 a_k^2F_1}{F_2^2}\right)
+\frac{2}{F_0}( W a_k'+a_0')^2=0,
\end{eqnarray}
\begin{eqnarray}
\label{eqF1}
\nonumber
&&
\frac{F_2'^2}{4F_2^2}
+\frac{F_0'F_2'}{2F_0F_2}
+\frac{F_0'F_3'}{4F_0F_3}
+\frac{F_2'F_3'}{2F_2F_3}
-\frac{4F_1}{F_2}
+ \frac{F_1F_3}{F_2^2}
+\frac{F_3W'^2}{4F_0}
\\
&&
\nonumber
{~~~~~~~~~~~~~~~~~~~~~~~~~}
-\frac{6 F_1} {L^2} 
+ \left(-\frac{a_k'^2}{F_3}+\frac{4 a_k^2F_1}{F_2^2}\right)
+\frac{1}{F_0}( W a_k'+a_0')^2=0,
\end{eqnarray}
\begin{eqnarray}
%
\label{eqF2}
\nonumber
&&
\frac{F_0''}{2F_0}
+\frac{F_2''}{2F_2} 
+\frac{F_3''}{2F_3}
-\frac{1}{4}(\frac{F_0'^2}{F_0^2}+\frac{F_3'^2}{F_3^2}+\frac{F_2'^2}{F_2^2})
-\frac{F_0'F_1'}{4F_0F_1}
+\frac{F_0'F_2'}{4F_0F_2}
-\frac{F_1'F_2'}{4F_1F_2}
+\frac{F_0'F_3'}{4F_0F_3}
-\frac{F_1'F_3'}{4F_1F_3}
+\frac{F_2'F_3'}{4F_2F_3}
\\
\nonumber
&&
{~~~~~~~~~~~~~~~~~~~}
- \frac{F_1F_3}{F_2^2}
-\frac{F_3W'^2}{4F_0 }
-\frac{6 F_1} {L^2} 
+ \left(\frac{a_k'^2}{F_3}-\frac{ 4 a_k^2F_1}{F_2^2}\right)
-\frac{1}{F_0}( W a_k'+a_0')^2=0,
\end{eqnarray}
\begin{eqnarray}
\label{eqs}
&&
\frac{F_0''}{2F_0}
+\frac{F_2''}{F_2}
-\frac{1}{4}(\frac{F_0'^2}{F_0^2}+\frac{F_2'^2}{F_2^2})
-\frac{F_0'F_1'}{4F_0F_1}
+\frac{F_0'F_2'}{2F_0F_2}
-\frac{F_1'F_2'}{2F_1F_2}
-\frac{4F_1}{F_2}
+\frac{3 F_1F_3}{F_2^2}
\\
\nonumber
&&
{~~~~~~~~~~~~~~~~~~~~~~~~~~~~~~~~~~~~~}
-\frac{3F_3W'^2}{4F_0}
-\frac{6}{L^2}F_1 
+ \left(-\frac{a_k'^2}{F_3}+\frac{ 4 a_k^2F_1}{F_2^2}\right)
-\frac{1}{F_0}( W a_k'+a_0')^2=0,
%
\end{eqnarray}
\begin{eqnarray}
\label{eqW}
&&
W''+
\left(
-\frac{F_0'}{2F_0}
-\frac{F_1'}{2F_1} 
+\frac{F_2'}{F_2} 
+\frac{3F_3'}{2F_3}
\right)W'
-\frac{4a_k}{F_3}( W a_k'+a_0')=0,
\end{eqnarray}
\begin{eqnarray}
\nonumber
\label{eq0}
&&
a_0''+  W a_k''+ a_k'W' 
+
\left(
-\frac{F_0'}{ F_0}
-\frac{F_1'}{ F_1}
+\frac{2F_2'}{F_2}
+\frac{F_3'}{F_3}
\right)\frac{1}{2} ( W a_k'+a_0') 
-\frac{8\lambda a_k' a_k}{\sqrt{3}F_2} %
\sqrt{\frac{F_0F_1}{F_3}}=0,
\end{eqnarray}
\begin{eqnarray}
\nonumber
&&
Wa_0''+  (W^2-\frac{F_0}{F_3}) a_k''
+
\left(
-\frac{F_0'}{F_0}
+\frac{F_1'}{F_1}
-\frac{2F_2'}{F_2}
+\frac{F_3'}{F_3}
\right) 
\frac{F_0}{2F_3}a_k'
+\left(a_0'W'+2WW'a_k'+\frac{4 a_k F_0F_1}{F_2^2}\right)
\\ 
&&
\nonumber
\label{eqafi}
+
\left(
-\frac{F_0'}{F_0}
-\frac{F_1'}{F_1}
+\frac{2F_2'}{F_2}
+\frac{F_3'}{F_3}  
\right)
( W a_k'+a_0') \frac{1}{2}W
+\frac{8\lambda  a_k a_0'}{\sqrt{3}F_2}
\sqrt{\frac{F_0F_1}{F_3}}=0,
\end{eqnarray}
where a prime denotes a derivative $w.r.t.$
the radial coordinate $r$.
Also, we notice the existence of the scaling symmetry
\begin{eqnarray}
\label{scale1}
 &&
F_0\to p^2 F_0,~~a_0 \to p a_0,~~W\to p W,
\end{eqnarray}
with $p$ an arbitrary nonzero constant.  

We remark that one cannot take $a_0=0$,
unless the magnetic potential also vanishes, $a_k=0$.
Also, the  equations of motion possess two first integrals\footnote{The origin of these first integrals 
can be traced back to the fact that 
the Einstein equation $E_\psi^t$ and the
 MCS equations possess a total derivative structure.
} 
\begin{eqnarray}
\label{1stInt-gen}
 &&
\frac{1}{2}F_2\sqrt{\frac{F_3} {F_0F_1}}
(W a_k'+ a_0')-\frac{2 \lambda a_k^2}{\sqrt{3}}=c_t,
\\
\nonumber
&&
\frac{1}{8}F_2F_3\sqrt{\frac{F_3} {F_0F_1}} W'-\left( a_k c_t+\frac{2}{9}\sqrt{3}  \lambda a_k^3\right)=c_W,
\end{eqnarray}
with $c_t,c_W$ two constants of integration.

The CLP BHs are a solution of the above equations, 
the corresponding expression of $(F_i,W)$ and $(a_0,a_k)$ being given $e.g.$
in the Appendix A of Ref. \cite{Blazquez-Salcedo:2016rkj}.
In practice, the non-supersymmetric solutions 
are found for a  reparametrization
of (\ref{metric})
which fixes the metric gauge and enforces the far behaviour, with
\begin{eqnarray}
\label{param}
 F_0(r) =f(r)\left(1+\frac{r^2}{L^2}\right),~
 F_1(r) =\frac{m(r)}{f(r)}\frac{1}{1+\frac{r^2}{L^2}},~
 F_2(r) =\frac{m(r)}{f(r)}r^2 ,~
F_3(r) =\frac{n(r)}{f(r)}r^2 ,~
W(r)=\frac{\omega(r)}{r}.~{~~~}
\end{eqnarray}
The  supersymmetric solutions are found for a more complicated
 parametrization of (\ref{metric}),
which is discussed in Section 4.

\subsection{Asymptotics } 

\subsubsection{The solutions in the far field } 

The far field expression of the solutions is found assuming that
$(i)$ they approach at infinity a locally AdS spacetime, with a
conformal
 boundary metric given by\footnote{In fact, the form
(\ref{g0}) is found only after a suitable rescaling, see the discussion in Section 3.2.
 } 
(\ref{g0}), and,
$(ii)$ they possess a boundary magnetic field.
As such,   as $r\to \infty$,
the metric functions 
$1/F_1(r)$ 
and $F_0(r)$
behave as $r^2/L^2$,
$F_2(r)$ and $F_3(r)/v^2$ as $r^2$,
while $W(r)$ vanishes.
Also we assume that
 $a_k(r) \to c_m$ in the same limit, 
$(c_m,v)$  
being input parameters.
This implies the existence of a nonvanishing asymptotic magnetic field,
$F_{\theta \phi} \to -\frac{1}{2}c_m \sin\theta $, such that the parameter $c_m$
can be identified\footnote{Static magnetized squashed BHs in $D = 5$ Kaluza-Klein theory 
were constructed in Ref. \cite{Nedkova:2011aa}.
However, the properties of those solutions are very different
as compared to the AlAdS case.} 
with the magnetic flux at infinity
through the base space $S^2$ of the $S^1$ fibration \cite{Blazquez-Salcedo:2017cqm},
\begin{eqnarray}
\label{flux} 
\Phi_m=\frac{1}{4\pi}  \int_{S^2_\infty} F =-\frac{1}{2}c_m.
\end{eqnarray} 

One should remark that  the assumptions $(i)$ and $(ii)$
above are not related.
There exist 'magnetized' 
solutions possessing a round sphere at infinity
\cite{Blazquez-Salcedo:2017cqm}, 
and also 
vacuum BHs with a conformal boundary geometry (\ref{g0}).
However, as we shall see in Section 4,
the existence of a Killing spinor
 imposes 
that both $(i)$ and $(ii)$ should hold,
$i.e.$ 
$c_m\neq 0$ $and$ $v\neq 1$,
with a special relation between these two constants\footnote{Here we exclude
 the 
supersymmetric Gutowski-Reall BHs, which 
have $c_m=0$ and $v= 1$, 
being recovered as a limit of the new solutions in this work.}.

An expression of the solution compatible with above assumptions
can be constructed in a systematic way, 
being shared by both (extremal and nonextremal) BHs and solitons.
The first terms  in the large-$r$ expansion read\footnote{
The occurrence of $log-$terms in this asymptotic expansion makes 
the existence of an analytic solution unlikely.
Moreover, this applies also in the supersymmetric case.
} 
\begin{eqnarray}
\nonumber
&&
f(r)=1+\frac{4}{9}(1-v^2) \left( \frac{L}{r}\right)^2
+\left[
\frac{\hat \alpha}{L^4}-\frac{4}{15} \left(\frac{9c_m^2}{L^2}+(1-v^2)(4v^2-3) \right)
\log \left(\frac{r}{L}\right)
\right] \left( \frac{L}{r}\right)^4+\dots,
\\
\nonumber
&&
m(r)=1-\frac{1}{9}(1-v^2) \left( \frac{L}{r}\right)^2
+\left[
\frac{\hat \beta}{L^4}-\frac{4}{15} \left(\frac{3c_m^2}{L^2}-(1-v^2)(2v^2+1) \log \left(\frac{r}{L}\right) \right)
\right]\left( \frac{L}{r}\right)^4+\dots,
\\
&&
\label{inf}
n(r)=v^2 
            \bigg(
1+\frac{17}{9}(1-v^2) \left( \frac{L}{r}\right)^2
+\bigg[
\frac{3(\hat \alpha-\hat \beta)}{L^4}+\frac{4c_m^2}{15L^2}
+\frac{1}{405} (389-497v^2)(1-v^2)
\\
\nonumber
&&
{~~~~~~~~~~~~~~~~~~~~~~~~~~~~~~~~~~~~~}
+\frac{8}{5}\left(8-\frac{3c_m^2v^2}{L^2}-(1-v^2)(7-3v^2) \log \left(\frac{r}{L}\right) \right)
\bigg]\left( \frac{L}{r}\right)^4
       \bigg)+\dots,
\\
&&
\nonumber
w(r)=\hat j\frac{1}{r^3}+\dots,
~~~
a_0(r)= -\frac{q}{r^2}+\dots,
~~~
a_k(r)=c_m+\left(\mu-2c_m L^2v^2 \log \left(\frac{r}{L}\right)\right)\frac{1}{r^2}+\dots,
\end{eqnarray}
containing, in addition to $(v,c_m)$, the free parameters
$
\{ \hat \alpha, \hat \beta, \hat j, q, \mu\}.
$
{In principle, $f(\infty)=f_{\infty}$ is also a free parameter of the far field expansion, but we can always fix it to one by means of the scaling symmetry (\ref{scale1})}
We observe that
the first integrals 
(\ref{1stInt-gen}),
evaluated for these asymptotics,
imply the following relations:
\begin{eqnarray}
&&
c_t=-\frac{2c_m^2\lambda}{\sqrt{3}}+qv,
~~
c_W=\frac{4c_m^3\lambda}{3\sqrt{3}}-c_mq v-\frac{1}{2}\hat j v^3.
\label{c_t_c_W}
\end{eqnarray}
The CLP BHs (as well as their $\lambda \neq 1$ generalizations in 
\cite{Blazquez-Salcedo:2016rkj})
have $c_m=0$, $v=1$, in which case no $log-$terms are present in the far field asymptotics.

 \subsubsection{The near-horizon expansion} 
In this work we shall restrict our study of solutions to the region outside the (outer)
BH  horizon.
For non-extremal solutions,
this horizon resides at $r=r_H > 0$,
 where the function $f(r)$ vanishes.
There the solutions possess the following expansion:
\begin{eqnarray}
\nonumber 
f(r) &=& f_2 (r-r_H)^2 -f_2\left(\frac{1}{r_H}+\frac{3r_H}{L^2+r_H^2}\right) (r-r_H)^3+ O\left(r-r_H\right)^4,
\\
\nonumber
 m(r) &=&  m_2 (r-r_H)^2 -3m_2\left(\frac{1}{r_H}+\frac{r_H}{L^2+r_H^2}\right) (r-r_H)^3+ O\left(r-r_H\right)^4,
\\
\label{eh-expansion}
n(r)  &=&  n_2 (r-r_H)^2  -3n_2\left(\frac{1}{r_H}+\frac{r_H}{L^2+r_H^2}\right) (r-r_H)^3+ O\left(r-r_H\right)^4,
\\ 
\nonumber
\omega(r)  &=&  \omega_0 +\frac{\omega_0}{r_H}  (r-r_H)   + O\left(r-r_H\right)^2,
\\ 
\nonumber
a_{0}(r)  &=&  a_{0 }^{(0)} + a_{0 }^{(2)} (r-r_H)^2+ O\left(r-r_H\right)^3, 
\\ 
\nonumber
a_{k}(r)  &=&   a_{k }^{(0)} +   a_{k }^{(2)} (r-r_H)^2+ O\left(r-r_H\right)^3, 
\end{eqnarray}
where $\{f_2,m_2,n_2,\omega_0;a_{0}^{(0)},a_{0}^{(2)},a_{k }^{(0)},a_{k }^{(2)}\}$ are free coefficients.


{In the quasi-isotropic coordinates we are using,
the horizon of extremal black holes is located at $r_H=0$.
As a result, the behavior of the functions near the horizon changes with respect to the non-extremal case,
with the occurrence\footnote{This is a consequence of the metric gauge choice used in this work.
It is worth to mention that this feature occurs already for the CLP solution,
when written within the metric Ansatz (\ref{param}).} of non-integer powers 
 of $r$.
The first terms in the near-horizon
expression of the solutions are}
\begin{eqnarray}
\nonumber
&&
f(r) = \bar f_4 r^{2k} + \bar f_4^{(s)} r^{3k} + \dots, ~~ 
m(r) = \bar m_2 r^{2k-2} +\bar m_2^{(s)} r^{3k-2} +\dots, ~~
\\
\label{nh-extremal}
&&
n(r) =\bar n_2 r^{2k-2} +\bar n_2^{(s)} r^{3k-2} +\dots, ~~
\omega(r) = \omega_{0}^{(1)} r +\omega_{0}^{(2)} r^{k+1} +  \dots,  
\\
&&
\nonumber
a_0(r) =a_{0}^{(0)} + \bar a_{0 }^{(2)} r^{k}\dots,  ~~~
a_{k}(r) = a_{k}^{(0)} + \bar a_{k }^{(2)} r^{k}\dots, 
\end{eqnarray}
{ 
with $k>2$ a number fixed by numerics.
The coefficients in  near-horizon solutions are determined order by order by
 $\{\bar f_4,\bar m_2, \omega_{0}^{(1)};a_{0}^{(0)},a_{0}^{(2)},a_{k }^{(0)},a_{k }^{(2)}\}$,
the corresponding expressions 
for 
$\{\bar n_2, \bar n_2^{(s)}, \bar f_4^{(s)}, \bar m_2^{(s)}, \omega_{0}^{(2)}, etc...\}$
being, however, very complicated.
Let us also notice that 
the near-horizon
expression of the solutions
takes a simpler form 
when written in terms of a 
 new radial coordinate 
$x=r^{k}$.
As such, the existence of squashed $AdS_2\times S^3$ solutions
(described by the leading order terms in (\ref{nh-extremal}))
becomes transparent.
They form a particular class of the  EMCS-$attractors$ 
 discussed in a 
more general context in Ref. \cite{Blazquez-Salcedo:2016rkj}.
 }
%
%
%
%

\subsubsection{Solitons: the small$-r$ expansion} 
As a new feature in contrast to the CLP case ($c_m=0$, $v=1$), 
the zero horizon size limit of the generic solutions
is nontrivial. 
For a given $v$,
this corresponds to a one-parameter family of spinning charged solitons
with nonzero global charges. 
Such solutions possess no horizon, 
while the size of both parts of the $S^3$-sector of the metric shrinks to zero\footnote{This 
contrasts with the case of topological solitons which exist inside the general solution in \cite{Cvetic:2005zi}.} 
as $r\to 0$.

A  small-$r$ approximate form of the solitonic solutions 
can be constructed 
as a power series in $r$, 
being
compatible with the assumption of regularity at $r=0$.
The first terms in this expansion
are
\begin{eqnarray}
\nonumber
&&
f(r) = f_0 + \left(\frac{m_0-f_0}{L^2}+\frac{4 u^2 f_0^2}{3m_0} \right)r^2 +\dots, 
~
m(r) =m_0+\tilde{m}_2 r^2 +\dots, ~ 
\omega(r)=w_1 r+\frac{8u^3f_0^{5/2}\lambda}{3\sqrt{3}m_0^2}r^3+\dots, 
\\
\nonumber
&&
n(r) = m_0+ \left(\frac{3m_0(m_0-f_0)}{f_0 L^2}-\tilde{m}_2+\frac{4 u^2 f_0}{3} \right)r^2 +\dots,~
a_{0}(r) =  v_0-\left(\frac{2u^2 f_0^{3/2}\lambda}{\sqrt{3} m_0}+u w_1 \right)r^2+\dots,  
\\
\label{zero}
&&
a_{k}(r) = u r^2+ \frac{u}{9f_0L^2 m_0}\bigg( 4u^2f_0^2L^2(1+2\lambda^2)+3(4m_0^2-3f_0(2m_0+L^2M_2)) \bigg)r^4+\dots, 
\end{eqnarray}
with the free parameters
$
\{
 f_0, m_0,\tilde{m}_2, w_1; u,v_0
  \}.
	$

Finally, let us remark that
after  evaluating the first integrals (\ref{1stInt-gen})
for the above asymptotics, 
one finds that the constants $c_W, c_t$ vanish for solitons,
\begin{eqnarray}
\label{cwct-sol}
c_W=c_t=0.
\end{eqnarray}

\subsection{Physical quantities} 

\subsubsection{Event-horizon quantities} 
The horizon is a squashed $S^3$ sphere,
with different sizes for the 
$S^1$ and the round $S^2$  
parts of it.
There the Killing vector 
\begin{eqnarray}
\nonumber
\zeta = \partial_t + 2\Omega_H   \partial_{\psi} 
\end{eqnarray}
becomes null  and  
orthogonal to the other Killing vectors on it.
For non-extremal BHs, 
the induced horizon metric reads
\begin{equation} 
\label{horizon-metric}
d\sigma_H^2=\frac{r_H^2 }{4f_2}
\left[
m_2(d\theta^2+\sin^2\theta d\phi^2) +n_2(d\psi+\cos\theta d\phi)^2
\right],
\end{equation} 
which leads us to define the horizon deformation parameter 
\begin{eqnarray}
\label{deform}
\varepsilon^2 =\frac{n(r)}{m(r)}\bigg|_{r=r_H}=\frac{n_2}{m_2},
\end{eqnarray}  
with $m_2$, $n_2$ and $f_2$ the coefficients in (\ref{eh-expansion}).
The area of the horizon $A_H$, the Hawking temperature $T_H$ 
and the horizon angular velocity $\Omega_H$
of these solutions are given by
\begin{equation} 
A_H= 
2\pi^2r_H^3 \frac{m_2}{f_2}\sqrt{\frac{n_2}{f_2}},~~~~
T_H= \frac{1}{2\pi}\left(1+\frac{r_H^2}{L^2} \right)\frac{f_2}{\sqrt{m_2}},~~~~
\Omega_H=\frac{\omega_0}{r_H}.
\end{equation}   
The horizon electrostatic potential $\Phi_H$ 
as measured in a co-rotating frame on the horizon is
\begin{equation}
\Phi_H 
= a_{0 }^{(0)}+\Omega_H a_{k}^{(0)}.
\label{Phi} 
\end{equation} 
In the extremal case,
the induced horizon metric is
\begin{equation} 
d\sigma_H^2=\frac{\bar m_2}{4 \bar f_4}
 (d\theta^2+\sin^2\theta d\phi^2)
+ \frac{\bar n_2}{4 \bar f_4}(d\psi+\cos\theta d\phi)^2,
\end{equation} 
while the horizon quantities are
\begin{equation} 
A_H= 
2\pi^2 \frac{\bar m_2}{\bar f_4}\sqrt{\frac{\bar n_2}{\bar f_4}},~~
\Omega_H=\omega_0^{(1)},~~
\Phi_H  
= a_{0 }^{(0)}+\Omega_H a_{k}^{(0)},
\end{equation} 
in terms of the constants
which enter the near-horizon expansion 
(\ref{nh-extremal}).

\subsubsection{Holographic renormalization and global charges}

The global charges
of the  solutions 
are encoded 
in the constants 
$\hat \alpha$, 
$\hat \beta$, 
$\hat j$
and $q$
which enter the large$-r$ expansion of the solutions (\ref{inf}).
In computing them,  
we use the holographic renormalization of the $D=5$
EMCS system as discussed $e.g.$ in Ref. \cite{Sahoo:2010sp}.
 The first step is to rewrite the
solution in the
standard
 Graham-Fefferman coordinate system 
\cite{FG},
by defining a new  radial coordinate,
\begin{eqnarray}
&&
\label{def-x}
x=r+\frac{L^2}{18 r}\left(7-\frac{5}{2}v^2\right)
+\bigg(
\frac{\hat \alpha-\hat \beta}{8}
-\frac{c_m^2L^2}{20}
+\frac{L^4}{2160}(83v^4-94v^2-124)
\\
\nonumber
&&
{~~~~~~~~~~~~~~~~~~~~~~~~}
+\frac{1}{15}(3c_m^2L^2-L^4(1-v^2)(1-3v^2))\log \left(\frac{L}{r}\right)
\bigg)\frac{1}{r^3}+\dots~.
\end{eqnarray}
This results in an equivalent asymptotic form of the line element 
\begin{eqnarray}
\label{metric-x}
ds^2=\frac{dx^2}{\frac{x^2}{L^2}}+\frac{x^2}{L^2} 
\left [ 
g_{ab}^{(0)}+\frac{1}{x^2}g_{ab}^{(2)}
+\frac{1}{x^4}\left(g_{ab}^{(4)}+h_{ab}^{(4)}\log\left(\frac{x^2}{L^2}\right)\right)
+\dots
\right ]dy^a dy^b,
\end{eqnarray}
and of the gauge field 
\begin{eqnarray}
A_{a}=A_{a}^{(0)}+\frac{1}{x^2}\left(A_{a}^{(2)}+ B_{a}^{(2)}\log\left(\frac{x^2}{L^2}\right) \right)+\dots~~.
\end{eqnarray}
The boundary metric tensor
$h_{ab}$ 
is found by taking $x=x_0$ in (\ref{metric-x}),
with $x_0$ sent to infinity in the final relations.
Also, 
$g^{(0)}$ and $A^{(0)}$ are imposed as boundary conditions,
providing the background metric and the external
gauge potential for the four dimensional dual theory. 
For the solutions in this work one takes
\begin{eqnarray}
ds^2=g_{ab}^{(0)}dy^a dy^b=-dt^2+ \frac{L^2}{4}
(\sigma_1^2+\sigma_2^2+v^2 \sigma_3^2)~,~~
A^{(0)}_a dy^a=\frac{1}{2}c_m \sigma_3.
\end{eqnarray}
The terms $g^{(2)}$, $h^{(4)}$  and $B^{(2)}$
are fixed by the equations of motion,
while
the terms 
$g^{(4)}$ 
and
$A^{(2)}$
 are not determined  by the field equations.
Their expression can easily be found from (\ref{inf})
together with (\ref{def-x}), 
in practice they being extracted from the numerical output.
 
In the next step one defines a regularized
 total action
$I_{tot}{=}I{+}I_{\rm ct}$,
which
is the sum of (\ref{EMCSac})
and a  counterterm
 $I_{ct}$,
with 
\cite{Cassani:2014zwa},
\cite{Sahoo:2010sp},
\cite{Bernamonti:2007bu}
\begin{eqnarray}
\label{ct}
I_{\rm ct}=-\frac{1}{8 \pi } \int_{\partial {\cal M}}d^{4}\mathrm{x}\sqrt{-h}\Biggl[
\frac{3}{L}+\frac{L}{4}{\rm R}
+\log \left(\frac{x}{L} \right) 
\frac{L^3}{8}
\left(
{\rm R}_{ab}{\rm R}^{ab}-\frac{1}{3}{\rm R}^3-\frac{4}{L^2}{\rm F}_{ab}^{(0)}{\rm F}^{(0)ab}
\right)
\Bigg],
\end{eqnarray}
where 
$\rm{R}_{abcd}$,
$\rm{R}_{ab}$,  
${\rm E}_{ab}$
denote the
Riemann, Ricci and Einstein tensors, respectively, $\rm{R}$  is the Ricci scalar for the boundary metric $h$
and $\rm{F}^{(0)}=dA^{(0)}$ is the boundary U(1) field.
{Note that in equation (\ref{ct}) we are only considering the counterterms that cancel the power-law and logarithmic divergences at the boundary, which provide finite expressions for the charges.
 However, additional counterterms can be added to the action \cite{Cassani:2014zwa} (see also the discussion in \cite{Genolini:2016ecx} for AlAdS solutions in supergravity). 
These additional counterterms in general introduce an ambiguity in the definition of the charges, but they can be useful in order to restore some lost symmetries at the boundary.}

As usual in AdS/CFT, one defines
 the expectation value of the stress tensor and current in the dual 
theory as 
\begin{eqnarray}
<T_{ab} > =-\frac{2}{\sqrt{-g^{(0)}}} \frac{\delta I_{tot}}{ \delta g^{(0)ab}}~,
~~~
<J^{a}  >= \frac{1}{\sqrt{-g^{(0)}}} \frac{\delta I_{tot}}{ \delta A^{(0)}_a}~,
\end{eqnarray} 
which results in  
\begin{eqnarray}
\label{s12}
&&
<T_{ab} >=  
-\frac{1}{8\pi }
\lim_{x\to \infty}
 \left( \frac{x}{L}\right)^4
\bigg\{
(K_{ab}-Kh_{ab}+\frac{3}{L}h_{ab}-\frac{L}{2} E_{ab})
\\
&&
\nonumber
{~~~~~~~~~~~~~~~~~~~~~~~~~~}
-\log \left(\frac{x}{L} \right)
\frac{L^3}{2}
\bigg[
\frac{1}{12}h_{ab}{\rm R}^2-\frac{1}{4}h_{ab}{\rm R}_{cd} {\rm R}^{cd}-\frac{1}{3} {\rm R}{\rm R}_{ab}
+ {\rm R}_{acbd} {\rm R}^{cd}
\\
&&
\nonumber
{~~~~~~~~~~~~~~~~~~~~~~~~~~}
+\frac{1}{2}\nabla^2  {\rm R}_{ab}
-\frac{1}{12}(h_{ab}\nabla^2+2\nabla_a \nabla_b){\rm R}
-\frac{4}{L^2}
\left(
{\rm F}_{ac}^{(0)}{\rm F}_{b}^{(0)c}-\frac{1}{4}h_{ab}{\rm F}^{(0)2}
\right)
\bigg]
\bigg\}
,
\end{eqnarray}
and
\begin{eqnarray}
\label{js13}
<J^{a} >=\frac{1}{8\pi} 
\left[
g^{(0)ab}
(
A_{b}^{(2)}+B_b^{(2)}
)
-\lambda \epsilon^{abcd}A_{b}^{(0)}F_{cd}^{(0)}
\right]~.
\end{eqnarray}

Then provided that the boundary geometry has an isometry generated by a
Killing vector $\xi$, a conserved charge
\begin{eqnarray}
{\mathfrak Q}_{\xi }=\int_{\Sigma }d^{3}S ~u^a\xi^{b}<T_{ab}> ,
\label{charge-gen}
\end{eqnarray}
can be associated with a closed surface $\Sigma $ \cite{Balasubramanian:1999re},
with $u_a=\delta_{a}^t$ a unit timelike vector
{(in general the expression (\ref{charge-gen}) will contain a contribution from the flux $<J^{a} >$ \cite{Genolini:2016ecx}, but this extra-term is not relevant for the particular solutions we are considering)}. 
The mass/energy $M$
is the conserved  charge associated with 
 $\xi =\partial /\partial t$; 
there is also an angular momentum $J$ associated
with the Killing vector  $\partial/\partial \psi$.
A similar expression holds for the electric charge $Q$
\begin{eqnarray}
\label{Qn}
Q=\int_{\Sigma }d^{3}S ~u^{a}<J_{a}>.
\end{eqnarray}

It is straightforward to apply this formalism to the solutions
with the asymptotics (\ref{inf}) (together with (\ref{def-x})). 
The nonvanishing components of the
boundary stress tensor (\ref{js13})   are
\begin{eqnarray}
&&
\nonumber
 <T_\theta^\theta>=<T_\phi^\phi >=
\frac{v}{8\pi L} 
\left(
 \frac{1}{8}
+\frac{(1-v^2)(3479-11057v^2)}{3240}
-\frac{5(\hat\alpha-\hat\beta)}{2L^4}-\frac{32 c_m^2}{15L^2}
\right),~
\\
\label{dual-tik}
&&
<T_\psi^\psi >=
\frac{1}{8\pi L}
\left(
 \frac{1}{8}
-\frac{(1-v^2)(2537-15911 v^2)}{3240}
+\frac{7\hat\alpha-11\hat\beta }{2L^4}+\frac{2c_m^2}{ 5L^2}
\right),~~
\\
\nonumber
&&
<T_\phi^\psi >= \cos\theta \big (<T_\psi^\psi > -<T_\phi^\phi > \big),~
<T_\psi^t >=\frac{1}{\cos \theta}<T_\phi^t >=-\frac{1}{4}L^2v^2<T_t^\psi > 
=\frac{\hat j}{8\pi L^3}  ,  
\\
\nonumber
&&
<T_t^t >=
\frac{1}{8\pi L}
\left(
 -\frac{3}{8}
-\frac{(1-v^2)(101-1883 v^2)}{3240}
+\frac{3\hat\alpha+\hat\beta }{2L^4}-\frac{2c_m^2}{ 15L^2}
\right).
\end{eqnarray}
while the boundary current is
\begin{eqnarray}
<J^{a} >= - \frac{1}{2\pi v}\left(q v-\frac{4\lambda}{3\sqrt{3}}c_m^2 \right)\delta_0^a.
\label{Ja-gen}
\end{eqnarray}
After replacing these expressions in 
(\ref{charge-gen}) and (\ref{Qn})
one finds the global charges of the solutions
\begin{eqnarray}
\label{MJ}
&&
M=\frac{\pi v}{8}
\left(
-\frac{(3\hat \alpha+\hat \beta)}{L^2}
+\frac{4c_m^2 }{15}
+\frac{3L^2}{4}
+\frac{(1-v^2)L^2}{1620}(101-1883v^2)
\right),
~
J=-\frac{\hat j \pi v^3}{4}~,~{~~}
\\
\label{R-charge}
&&
Q 
 =-\pi \left(q v-\frac{4\lambda}{3\sqrt{3}}c_m^2 \right).
\end{eqnarray}
Note also that 
 the total derivative structure of the MCS equations
implies the existence of
  a conserved  {\it Page charge} 
\begin{eqnarray}
\nonumber
&&
Q^{(Page)}=-\frac{1}{4\pi }\int_{S^3_{\infty} } (*_5 F+\frac{2\lambda}{\sqrt{3}}A\wedge F)
=\frac{1}{4\pi }\int_{S^3_{\infty} }d\Sigma_3
(\sqrt{-g}F^{rt}-\frac{\lambda}{\sqrt{3}}\varepsilon^{abc}A_{a}F_{bc})
\\
\label{Page-charge}
&&
{~~~~~~~~~~~~~~~~~~~~~~~~~~~~~~~~~~~~~~~~~~}
=-\pi \left(q v-\frac{2\lambda}{\sqrt{3}}c_m^2 \right).
\end{eqnarray}   
The Page charge is proportional to the conserved charge $c_t$ of (\ref{c_t_c_W}), with $Q^{(Page)}=-\pi c_t$.
In the standard  $c_m=0$ case,
the  (holographic) electric charge $Q$ and $Q^{(Page)}$ are the same.
However, (\ref{R-charge}) and (\ref{Page-charge}) 
do not coincide 
for solutions with a boundary magnetic field
(and, in fact, the Page-charge vanishes for solitons, while $Q\neq 0$).

One also notices that the stress tensor 
(\ref{dual-tik}) 
is not traceless,
\begin{eqnarray}  
\label{r3}
<T _{a}^a> =
\frac{(1-v^2)^2}{6\pi L} 
-\frac{c_m^2}{2L^3 \pi},
\end{eqnarray}
its trace consisting of two parts.
The first part is due to  the conformal anomaly of the boundary CFT
coming from the background curvature
\cite{Skenderis:2000in},
\cite{deHaro:2000vlm}
\begin{eqnarray} 
{\cal A}_{(g)}=-\frac{ L^3}{8\pi }\left(-\frac{1}{8}
\mathsf{R}_{ab}\mathsf{R}^{ab}+\frac{1}{24}\mathsf{R}^2\right).
\end{eqnarray}
The  part of (\ref{r3})  proportional
to $c_m^2$ results from the coupling of the CFT to a background gauge field
\cite{Taylor:2000xw}
\begin{eqnarray} 
{\cal A}_{(em)}=-\frac{L}{16 \pi } {\rm F}_{ab}{\rm F}^{ab}.
\end{eqnarray}
Moreover, one can verify that the following Ward identities are
satisfied \cite{Sahoo:2010sp}
\begin{eqnarray} 
\label{ward}
&&\nabla_b<T^{ab}>=F^{(0)ab}<J_b>-\frac{\lambda}{16\pi \sqrt{-g^{(0)}}}
\epsilon^{bcde}F_{b}^{(0)a}A_{c}^{(0)}F_{de}^{(0)},
\\
\nonumber
&&
\nabla_a<J^{a}>=\frac{\lambda}{64 \pi \sqrt{-g^{(0)}}}
\epsilon^{abcd}F_{ab}^{(0)}F_{cd}^{(0)}.
\end{eqnarray}

\section{Nonsupersymmetric solutions} 

\subsection{Numerical procedure }

The equations (\ref{eqs})
do not seem to possess closed-form solutions  
with $v\neq 1$ and/or $c_m\neq 0$.
Therefore all new configurations reported in this work
are found numerically. 
The methods
we have used are similar to those used in 
\cite{Blazquez-Salcedo:2015kja},
\cite{Kunz:2006eh},
\cite{Kunz:2006yp},
\cite{Kunz:2007jq}
to find other
 numerical 
 solutions with equal-magnitude angular momenta in $D=5$ EM(CS) theory.

By making use of all the available symmetries, 
the set (\ref{eqs}) of field equations can be reduced
to a
 system of four second-order  differential equations (ODEs) for the functions 
$(f,~m,~n,~a_k)$
together with  two first-order ODEs for $(\omega,a_0)$.
A relation between first-order derivatives of the functions 
$f$, $n$, $m$, $\omega$ and $a_k$ can be used as a constraint,
which the numerical solutions must satisfy with a given precision.

In our numerical scheme,
the input parameters are: $i)$ the AdS length scale $L$,
$ii)$ the magnetic parameter $c_m$, $iii)$ the boundary squashing $v$,
$iv)$ the constants $\hat j,q$ in the far field asymptotics, and, for
non-extremal BHs, $v)$ the event horizon radius $r_H$.  
The event horizon data 
 and the coefficients at infinity $\alpha$, $\beta$ and $\mu$
are read from the numerical output.
In practice, we fix the AdS length scale $L= 1$ and construct families of solutions by varying
the other input parameters.

The equations are solved by 
using a professional software package 
\cite{COLSYS}
which employs a collocation method for boundary-value ordinary differential equations
and a  damped 
Newton method of quasi-linearization.
The number of mesh points used in our calculation
was  around $10^4$, distributed non-equidistantly on 
$x$, where $x=1-r_H/r$ is a compactified radial coordinate
employed in the BH non-extremal case;
for solitons and extremal BH solutions one takes $x=r/(1+r)$
(with $0\leq x\leq 1$ in both cases).
One should remark that the computation of 
global charges for these solutions is a nontrivial problem
which requires a very good numerical accuracy, since the coefficients 
$\hat \alpha$,
$\hat \beta$, 
 appear as subleading terms in the
asymptotic expansion (\ref{inf}).
The typical relative accuracy of the solutions here 
is around $10^{-10}$.

Finally, let us mention that  
all solutions reported in this work are 
regular
on the horizon or outside of it\footnote{
For example, the Ricci or Kretschmann scalars  were monitored for most of the solutions
and we did  not find any sign of a singular behaviour.}. 
Also, 
since
$
g^{tt}=-f(r)<0~ 
$
for any $r>r_H$, {while the metric functions $m(r)$ and $n(r)$ remain strictly positive
(in particular $v^2>0)$,
the solutions are also free of closed timelike (or null) curves, 
 $t$ being a time function
(see the general discussion in
\cite{Cvetic:2005zi}, which covers also the framework here)}.
Similar to the well known CLP case, the generic BHs possess, however, an ergoregion
located between the horizon and the ergosurface $r=r_c$ (with $g_{tt}(r_c)=0$).

\subsection{Black holes}

In the generic case,
given $(v,c_m)$, 
the solutions possess three independent charges 
$M,J$ and $Q$.
Therefore
finding their domain of existence 
is a considerable task
which is beyond the purposes of this paper. 
Instead, we shall analyze several particular classes  of solutions, 
hoping that they capture a part of the general pattern.
%

\subsubsection{$v=1$: a globally AdS background} 
Let us start with the simplest case of solutions possessing
a round $S^3$-part in the boundary metric.
For $c_m=0$, these are the
Cveti\v c, L\"u and Pope (CLP)
 BHs \cite{Cvetic:2004hs}.
	However, as found 
 in the recent work
 \cite{Blazquez-Salcedo:2017cqm},
they possess  a generalization
with  
a nonvanishing magnetic field in the far field,
which can also be described within the framework in Section 2. 

The results in
 \cite{Blazquez-Salcedo:2017cqm}
 show that
the qualitative behaviour of the BHs with small $|c_m|$ resembles that of the unmagnetized CLP case.
However, a different picture is found for large enough values of  $c_m$,
with
a monotonic behaviour of mass and horizon area as a function of temperature
(also the solutions do not appear to possess 
an upper bound on $|c_m|$).
In contrast to the CLP case,
one finds BHs which have $J=0$
but still rotate in the bulk,
with a nonvanishing angular momentum density, $T_{\psi}^t\neq 0$.
Extremal BHs with $c_m\neq 0$ exist as well,
possessing generically a nonzero 
horizon area.
Moreover, the BH solutions 
satisfying a certain relation between $J,Q$ and $c_m$
do not  trivialize as $r_H\to 0$,
becoming solitonic deformations of the AdS background.

\subsubsection{$v\neq1$: static, vacuum configurations }
As expected, the $v=1$
solutions in 
\cite{Blazquez-Salcedo:2017cqm}
possess generalizations with  a squashed sphere at infinity,
and
new qualitative features occur as well.  
To simplify the problem, let us consider first 
the static, vacuum configurations,
in which case it is possible to 
perform a systematic study of the solutions
together with their relevant limits.
These BHs are found within a consistent truncation of the general Ansatz 
(\ref{metric}), (\ref{U1})
with $W=0$, $a_k=a_0=0$.
Our numerical results clearly indicate the existence of  $v\neq 1$
(static, vacuum)
BH solutions 
of the equations (\ref{eqs}),
smoothly
interpolating between the asymptotics (\ref{eh-expansion}) and (\ref{inf}).

These solutions are most naturally interpreted as squashed\footnote{
Their basic properties have been discussed in a different context
in
\cite{Brihaye:2009dm},
\cite{Murata:2009jt}.}
BHs,
being in some sense 
the AdS counterparts of the ($\Lambda=0$) Kaluza-Klein solutions in 
\cite{Ishihara:2005dp}.
Such configurations exist for an arbitrary value
 of the horizon size,
without an extremal limit.
	In fact, their thermodynamics is similar to the one of the Schwarzschild-AdS BHs
	\cite{Brihaye:2009dm}.
For any $v$,	their temperature is bounded from below, 
and one finds two branches consisting of small (unstable) and large (stable)  BHs. 
	As $r_H\to 0$,
	a singularity-free solitonic configuration is approached,
the size of both parts of
the $S^3$-sector of the metric shrinking to zero.
The properties of these solutions are discussed in the next subsection.

Let us now explore\footnote{
In understanding the limiting-$v$ behaviour,
some useful hints are provided by
the nutty-instanton toy model in Appendix A.} 
the behaviour of the squashed BHs as a function of $v$.
For a given value of $r_H$,
the parameter $v$ can take
arbitrary values.
Apparently, as $v\to 0$,
 the size in the far field of the U(1) fiber over $S^2$ 
shrinks to zero,
such that this limit does not seem to be well defined.
However, this is not the case.
Following the discussion in the Appendix A,
we consider an equivalent form of (\ref{metric})
which absorbs the $v^2$ factor in the asymptotic value of $n(r)$
via a redefinition of $\psi$, 
with
\begin{eqnarray}
\label{metric1}
\nonumber
ds^2 =
\frac{1}{f(r)}
\left[
m(r)
\left(\frac{dr^2}{1+\frac{r^2}{L^2}}
  + \frac{1}{4} r^2 
  (d\theta^2+\sin^2\theta d\phi^2 )
	\right)
 +  \frac{1}{4} n(r)r^2\big(d \bar {\psi} + v \cos\theta d \phi \big)^2
	\right]
	\\
\label{metric-new}
{~~~~~~~~~~~~~~~~~~~}
	-f(r)(1+\frac{r^2}{L^2})dt^2
,
\end{eqnarray} 
(where $n(r) \rightarrow v^2 n(r)$, $\psi\rightarrow\bar\psi/v$)
in which case $n(r)\to 1$ as $r\to \infty$.
Also, the numerics shows that for small $v$,
the size of the $S^1$-circle in the horizon metric (\ref{horizon-metric})
becomes proportional to $v$ 
\cite{Blazquez-Salcedo:2017kig}.
As such, the limit $v\to 0$
is smooth for the $\bar \psi$-parametrization,
and describes the 
AdS$_5$
black strings and vortices,
originally found in 
\cite{Copsey:2006br}
for a different metric Ansatz. 
The black strings'
horizon metric reads
\begin{eqnarray}
\label{metric1n} 
ds^2 =
\frac{r_H^2}{4f_2}
\left[
m_2
  (d\theta^2+\sin^2\theta d\phi^2 )
 +  n_2  d \bar {\psi} ^2
	\right ] ,
\end{eqnarray} 
the horizon topology being $S^2\times S^1$,
while
the conformal boundary metric is the product of time and 
a line element of the form (\ref{metric1n}).
Also, the solutions possess a nontrivial $r_H\to 0$ limit
describing AdS vortices.
We note that for both black strings and vortices,
the range of $\bar \psi$ (usually denoted as $z$-coordinate in the literature)
is not fixed a priori\footnote{However, for black strings,
 the Gregory-Laflamme instability 
\cite{Gregory:1993vy}
implies the existence of a critical periodicity of the 
$\bar \psi$-coordinate for a given value of the mass \cite{Brihaye:2007ju}.}.

No upper bound seems to exist for the value of $v$, although the numerical
integration becomes more difficult as we increase this parameter,
with the mass $M$ diverging as $v\to \infty$.
However, a careful analysis of this limit reveals the existence of a different solution
of the field equations. 
Following again the discussion in the nutty instanton case (see Appendix A),
we define the scaled coordinates 
\begin{eqnarray}
\label{scale1i}
r=\lambda \bar r,~~\theta= \frac{\Theta}{\lambda},~~\psi= -\frac{\Psi}{v\lambda}-\phi,
\end{eqnarray}
together with
\begin{eqnarray}
\label{scale2}
v=\lambda N~.
\end{eqnarray} 
Then as $\lambda\to \infty$
 the  line element (\ref{metric}) (with $W=0$)
becomes
\begin{eqnarray}
\label{twist1}
ds^2=\bar F_1(\bar r)d\bar r^2+\bar F_2(\bar r)\frac{1}{4}(d\Theta^2+ \Theta^2 d\phi^2)
+\bar F_3(\bar r)\frac{1}{4}\left(d\Psi+  2 N  \left(\frac{\Theta}{2}\right)^2 d\phi \right)^2
-\bar F_0(\bar r)dt^2~,
\end{eqnarray} 
which corresponds to a 'twisted' black brane configuration. 
The horizon is located again at some $\bar r=\bar r_H$, with an induced horizon geometry 
\begin{eqnarray}
\label{h-twist1}
ds^2= \bar F_2(\bar r_H)\frac{1}{4}(d\Theta^2+ \Theta^2 d\phi^2)
+\bar F_3(\bar r_H)\frac{1}{4}\left(d\Psi+  2 N  \left(\frac{\Theta}{2}\right)^2 d\phi \right)^2.
\end{eqnarray} 
In the absence of an analytical solution, the expression of 
$F_i(\bar r)$ is found numerically\footnote{Note that the limit 
$N=0$
 corresponds to a Schwarzschild black brane,
with
$F_0=1/F_1=\bar r^2/L^2-\bar r_H^2/L^2$,
$F_2=F_3=\bar r^2$.
}. 
We also remark  that they do not possess a  solitonic limit.
Their conformal boundary metric is\footnote{
It is interesting to note that 
(\ref{twist2}) corresponds to
an analytical continuation of the 
Som-Raychaudhuri spacetime \cite{Som}.
}
\begin{eqnarray}
\label{twist2}
ds^2=
\frac{1}{4}L^2
\bigg[
d\Theta^2+ \Theta^2 d\phi^2 
+ \left(d\Psi+  2 N  \left(\frac{\Theta}{2}\right)^2 d\phi \right)^2
\bigg]
-dt^2~,
\end{eqnarray} 
(with $0\leq \Theta<\infty$, $0\leq \phi<2 \pi$
and an arbitrary periodicity for $\Psi$).
Although a $t=const.$
surface is
topologically a direct product of $\Psi$ and the $(\Theta,\phi)$- plane, 
the product is "twisted" (or warped), and the boundary
is not flat (its Ricci scalar is proportional to $N^2$).  
More details of the limiting solutions can be found in Appendix B.

\begin{figure}[t]
\centering
\subfigure[]{
\includegraphics[scale=0.28,angle=-90]{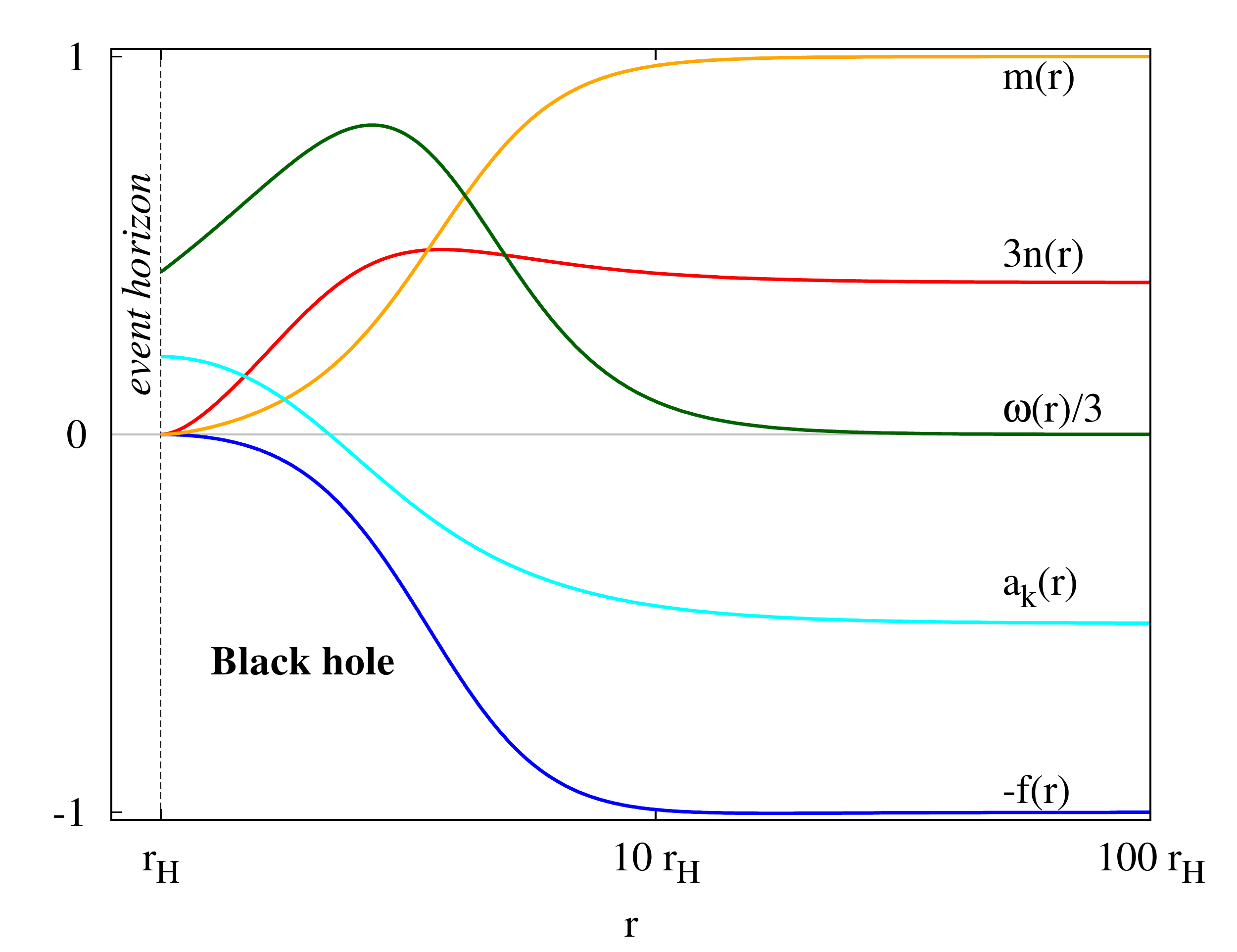}
\label{profiles_BH}}
\subfigure[]{
\includegraphics[scale=0.28,angle=-90]{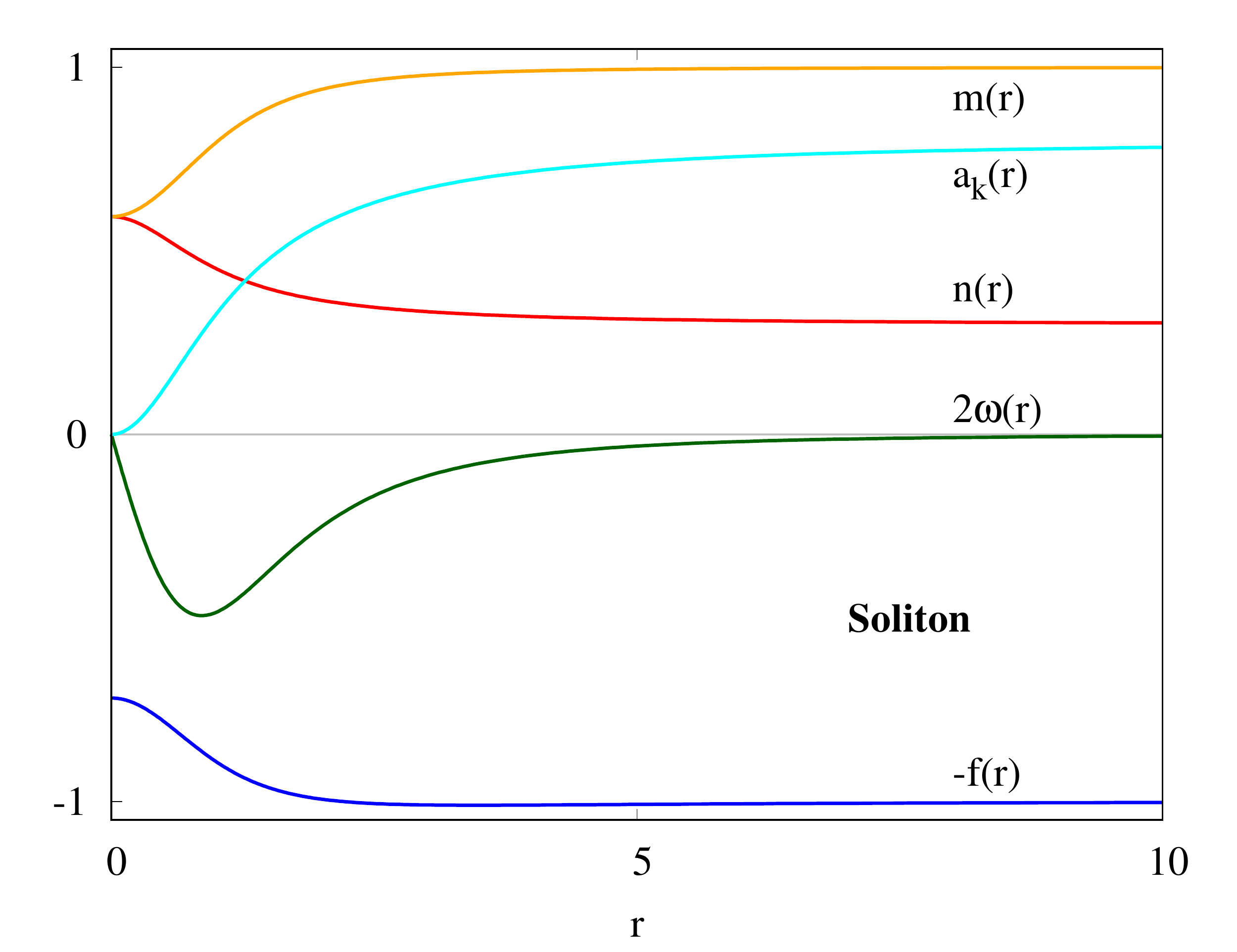}
\label{profiles_soliton}}
\caption{(a) Metric and gauge functions are shown for a typical 
squashed magnetized black hole.
The input parameters here are 
$L=1$, $r_H=0.492$, $Q=-3.19$, $J=-1.33$, $c_m=-0.5$ and $v^2=0.133$. 
(b) Same functions for a squashed magnetized soliton with parameters $L=1$, $c_m=0.8$ and $v^2=0.3$.}
\end{figure}

\subsubsection{$v\neq1$: the generic case}

Increasing the complexity of the solutions,
we first notice that the (vacuum, static) BHs of the previous subsection
possess  spinning generalizations.  
They are interpreted as squashed rotating BHs, their 
 thermodynamics being
similar to that of the ($v=1$) Myers-Perry-AdS
BHs with two equal angular momenta.

Static, electrically charged 
BHs with $v\neq 1$
exist as well.
These solutions are  generalizations of the 
Reissner-Nordstr\"om-AdS
BHs,
 with a squashed horizon (and a squashed sphere at infinity)
and possess similar thermal properties.

What 
$J\neq 0$  
or
$Q\neq 0$
brings new  
is the absence of a smooth particle-like solitonic limit\footnote{
This feature can  be understood from the 
results in Section 2.
The constants $c_t$ and $c_W$
necessarily
vanish for solitons.
Then the first integrals (\ref{1stInt-gen}),
imply that $W=0$  and also $a_0=0$.
}
(we recall $c_m=0)$.
Instead, one notices the existence of extremal BHs
with a nonzero  horizon area,
which are smooth on and outside the horizon.

Moreover, as expected, spinning 
$v\neq1$ solutions with 
$Q\neq 0$ and $c_m=0$
exist as well.
 They can be interpreted as squashed counterparts of the CLP
BHs and appear to share all their basic properties. 
Again, these unmagnetized solutions do not possess a smooth solitonic limit.
Also, their behaviour in terms of 
the squashing parameter
$v$
is similar to the one in the vacuum static case. 
In particular, the limit $v=0$
describes a generalization of the AdS black strings in
\cite{Copsey:2006br}
with a nonzero electromagnetic field and a momentum along the $\psi$-direction.

\medskip

However, these limits are in some sense less interesting, since, as we shall see in the next
Section, they do not allow for supersymmetric solutions.
Therefore let us now consider the general case
of spinning, magnetized solutions with a squashed sphere at infinity. 
As a general remark, our numerical results show that they share a number of basic properties of the 
$v=1$ solutions with $c_m\neq 0$  in  \cite{Blazquez-Salcedo:2017cqm}.
For example, for large enough $J,Q$, the magnetic field induces
 subleading effects only, and we recover the general pattern found for the CLP BHs.
Also, these generic solutions do not allow for a smooth black string limit as $v\to 0$.
However, the limit $v\to \infty$ is well defined, describing   
(after a suitable rescaling)
a new family of twisted charged black branes.
Although the asymptotics of these solutions is very similar to 
(\ref{eh-expansion}), (\ref{inf}) 
they possess a number of distinct properties,
see the discussion in Appendix B.

The typical profiles of the metric and gauge functions of squashed 
magnetized black holes are presented in Fig. \ref{profiles_BH}, 
for the a typical magnetized spinning BH
(note the absence of nodes in the profile of the magnetic potential, a feature which holds for 
all configurations in this work, including the solitonic ones). 
%

\begin{figure}[t]
\centering
\subfigure[]{\includegraphics[scale=0.28,angle=-90]{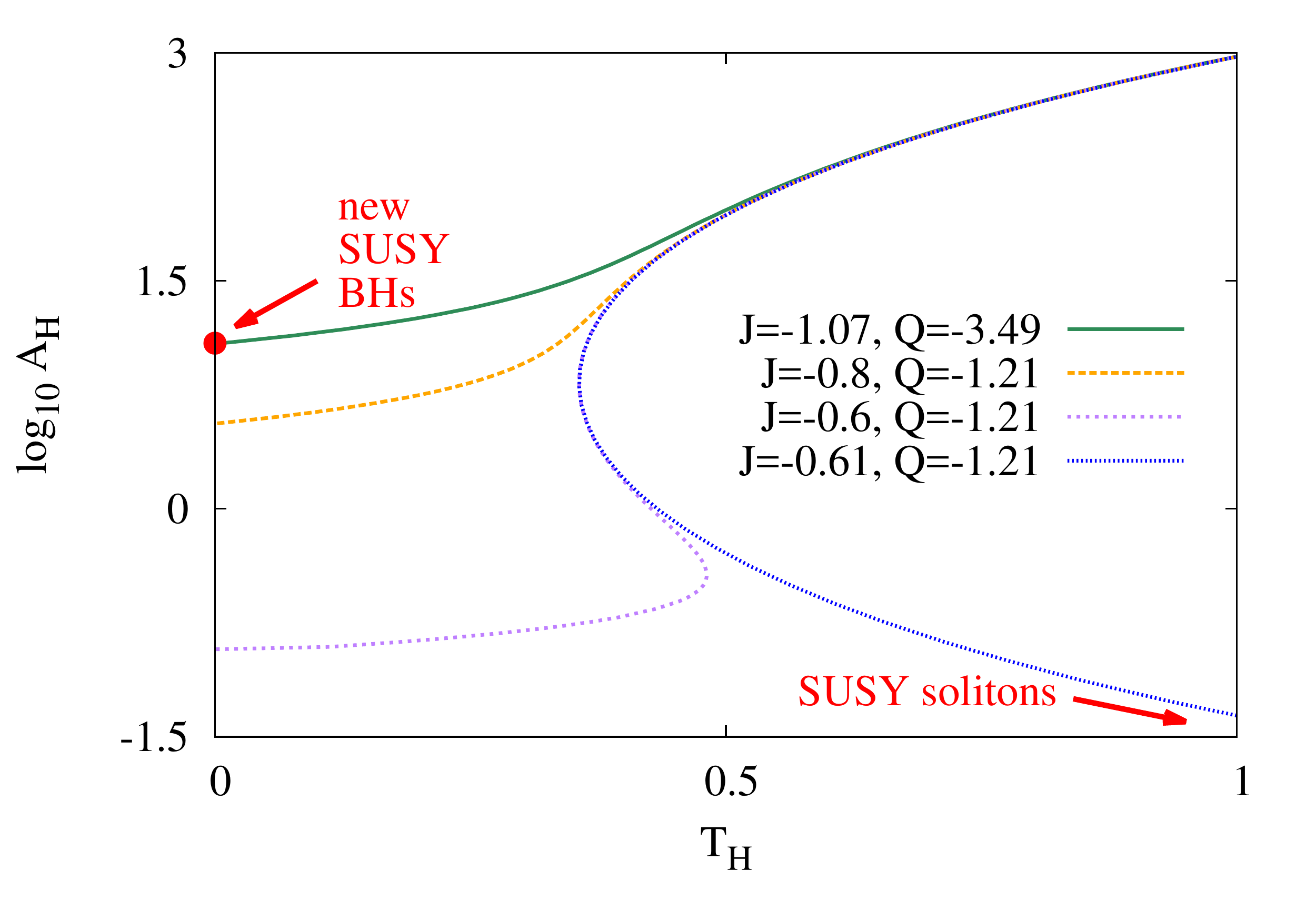}
\label{plot_AhvsT}}
\subfigure[]{\includegraphics[scale=0.28,angle=-90]{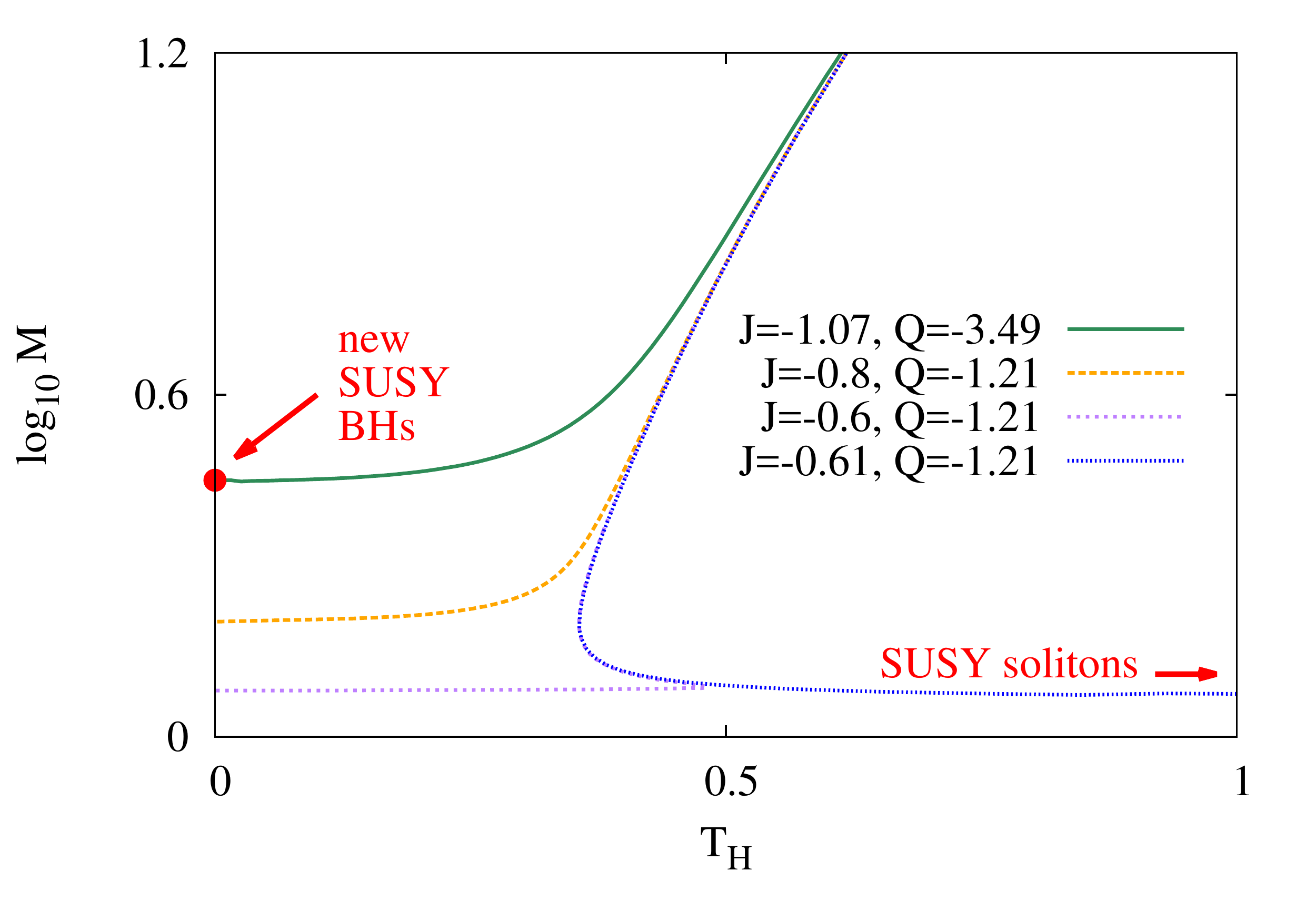}
\label{plot_MvsT}}
\subfigure[]{\includegraphics[scale=0.28,angle=-90]{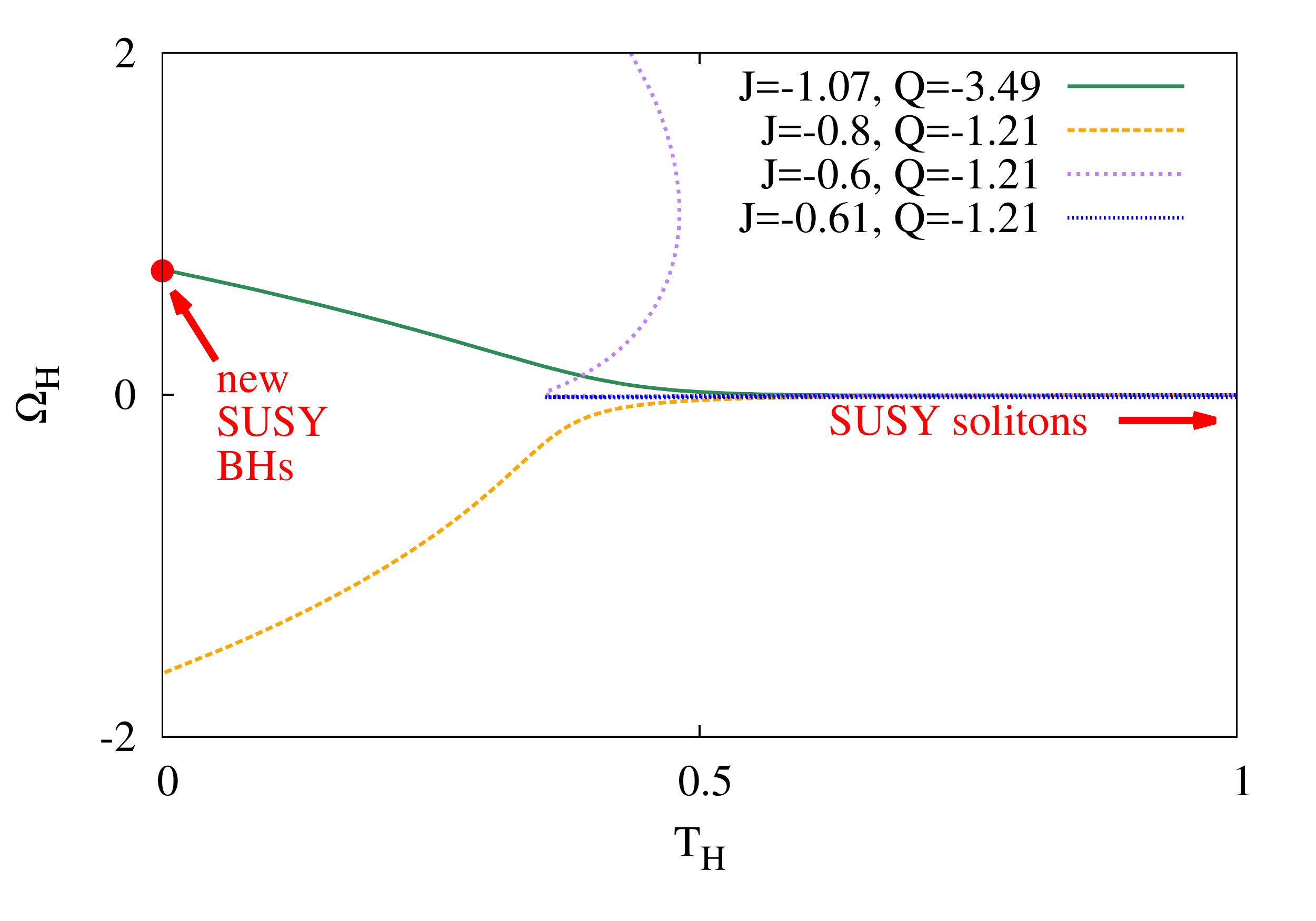}
\label{plot_OmHvsT}}
\subfigure[]{\includegraphics[scale=0.28,angle=-90]{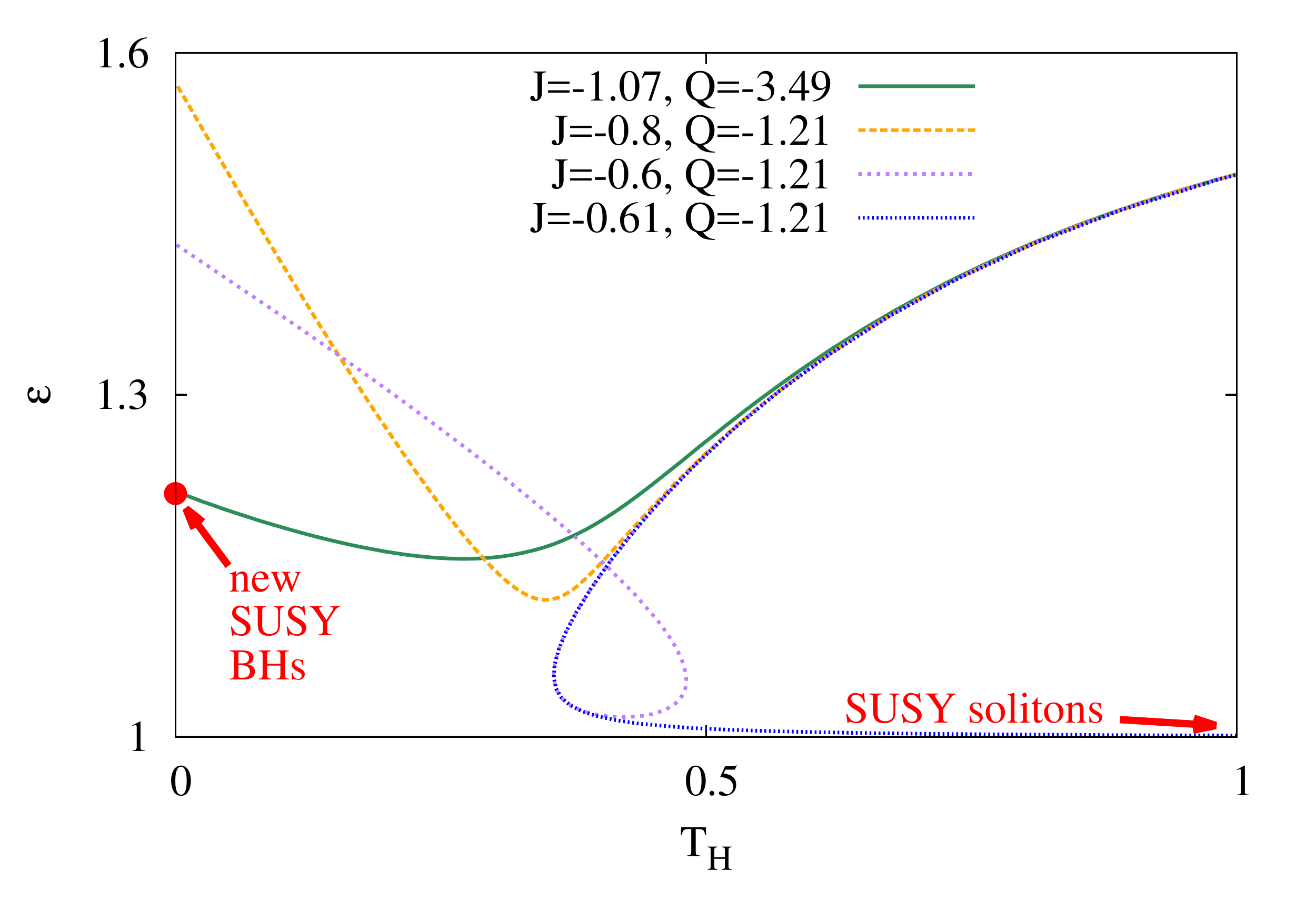}
\label{plot_defvsT}}
\caption{Several properties of squashed magnetized black holes
are shown as a function of the horizon temperature.  
The configurations have $v=1.65$, $c_m=-1$, $L=1$ 
 and different values of $Q$ and $J$. 
The generic solutions possess an extremal (non-susy) limit  (cyan, orange and purple lines). 
However, a particular set of solutions (red line in this figure) possess an extremal limit resulting in a supersymmetric black hole
 (red point).
A different set of solutions  (blue line)   form a soliton in the limit of a vanishing horizon,
$T_H\rightarrow \infty$,
which corresponds to the susy configuration in Ref. \cite{Cassani:2014zwa}.}
\label{plot_2d_T}
\end{figure}

\medskip

Let us now discuss some thermodynamical 
properties of these solutions, as shown in Figure \ref{plot_2d_T}.
 The configurations there have a fixed value of the squashing parameter, $v=1.56$, 
and of the magnetic parameter, $c_m=-1$.
This choice implies that the trace (\ref{r3}) of the boundary stress tensor vanishes, $<T _{a}^a> =0$,
a condition which is a requirement for the existence of supersymmetric solutions (see the discussion in the next Section). 
Several families of BHs are shown in that Figure (with different  color lines) as a function of
the  temperature $T_H$, 
each one possessing different values of the charges $Q$ and $J$. 

\begin{figure}[t]
\centering
\includegraphics[width=80mm,angle=-0]{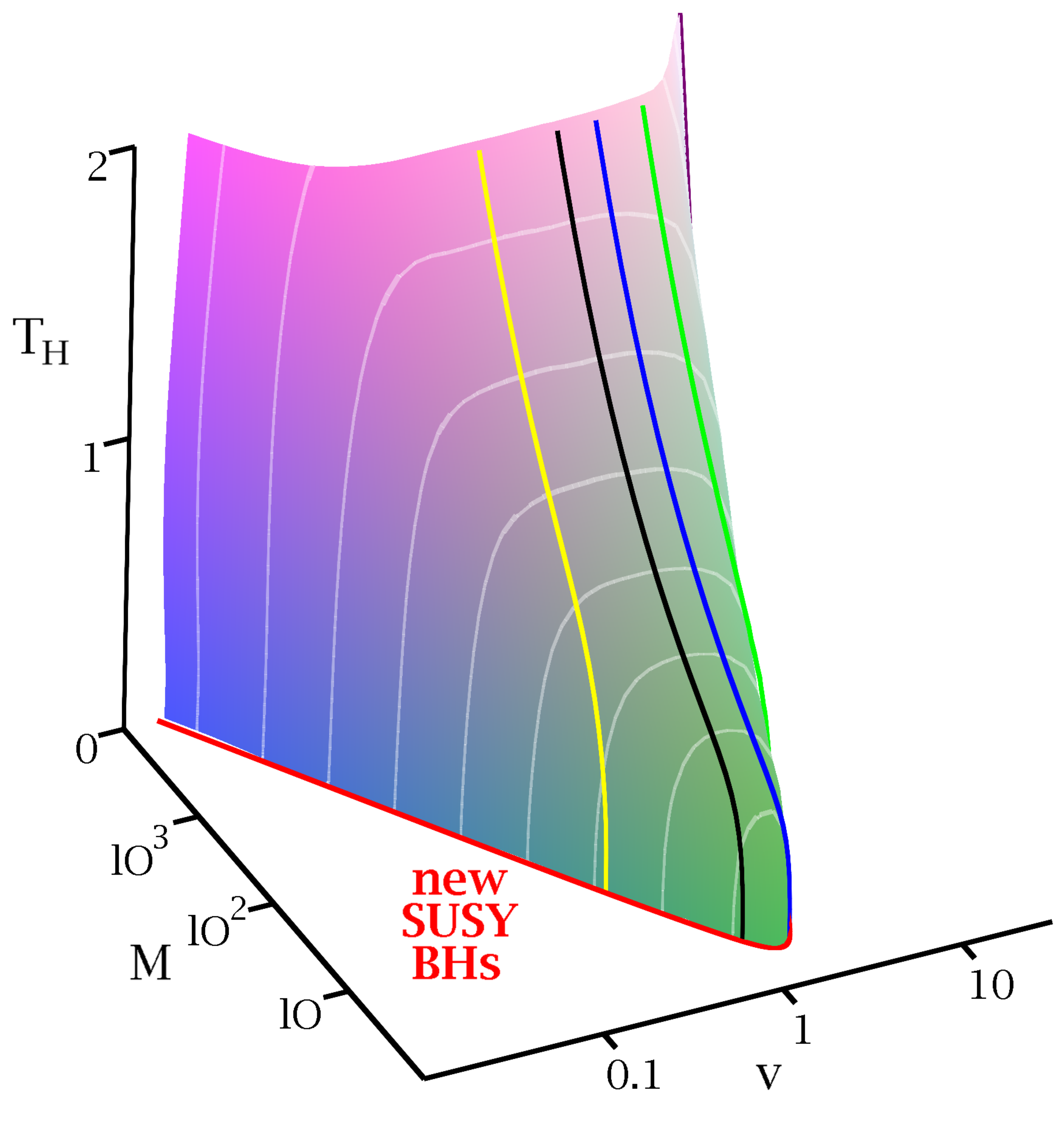}
\caption{Mass $M$ 
 vs.  squashing parameter $v$ 
 vs.  temperature $T_H$ for a particular set of black holes.
Their magnetization parameter $c_m$, electric charge $Q$ and angular momentum $J$ are given functions of $v$
such that 
  the extremal limit, $T_H = 0$, corresponds to the susy black holes in Section 4 (red line).
}
\label{plot_M_bh3d}
\end{figure}

\begin{figure}[t]
\centering
\includegraphics[width=80mm,angle=0]{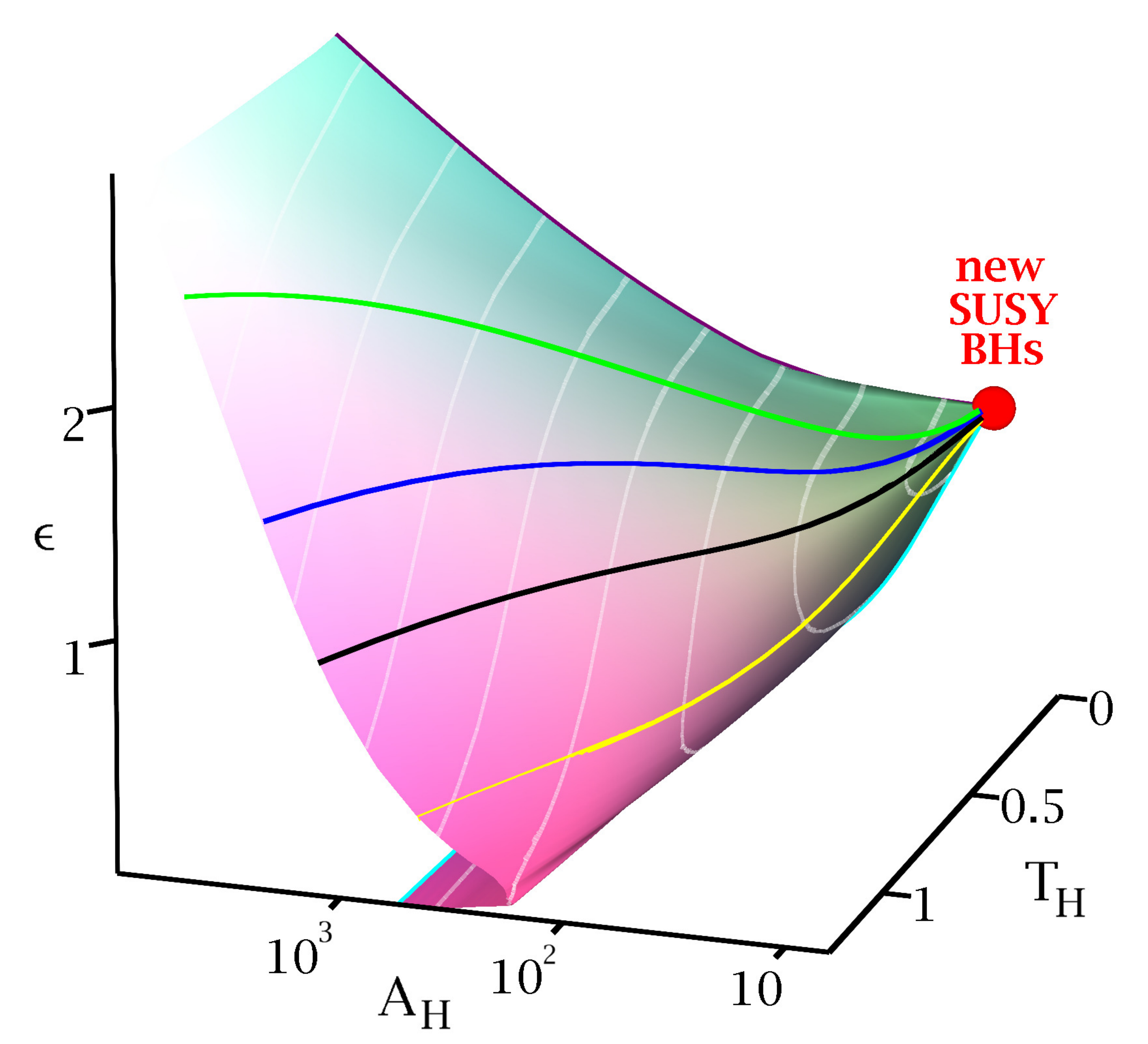}
\caption{Horizon area $A_H$ 
vs. horizon deformation $\epsilon$ 
vs. temperature $T_H$ for the same set of solutions in
Figure \ref{plot_M_bh3d}. 
In the extremal limit, $T_H=0$, the susy black holes in Section 4 are obtained, 
all of them possessing the same horizon properties and being  represented by
a red point in this Figure.
}
\label{plot_Ah_bh3d}
\end{figure}
%
One can see that
in general, the BHs possess an extremal limit, $T_H=0$. 
However, the situation is different 
for BHs with a specific relation between $J$ and $Q$ as given by (\ref{cond-sol}) 
 (shown with the blue line in Fig. \ref{plot_2d_T}).
For these solutions, the temperature diverges as the horizon size decreases,
a smooth solitonic configuration being approached as  
 the  horizon size shrinks to zero.
Anticipating the discussion in Section 4,
we mention that in general, the extremal BHs  are not supersymmetric.
 However, for particular values of the charges and the squashing, the limiting solutions satisfy the susy equations. 
This is the case of the extremal black hole marked with a red point (and also the solitonic limit of the blue line). 

In Fig. \ref{plot_AhvsT} we show the event horizon area $A_H$ vs. the temperature $T_H$. 
Depending on the values of $Q$ and $J$, the area can be a monotonic function of $T_H$
 (red, cyan and orange lines); 
a more complicated picture is also allowed,
with regions where $A_H$  decreases with increasing temperature (purple and blue lines). 
In the solitonic limit of the blue line, the area vanishes as $T_H \to \infty$, although the limiting solitons 
are regular everywhere. 
Also note that this family of BHs is special, with a finite minimum temperature.

In Fig. \ref{plot_MvsT} we present a similar figure for the mass $M$ vs the temperature $T_H$. 
In general, the behaviour of $M$ is similar to the area. 
However, in the solitonic limit the mass does not vanish, reaching a finite value, 
which, as we shall see, depends on the soliton charges and squashing. 

The horizon 
angular velocity $\Omega_H$ is shown in Fig. \ref{plot_OmHvsT},
again as a function of $T_H$. 
As an interesting feature,
we note that the configurations can present a counter-rotating horizon, depending on the specific combination of the charges.
 In the solitonic limit, the angular velocity vanishes since there is no horizon.

\begin{figure}[t]
\centering
\includegraphics[width=80mm,angle=0]{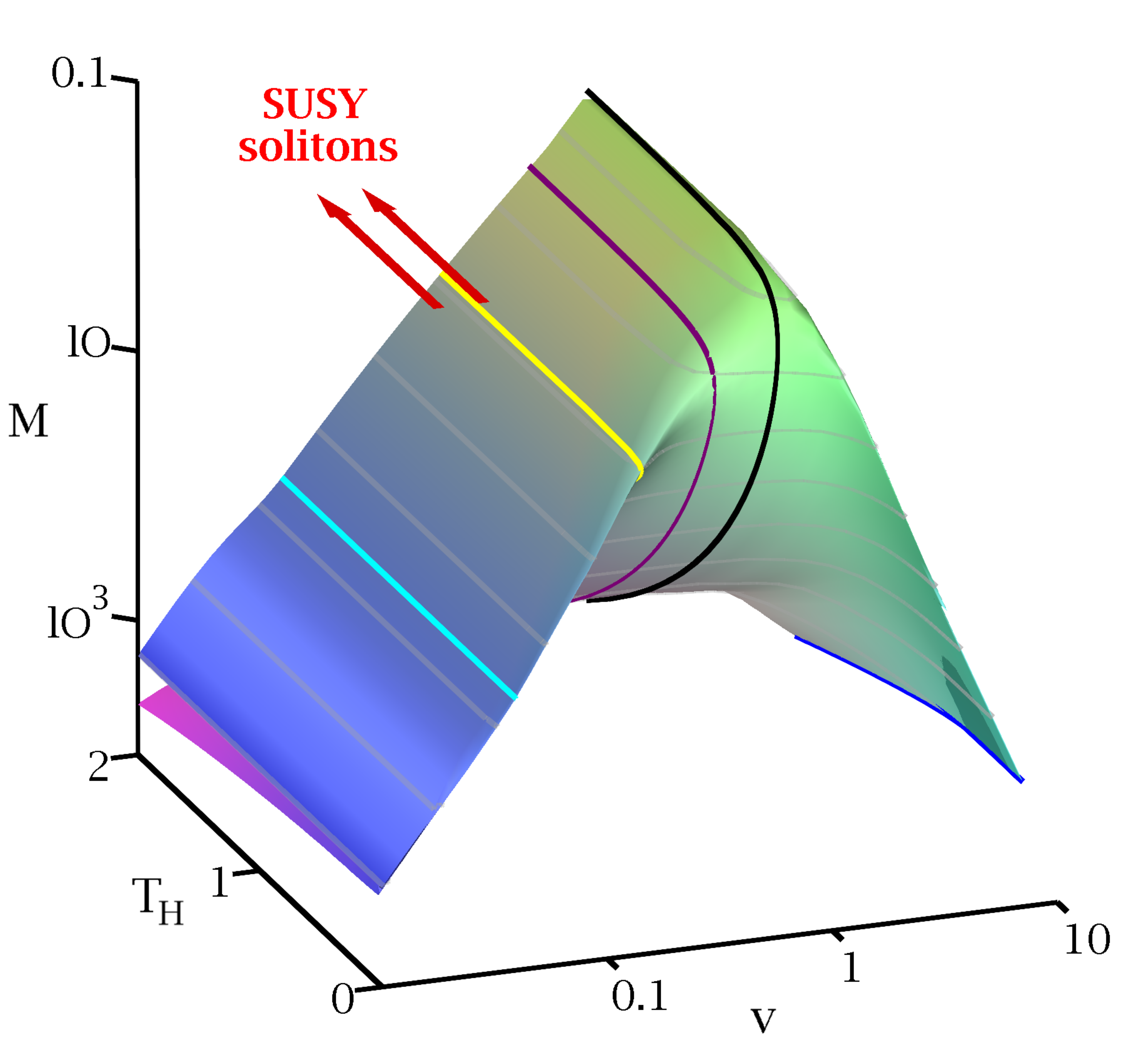}
\caption{Mass $M$ vs. squashing $v$ vs temperature $T_H$ 
for a particular set of solutions.
Their magnetization parameter $c_m$, electric charge $Q$ and angular momentum $J$ are given functions of $v$
such that no extremal black holes are obtained.
Instead, the limit  $T_H \to \infty$ describe susy solitons. 
}
\label{plot_Ah_sol3d}
\end{figure}
\begin{figure}[t]
\centering
\includegraphics[width=80mm,angle=-0]{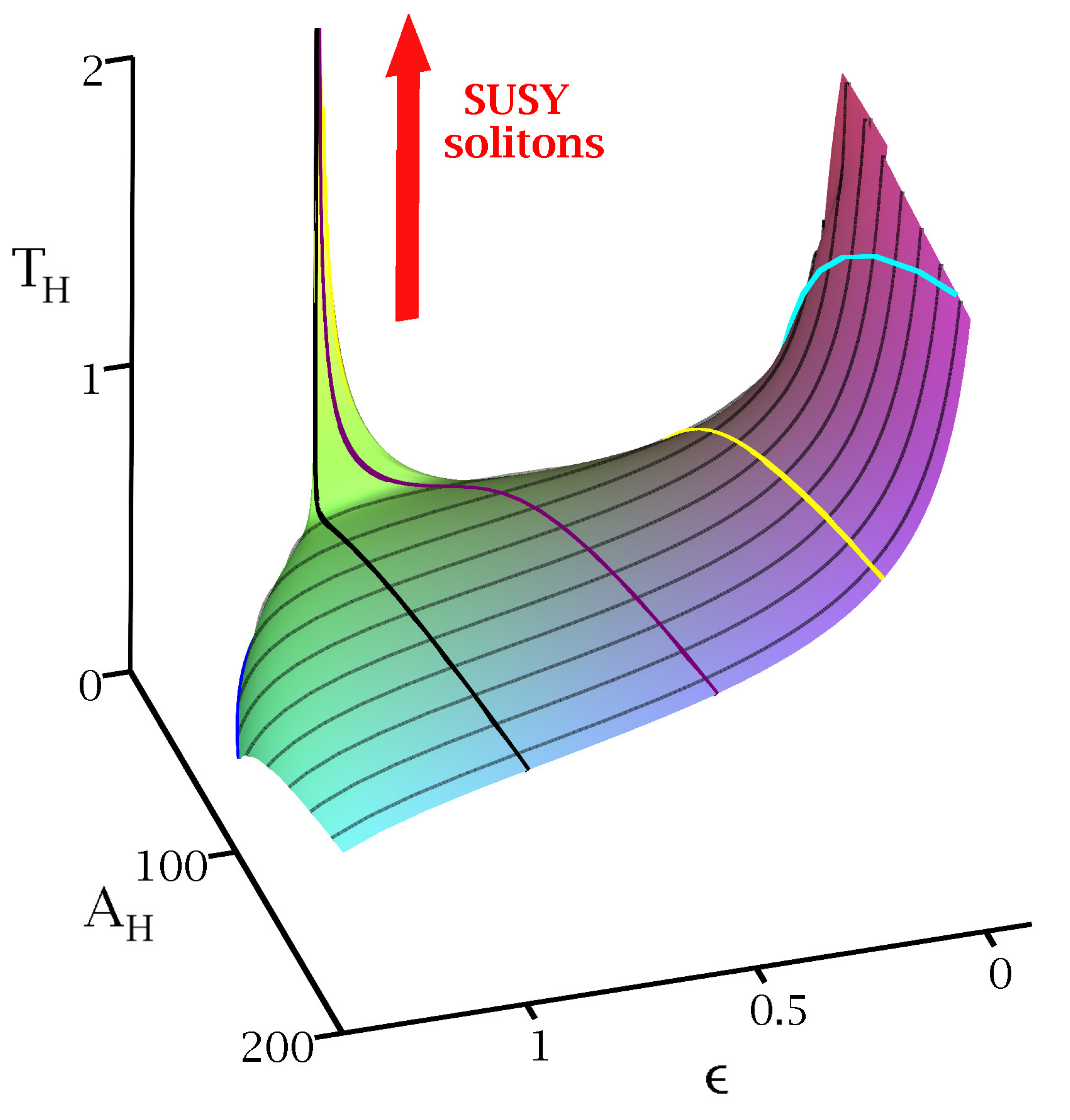}

\caption{Area $A_H$ vs.
 horizon deformation $\epsilon$ 
vs. temperature $T_H$ for black holes
is shown for the same solutions in Figure \ref{plot_Ah_sol3d}. 
As the temperature diverges, the horizon area vanishes and the susy solitons are recovered.
}
\label{plot_M_sol3d}
\end{figure}

In Fig. \ref{plot_defvsT} we present the horizon deformation $\epsilon$ (as given by rel. (\ref{deform}))
as a function of $T_H$. Although the squashing $v$ of these particular sets is fixed to $v=1.56$, the horizon deforms depending on the black hole electric charge $Q$, and angular momentum $J$ 
(and in general, on the magnetization parameter, which in these Figures is also fixed). 
Although $\epsilon \geq 1$ for the solutions there,
 the horizon deformation can also be smaller than one for a different choice of the input parameters.
 \medskip

Other properties of the solutions
are shown in Figs. \ref{plot_M_bh3d}, \ref{plot_Ah_bh3d}.
The set of solutions presented there have the value of the 
magnetization parameter fixed by the squashing $v$,
with
$c_m=L(1-v^2)/\sqrt{3}$,
such that  the trace (\ref{r3}) of the boundary stress tensor vanishes.
Also, they have fixed values of the integration constants 
which enter the 1st integrals (\ref{1stInt-gen}),
 $c_t=\frac{203\sqrt{3}}{484}L^2$ 
and 
$c_W=-\frac{2303}{5324}L^2$.
As a result, one can see from  (\ref{1stInt-gen})  that the solutions' 
electric charge and angular momentum  possess 
a dependence on  the squashing $v$ as given in the corresponding
relations  in (\ref{MJQ-susy-BH}).
Therefore, for a given $v$, they form a one parameter family of solutions
which are constructed by varying the value of the horizon radius $r_H$  
(note that other quantities of interest  of the solutions ($e.g.$ mass, horizon area and temperature) 
are unconstrained, 
being read from the numerical  output).
	
	The reason for this special choice of $Q,J$ and $c_m$
	is that the extremal limit of these solutions
	possesses some special properties,  being supersymmetric, as 
 we will see in the next Section.
 Here they are constructed directly, as solutions of the second-order equations of motion.

In Fig. \ref{plot_M_bh3d}	
 we show the mass $M$ as a function of the squashing parameter $v$ and the Hawking temperature $T_H$.
We can see that $M$ strongly increases as $v$  decreases,
with the existence of a minimal value for a given $v$.
In Fig. \ref{plot_Ah_bh3d} 
we show the event horizon area $A_H$
as a function of the horizon deformation $\epsilon$ and the Hawking temperature $T_H$.
Note that in the extremal limit, all solutions converge to a single point (in red),
 meaning that all the extremal BHs possess the same horizon properties. 
An explanation of this feature will be provided in the next Section
when studying the susy squashed BHs  

Further properties of BHs  
 are shown in Figs. \ref{plot_Ah_sol3d}, \ref{plot_M_sol3d}.
There the choice of the magnetization parameter is the same as above,
$c_m=L(1-v^2)/\sqrt{3}$,
and $c_t=c_W=0$.
This implies that $Q$ and $J$
possess a different dependence on $v$
as given in relation (\ref{susy-sol}) below,
other quantities being determined by the value of $r_H$.
As such, these configurations provide a different cut in the parameter space of solutions,
the set of susy solitons being recovered in the 
vanishing horizon size limit (see the discussion below).

We mention that we have  also  studied families of solutions with $<{T_a^a}> \neq 0$
(which thus do not possess a supersymmetric limit).
The picture we found here is similar to the generic case in Figure \ref{plot_2d_T}  
(red, cyan and orange lines),
with the existence of an extremal limit possessing a nonzero horizon area.

\subsection{Solitons}

\begin{figure}[t]
\centering
\includegraphics[scale=0.38,angle=-90]{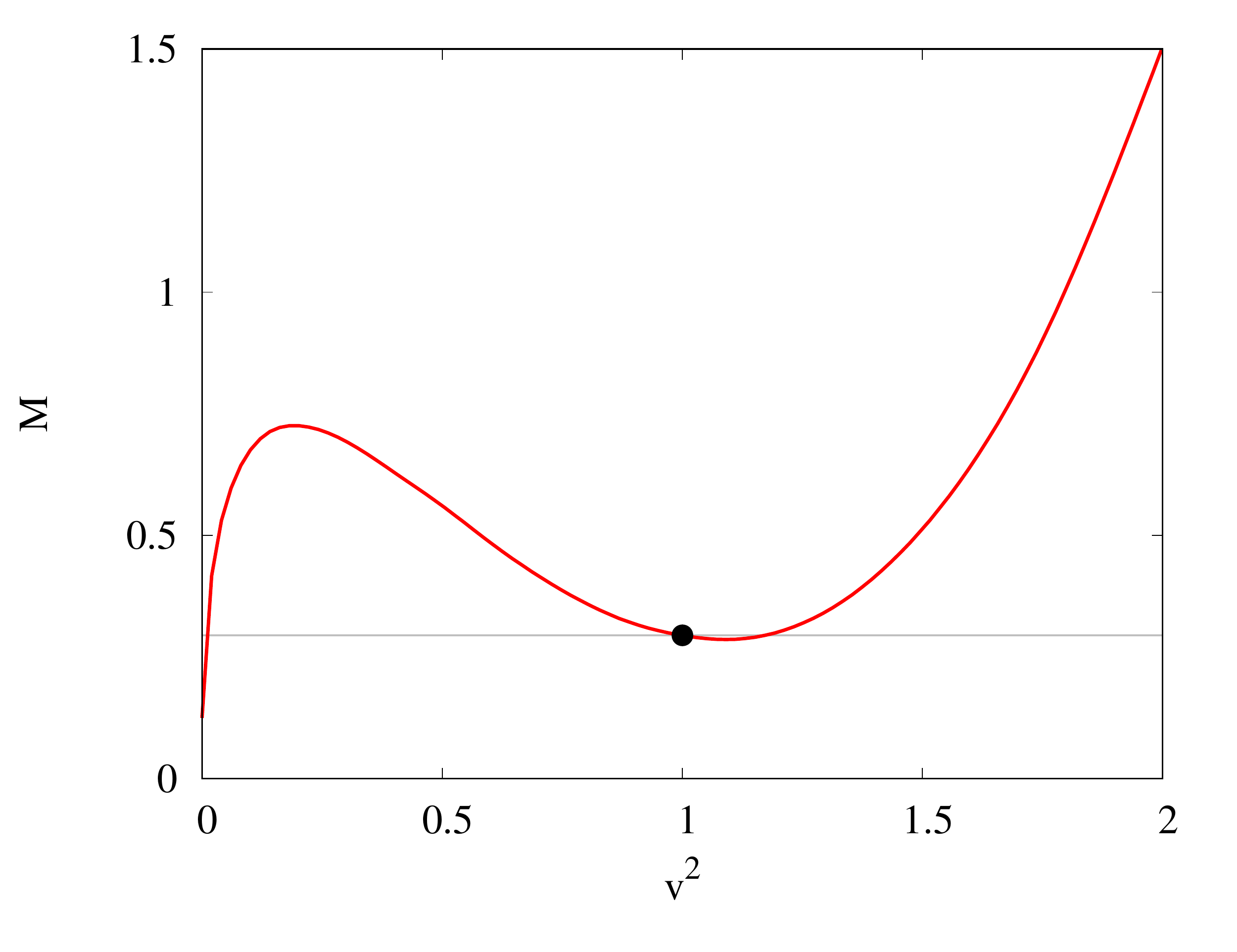}
\caption{Mass $M$ is shown $vs.$
 the deformation parameter $v^2$ 
for the squashed backgrounds (vacuum static spacetimes).
The horizontal line and the dot mark the $v=1$ mass, corresponding to the Casimir value $3\pi L^2/32$, with $L=1$.
}
\label{plot_vac_mass}
\end{figure}

We start the discussion of the solitonic solutions of the model
with the vacuum static case.
These configurations naturally emerge as the zero horizon size limit of the corresponding 
families of BHs discussed in the  Section 3.2.2,
representing deformations of the globally AdS$_5$ spacetime.
As such, their most natural interpretation is as providing a $background$ for  
models with given geometric parameter $v$.
As expected  their mass is nonvanishing,
being shown in Figure \ref{plot_vac_mass} as a function of $v$.
These configurations can also be viewed as the zero horizon size limit
of 
families of charged and/or spinning  
BHs with $c_m=0$.
However,
as already discussed above, both $J$ and $Q$ vanish as $r_H\to 0$.

One feature a nonzero boundary magnetic field brings new is the 
existence of a nontrivial limit of the solutions
which describes  spinning electrically charged solitons.
Similar to the vacuum case,
they possess no horizon, while the size of both parts of the $S^3$-sector of the metric shrinks to zero as $r\to 0$. 

An interesting property of the solitons is that their electric charge and angular momentum are proportional.
To prove it, we notice that
since $c_t=c_W=0$,  they satisfy the simple relations\footnote{
These relations can also be viewed as a consequence  of the
vanishing of the Page charge for solitons.
}
\begin{eqnarray}
qv=\frac{2c_m^2 \lambda }{\sqrt{3}},~~\hat j=-\frac{2}{3 v^2}c_m q.
\end{eqnarray}
Then, after expressing $\hat j$, $c_m$ and $q$ and in terms of
the angular momentum $J$ (as given by rel. (\ref{MJ})),
the magnetic flux at infinity  $\Phi_m$ (rel. (\ref{flux})),
and the electric charge $Q$ (rel. (\ref{R-charge}))
one finds 
\begin{eqnarray}
\label{cond-sol}
J=\Phi_m   Q, ~~{\rm with}~~~Q=-\frac{8\pi \lambda}{2\sqrt{3}}\Phi_m^2~. 
\end{eqnarray} 
These relations are universal, being satisfied for any 
value of the CS coupling constant
$\lambda\neq 0$.

As we have already seen in Fig. \ref{plot_2d_T}
 the solitons appear as the $r_H\to 0$ limit of  particular  black hole solutions which satisfy (\ref{cond-sol}).
Further insight on the properties of  soliton solutions 
can be found in Fig. \ref{plot_M_sol3d}. 
There the horizon area $A_H$ is plotted as a function of the horizon deformation $\epsilon$ 
and the temperature $T_H$
for a set of BH solutions with
  $c_t=c_W=0$
	and
$c_m=L(1-v^2)/\sqrt{3}$.
In this case, the BHs only depend on the squashing parameter $v$ and the temperature $T_H$. 
The lines in yellow, black, blue and purple represent sets of solutions with constant values of $v$. 
The non-extremal solutions present a limit in which $T_H \to \infty$, the area vanishes and the horizon deformation goes to one. 
This limit comprises a whole family of solitons.

In Fig. \ref{plot_Ah_sol3d} 
we represent the mass $M$ of BHs as a function of the squashing parameter $v$ and the  temperature
$T_H$  
for the same solutions as in Fig. \ref{plot_M_sol3d}.
The solitons are approached in 
 the limit of vanishing area and diverging temperature.
This results
 in a family of solutions with finite mass and varying squashing parameter $v$
(we recall that the magnetic parameter $c_m$ is determined by $v$  
such that the boundary stress tensor  (\ref{r3})  is traceless). 
We have verified that these 
 solutions correspond 
in fact to the supersymmetric configurations in \cite{Cassani:2014zwa}
(this provides another useful test of our numerical results). 
  More details on these special solutions is presented in the next Section.

Finally, let use remark that the considered choice 
$c_m^2=L^2(1-v^2)^2/3$
is not a necessary condition for the existence 
of solitons (while (\ref{cond-sol}) is mandatory).
In fact, solitonic solutions in a globally AdS background ($v=1$)
have been already reported in 
Ref. \cite{Blazquez-Salcedo:2017cqm},
most of their properties being recovered for other values of $v$.

\section{ Supersymmetric solutions} 
The only known supersymmetric BHs 
 within the framework in Section 2
are those found by  Gutowski and Reall 
in Ref. \cite{Gutowski:2004ez}.
They possess a round sphere at infinity ($v=1$)
and no boundary magnetic field ($c_m=0$),
being a special limit of the CLP solution. 
These solutions contain a single parameter, $\alpha>1/2$,
the global charges being given by
\begin{eqnarray} 
\nonumber
&&
M=\frac{\pi L^2}{216}(3\alpha^2-1)(31+76 \alpha^2+64 \alpha^4)+\frac{3\pi L^2}{32},
\\
\label{nGR2}
&&
J=-\frac{\pi L^3}{216}(1-4\alpha^2)^2(7+8\alpha^2),
\\
\nonumber
&&
Q=\frac{\pi L^2}{12\sqrt{3}}(1-4\alpha^2)(5+4\alpha^2),
\end{eqnarray}
while the horizon area, horizon angular momentum,
 electrostatic potential
and 
horizon angular velocity
are
\begin{eqnarray} 
\label{nGR3}
&&
A_{H}=\frac{\pi^2L^3}{3\sqrt{3}}(4\alpha^2-1)
\sqrt{(4\alpha^2+3)(4\alpha^2-1)},~~
\\
\nonumber
&&
J_{H}={\frac {\pi L^3   }{384 }}
\left( 4\,{\alpha}^{2}-1 \right) ^{2}
 \left( 4\,{\alpha}^{2}+3 \right) ,~~
\Omega_H=\frac{1}{L},~~\Phi=\frac{1+2\alpha^2}{\sqrt{3}}.
\end{eqnarray}

\medskip

It is natural to inquire if the more general
squashed magnetized solutions in this work 
also possess a
supersymmetric limit.
A hint in this direction
comes from the observation \cite{Cassani:2013dba}
that  the
two contributions
in the trace 
$<T _{a}^a>$ 
of the stress tensor (\ref{r3})
 exactly cancel for
\begin{eqnarray} 
\label{cancel}
 c_m= \pm \frac{L}{\sqrt{3}}(1-v^2).
\end{eqnarray}
In the extremal BH case, this requirement leads to a two parameter family of solutions
(the parameters can be taken as the squashing  $v$ and the electric charge $Q$).
However, the situation changes for solitons,
the above condition together 
with the charge-angular momentum relation 
(\ref{cond-sol})
leading to a single family of solutions which can be parametrized in terms of $v$.
We have found numerically that these solutions correspond\footnote{Although one cannot exclude
the existence of non-susy excitations of these solutions, so far we have no indication for that.} 
in fact to the supersymmetric solitons in
\cite{Cassani:2014zwa}.
Although they cannot be written in closed form, 
the supersymmetry 
allows for an  almost complete
description of the solution in terms of
the squashing parameter $v$.
The mass, angular momentum and electric charge of the supersymmetric solitons
are given by
\cite{Cassani:2014zwa}
\begin{eqnarray}
\label{susy-sol}
M =\pi L^2
\left(
\frac{5}{288}+\frac{2}{27v^2}-\frac{7}{36}v^2+\frac{89}{864}v^4
\right),
~
J =\frac{\pi L^3}{27}  \left( v^2-1 \right) ^{3} ,
~
Q=-\frac{ 2\pi L^2 } {9\sqrt{3}}\left( v^2-1 \right) ^{2}.~{~~~}
\end{eqnarray}
We have already shown in the previous section that these supersymmetric solitons can be connected with squashed magnetized black holes in the limit of vanishing size of the horizon.
In what follows, we show that, in addition 
to these solitons,
the EMCS equations 
possess as well a one-parameter family of supersymmetric BHs.

\subsection{The formalism} 
In constructing the  supersymmetric solutions which fit the framework in Section 2,
the most convenient approach is to use
the general formalism proposed in  \cite{Gutowski:2004ez}.
Then
such configurations have the following  line element\footnote{To agree  
with the standard notation in the literature,  in this Section  we use
$\rho$ for the
 radial coordinate, instead of $r$ as in (\ref{metric0}).}:
\begin{eqnarray}
\label{metric0}
ds^2=-f^2(\rho) \left( dy+\Psi(\rho)  {\hat\sigma}_3 \right)^2
+\frac{1}{f(\rho)}
\big[
   d\rho^2+a^2(\rho)( {\hat\sigma}_1^2+ \hat{\sigma}_2^2)+ b^2(\rho) \hat{\sigma}_3^2
\big] ,
\end{eqnarray}
with
\begin{eqnarray}
\nonumber
 \hat{\sigma}_1 = -\sin  \hat{\psi} d \theta + \cos \hat{\psi} \sin\theta d \phi,
~~
 \hat{\sigma}_2 =\cos\hat{\psi} d \theta + \sin \hat{\psi} \sin\theta d \phi ,
~~
 \hat{\sigma}_3 = d \hat{\psi} + \cos\theta d \phi,  
\end{eqnarray}
and the gauge potential
\begin{eqnarray} 
\label{A-susy}
A= \frac{\sqrt{3}}{2} \left[ f(\rho) \diff y + \left(f(\rho)\Psi(\rho)+\frac{L}{3}p(\rho) \right) \hat{\sigma}_3  \right ].
\end{eqnarray}
Thus the framework contains
 five functions
$\{a, b,p,f,\Psi\}$, instead of six as in the generic case.
However,
the expression of $\{b,p, f, \Psi\}$
is fixed by $a$, via the following relations
(which are found from the corresponding Killing spinor equations) \cite{Gutowski:2004ez}:
\begin{eqnarray} 
&&
\nonumber
b=2aa',
\\
&&
\label{set}
p=4a'^2+2aa''-1,
\\
&&
\nonumber
f^{-1} \ = \ \frac{L^2}{12 a^2 a'}[4 (a')^3 + 7 a\, a' a'' - a' + a^2 a'''],
\\
&&
\nonumber
\Psi\ = \ - \frac{L a^2}{4}\left( \nabla^2 f^{-1} + 8 L^{-2} f^{-2} - \frac{L^2 g^2}{18} + f^{-1} g \right),
\end{eqnarray}
where we denote
$
 g =-\frac{a'''}{a'} - 3 \frac{a''}{a} - \frac{1}{a^2} + 4 \frac{(a')^2}{a^2} .
$
The function $a$ is the solution of a sixth order equation
\begin{eqnarray} 
\label{eqa}
\Big(\nabla^2 f^{-1} +8 L^{-2}f^{-2} - \frac{L^2 g^2}{18} + f^{-1} g \Big)' + \frac{4a'g}{af} \ = \ 0\,,
\end{eqnarray}
where
$\nabla^2 = \frac{d^2}{d\rho^2}+(\frac{2a'}{a}+\frac{b'}{b}) \frac{d}{d\rho}$.
Any solution to this equation 
(together with (\ref{set}), (\ref{A-susy}))
 corresponds to a
 configuration 
which preserves at least one quarter of the supersymmetry.
Let us also remark that the $a-$equation (\ref{eqa})
is invariant under the transformation
\begin{eqnarray} 
\label{scale}
\rho \to \lambda  \rho ,~ a \to  a/\lambda,
\end{eqnarray}
with $\lambda>0$.
 
The only (known) closed form
solution of the eq. (\ref{eqa}) 
which describes a BH has been
found by
Gutowski and Reall (GR)
and has
\begin{eqnarray} 
\label{nGR1}
 a(\rho)=\alpha L \sinh (\frac{\rho}{L}),
\end{eqnarray}
 with $\alpha>1/2$ a real parameter (the value  $\alpha=1/2$
corresponding to the globally AdS background)).

 \medskip
Also, 
one notes that the line element  (\ref{metric0}) can be put into the form (\ref{metric})
by taking
\begin{eqnarray}
\label{transf2}
 F_1=\frac{1}{f},~~F_2=\frac{4a^2}{f},~ 
F_3=\frac{4}{f}(b^2-\Psi^2 f^3),~~F_0=\frac{b^2f^2}{b^2-\Psi^2 f^3},~ 
W=\frac{1}{2}
\left(
\frac{f^3\Psi}{b^2-f^3 \Psi^2}-u  
\right),~~~{~~~~}
\end{eqnarray}
with
\begin{eqnarray}
\label{transf1}
 t=y,~~\psi= \hat \psi-u t~.
\end{eqnarray}
The reason we introduce the constant $u$ in 
the above relations 
is that the supersymmetric solutions are 
found in a frame which rotates at infinity, with
\begin{eqnarray}
\label{limit}
 \lim_{\rho \to \infty}\frac{f^3\Psi}{b^2-f^3 \Psi^2}=u \neq 0.
\end{eqnarray} 
Then the transformation (\ref{transf1}) 
brings the solution to a static frame at infinity,
such that the solutions become a particular limit
of the general case in Sections 2, 3.
For example, 
the corresponding expression of the U(1) potential  (\ref{U1}) reads
\begin{eqnarray}
\label{newA2}
a_0= \frac{\sqrt{3}}{2}   \left[f(\rho)+ u\left ( f(\rho)\Psi(\rho)+\frac{L}{3}p(\rho)\right)    \right]  ,~~
a_k=
\frac{\sqrt{3}}{2}   \left(f(\rho)\Psi(\rho)+\frac{L}{3}p(\rho) \right)   .
\end{eqnarray}
 %

\subsection{The large$-\rho$ expansion} 

Despite the absence of an exact solution
 in the general squashed magnetized case,
it is still possible to find 
 an approximate solution at the limits of the domain of integration. 
Keeping the notation of Ref. \cite{Cassani:2014zwa},
the first terms in the far field expansion of $a(\rho)$ 
read
\begin{eqnarray} 
\label{inf-rho}
\frac{a(\rho)}{L} &=& 
 a_0 e^{\frac{\rho}{L}} + \left(  a_2 +  c \frac{\rho}{L}\right)
\frac{e^{-\frac{\rho}{L}}}{  a_0} + \left(  a_4 
+ \frac{2- 16a_2 -5  c}{12}c\frac{\rho}{L} 
-\frac{2}{3} c^2 \left(\frac{\rho}{L}\right)^2\right) \frac{e^{-\frac{3\rho}{L}}}{a_0^3} 
\\
&&
\nonumber
+\Big( a_6 + \frac{1}{972}(12 - 282a_2 + 1488a_2^2 -1548 a_4 -54 c +537 a_2 c + 59 c^2 ) c\frac{\rho}{L}
\\
&&
\nonumber
{~~~~~~}
-\frac{ 90 - 840a_2 - 197 c }{324}c^2
\left(\frac{\rho}{L}\right)^2 + \frac{70}{81}c^3\left(\frac{\rho}{L}\right)^3\Big)\frac{e^{-\frac{5\rho}{L}}}{a_0^5} \,+\, \mathcal O(e^{-6\rho/L}) \,
+\dots ,
\end{eqnarray} 
containing the free parameters 
$
\{ c,a_0,a_2,a_4,a_6\}
$
%
(with $a_0 \neq 0$).
Then it is straightforward to derive 
from (\ref{set}) 
the 
asymptotic 
form of the other functions
$ 
\{b,p,f,\Psi\}.
$
However, these expressions are rather complicated and 
we shall not include them here\footnote{The corresponding expression 
for supersymmetric solutions can be found in the Appendix A of 
Ref. \cite{Cassani:2014zwa}.
Since the generic solitons and BHs possess the same far field expansion,
the expressions there are valid also for the solutions in this work.
}.

Instead, it is interesting to give the asymptotic expansion of the metric functions
$F_i$, $W$
which enter the metric Ansatz (\ref{metric}).
The analysis is simplified by introducing a new radial coordinate
\begin{eqnarray}
r=Le^{\frac{\rho}{L}},
\end{eqnarray}
such that the far field expansion resembles (\ref{inf})
\begin{eqnarray}
&&
\nonumber
F_0(r)=\frac{4a_0^2}{(1-4c)}
\left(
\frac{r}{L}
\right)^2
+\frac{2}{3(1-4c)}\left(1+4a_2+4c+4c \log\left(\frac{r}{L}\right) \right)+\dots,
\\
&&
\nonumber
F_1(r)\frac{r^2}{L^2} 
=
1-\frac{1+16a_2+4c+16c\log\left(\frac{r}{L}\right)}{12a_0^2}
\left(
\frac{L}{r}
\right)^2+\dots,
\\
&&
\label{as-form1}
\nonumber
F_2(r)=4a_0^2 r^2
+\frac{L^2}{3}\left(-1+8a_2-4c+8c \log\left(\frac{r}{L}\right) \right)+\dots,
\\
&&
\label{large-r}
F_3(r)=4a_0^2(1-4c) r^2
-\frac{(1-4c)L^2}{3}\left(1-8a_2-20c-8c \log\left(\frac{r}{L}\right) \right)+\dots,
\\
&&
\nonumber
W(r)=\frac{1}{7776a_0^4(1-4c)^2L}
\bigg(
63
+576a_2(3+12a_2-128a_2^2)
-1152a_4(40c+516a_2-15)
\\
\nonumber
&&
{~~~~~~~~~~}
-373248a_6
+c(1860-5856a_2-51456a_2^2)
+8c^2(432a_2+3436c-1935)
\bigg)
\left(
\frac{L}{r}
\right)^4+\dots,
\end{eqnarray}
while  the asymptotic form of the gauge potentials is
\begin{eqnarray}
\label{as-form2}
&&
a_k(r)=-\frac{2cL}{\sqrt{3}}
+
 \frac{L}{96\sqrt{3} a_0^2}
\bigg(
1+256a_2^2+384a_4-32a_2(7c-1)
\\
\nonumber
&&
{~~~~~~~~~~~~~~~~~~~~~~~~~~~~~~~~~~~~~~}
+8c(17c-4)-96c(4c-1)\log\left(\frac{r}{L}\right) 
\bigg)
\left(
\frac{L}{r}
\right)^2+\dots,
\end{eqnarray}
\begin{eqnarray}
\nonumber
&&
a_0(r)=\frac{4c-3}{2\sqrt{3}(4c-1)}
+
\frac{1}{48\sqrt{3}a_0^2(4c-1)}
\bigg(
-5+256a_2^2+384a_4
\\
\nonumber
&&
{~~~~~~~~~~~~~~~~~~~~~~~~~~~~~~~~~~~~~~~~~~~~}
+32a_2(5c-2)
+8c(29c-4)
\bigg)
\left(
\frac{L}{r}
\right)^2+\dots~.
\end{eqnarray}
One can see that $W\to 0$ as $r\to \infty$,
such that the solution is  
 indeed  written in
 a nonrotating frame at infinity\footnote{
Here we have used
%
$
u= \lim_{\rho \to \infty}\frac{f^3\Psi}{b^2-f^3 \Psi^2}=\frac{2}{4c-1},
$
and replaced it in  (\ref{transf1}).
}.

Then, from  
(\ref{as-form2}),
one finds that
 the constant $c$ in the far field expansion corresponds to the magnetic flux parameter
\begin{eqnarray} 
c_m=\frac{c}{\sqrt 3}~.
\end{eqnarray}  
Also, one can see that
the solution necessarily possesses a squashed sphere at infinity\footnote{
Note that 
given the asymptotics  
(\ref{inf-rho}), 
(\ref{as-form1}),
the conformal boundary metric is
$ds^2_{(bdry)} =4 a_0^2 L^2(\sigma_1^2+\sigma_2^2+v^2 \sigma_3^2)-\frac{4a_0^2}{v^2}dt^2$,
which is slightly different from (\ref{g0}).
However, one can always set $a_0=1/2$
by rescaling the radial coordinate
(which implies a redefinition of other constants in (\ref{inf-rho})). 
Also, the scaling (\ref{scale1}) can be used to dispose of the $1/v^2$ factor in the above 
expression of $g_{tt}$, 
such that the standard form (\ref{g0}) is recovered. 
Moreover, let us remark that although the $a-$equation (\ref{eqa})
is invariant under the transformation (\ref{scale}),
that symmetry
is fixed by imposing the far field asymptotics (\ref{inf-rho}).
}, 
with
\begin{eqnarray} 
\lim_{r\to \infty}\frac{F_3}{F_2}=1-4 c=v^2~,
\end{eqnarray} 
 the zero-trace condition (\ref{cancel}) being satisfied.
 The squashing parameter $v$ takes arbitrary values, the solutions 
with $v^2<0$ possessing CTCs.

An important observation here is 
that 
after evaluating
the first integrals (\ref{1stInt-gen}),
for  
the asymptotic form (\ref{as-form1}),  
the constants $a_4$ and $a_6$ 
can be expressed
in terms of 
$c_t,c_W$
together with $a_2,c$:
\begin{eqnarray}
&&
\label{a4a6-inf}
a_4=-{\frac {13}{48}}{c}^{2}+ \left( -{ {5\,a_2}}
+1 \right)\frac{c}{12}-\frac{2}{3}\,{a_2}^{2}+\frac{a_2}{6}+{\frac {5}{384}}+\frac{1}{32}\,
{\frac {\sqrt {3}{\it c_t}}{{L}^{2}}},
\\
&&
\nonumber
a_6={\frac {1105\,{c}^{3}}{11664}}+ \left( {\frac {1913\,a_2
}{3888}}-{\frac {125}{1944}} \right) {c}^{2}+ \left( {\frac {197\,{a_2}^{2}}{324}}-{\frac {61\,a_2}{324}}+{\frac {25}{3456}}-{
\frac {19\,\sqrt {3}c_0}{2592\,{L}^{2}}} \right) c ~,
\nonumber 
\\ 
&&
{~~~~~~}
+{\frac {70\,{a_2}^{3}}{81}}-{\frac {5\,{a_2}^{2}}{18}}-{\frac {29\,a_2}{3456}}-{\frac {43\,\sqrt {3}a_2\,c_0}{864\,{L}^{2}}}+{
\frac {1}{1296}}+{\frac {5\,\sqrt {3}c_t}{3456\,{L}^{2}}}+{\frac 
{c_W}{384\,{L}^{3}}}~.
\end{eqnarray}

\subsection{The near-horizon expansion} 

Without any loss of generality,
the horizon is located at $\rho=0$.
There one assumes the existence of a power series expansion for the function $a(\rho)$, with
\begin{eqnarray}
\label{nh-susy}
a(\rho)= L \sum_{k\geq 0}  \alpha_k \left(\frac{\rho}{L}\right)^k~.
\end{eqnarray}
Then, after replacing the above expression in the sixth-order equation (\ref{eqa}) 
and solving order by order in $k$, one finds that
 the problem possesses 
(at least)
two independent  solutions describing the near-horizon of a
BH.
The argument goes as follows. 
First, the existence of a horizon requires $\alpha_0=0$.
Then, to lowest order the eq. (\ref{eqa})
implies the algebraic relation
\begin{eqnarray}
\label{nh-susy-r1}
\frac{(8+13 \alpha_1^2)\alpha_2}{\alpha_1^3}=0,
\end{eqnarray}
which implies $\alpha_2 =0$ and $\alpha_1\neq 0$.
The next order relation reads
\begin{eqnarray}
\label{nh-susy-r2}
\frac{(-8+11 \alpha_1^2)\alpha_4}{\alpha_1^3}=0,
\end{eqnarray}
which admits two independent solutions.
The first one has $\alpha_4=0$,
and leads to an expression for $a(\rho)$
containing odd powers of $\rho$ only,
with
\begin{eqnarray}
\label{sola0}
\frac{a(\rho)}{L}= \alpha_1 \frac{\rho}{L}
+ \alpha_3 \left(\frac{\rho}{L}\right)^3
+\frac{3\alpha_3^2}{10\alpha_1}\left(\frac{\rho}{L}\right)^5
+\frac{3\alpha_3^3}{70\alpha_1^2}\left(\frac{\rho}{L}\right)^7
+O(\rho^9),
\end{eqnarray}
in terms of two coefficients $\alpha_1$, $\alpha_3$.
However, after using the
scaling symmetry (\ref{scale}) (with $\lambda=\sqrt{\frac{\alpha_1}{6\alpha_3}}$),
one finds that this corresponds in fact to 
the small-$\rho$ expansion of the Gutowski-Reall solution (\ref{nGR1})
(where $\alpha=\alpha_1$).
 
The second choice to satisfy the Eq. (\ref{nh-susy-r2})  
is
$\alpha_1=2\sqrt{\frac{2}{11}}$,
which leads
to a second consistent small-$\rho$
expansion of $a(\rho)$
different from (\ref{sola0}).
{One should remark that this possibility has been noticed in  Ref. \cite{Cassani:2014zwa}, where the 
near-horizon expression of $f(\rho)$
has been already displayed.}

The first few terms in this alternative expression of $a(\rho)$ are
\begin{eqnarray}
\label{sola0n}
\frac{a(\rho)}{L}= 2\sqrt{\frac{2}{11}}\frac{\rho}{L}
+ \alpha_3 \left(\frac{\rho}{L}\right)^3
+ \alpha_4 \left(\frac{\rho}{L}\right)^4
+\frac{3}{20}\sqrt{\frac{11}{2}}\alpha_3^2 \left(\frac{\rho}{L}\right)^5
+\frac{1}{8}\sqrt{\frac{11}{2}} \alpha_3   \alpha_4 \left(\frac{\rho}{L}\right)^6+\dots,
\end{eqnarray}
in terms of two undetermined parameters
$ \alpha_3$ and $\alpha_4$.
The near-horizon expansion of other functions which enter 
the line element 
(\ref{metric0}) read
\begin{eqnarray}
\label{sola0n1}
&&
\frac{b(\rho)}{L}=  \frac{16}{11}\frac{\rho}{L}
+16\sqrt{\frac{2}{11}} \alpha_3\left(\frac{\rho}{L}\right)^3
+20\sqrt{\frac{2}{11}} \alpha_4\left(\frac{\rho}{L}\right)^4
+\frac{48 \alpha_3^2}{5}\left(\frac{\rho}{L}\right)^5+\dots,
\\
&&
\nonumber
f(\rho)=
\frac{32}{7 }\left(\frac{\rho}{L}\right)^2
-\frac{1424\sqrt{22} }{49 }\alpha_3\left(\frac{\rho}{L}\right)^4
-\frac{176\sqrt{22} }{3 }\alpha_4\left(\frac{\rho}{L}\right)^5+\dots ,
\\
&&
\nonumber
\frac{\Psi(\rho)}{L}=
-\frac{147 }{1408 }\left(\frac{L }{\rho}\right)^2
-\frac{3885  \alpha_3 }{128\sqrt{22}}
-\frac{1127}{128}\sqrt{\frac{11}{2}}\alpha_4  \frac{\rho}{L}
-\frac{92601}{1280} \alpha_3^2  \left(\frac{\rho}{L}\right)^2+\dots~.
\end{eqnarray} 
This solution translates into the following  near-horizon
expansion in terms of the metric Ansatz (\ref{metric}):
\begin{eqnarray} 
\label{nhl}
&&
F_1(\rho)=\frac{7L^2}{32 \rho^2}+\dots,~
F_2(\rho)=\frac{7L^2}{11}+\dots,~
F_3(\rho)=\frac{455L^2}{484}+\dots,~
\\
\nonumber
&&
F_0(\rho)=\frac{131072 }{3185 L^4} \rho^4+\dots,~
W(\rho)=-\frac{c}{2}+\dots,
\end{eqnarray}
while for the gauge potential one finds
\begin{eqnarray} 
\label{nh2}
&&
a_0(\rho)=\frac{7\sqrt{3}}{44(4c-1)} +\left(\frac{12\sqrt{\frac{6}{11}} \alpha_3}{1-4c}+\frac{16\sqrt{3}}{7L^2}\right)\rho^2+\dots,
\\
\nonumber
&&
a_k(\rho)=
\frac{7\sqrt{3}L}{88}
-6\sqrt{\frac{6}{11}} \alpha_3 L \rho^2
-175\sqrt{\frac{2}{33}} \alpha_4 L \rho^4+\dots~,
\end{eqnarray}
while
\begin{eqnarray}
A_\mu A^\mu= -\frac{3}{32}-\frac{63 \sqrt{\frac{11}{2}}}{32L^2}\alpha_3 \rho^2+\dots.
\end{eqnarray}
 
Also, this near-horizon expansion implies the following expressions for the constants $c_t$, $c_W$,
which enter the first integrals (\ref{1stInt-gen}):
\begin{eqnarray} 
\label{nh22}
 c_t=\frac{203\sqrt{3}L^2}{484},~~c_W=-\frac{2303L^3}{5324}.
\end{eqnarray}
Then, after replacing in (\ref{a4a6-inf}),
one finds an expression 
of the far field coefficients 
$a_4$
and $a_6$
in terms of $v$ and $a_2$.

\subsection{The   solutions } 
\subsubsection{Numerical approach } 

We have not succeeded in solving  analytically\footnote{
However, one cannot exclude the existence of a (partial) analytical solution.
For example,  a closed form (approximate) expression 
of the supersymmetric solitons
has been reported in
Ref. \cite{Cassani:2014zwa}.
The solution there has been found to first order 
in a 
perturbative expansion around the AdS background
in terms of $c_m$.
So far we did not succeed in finding a similar expression in the BH case.
}
the sixth-order Eq.(\ref{eqa}),
to find a solution
connecting the asymptotics (\ref{sola0n}) and (\ref{inf-rho}).
However, its numerical integration is straighforward.
In our approach,
we reformulate the problem in terms  of a new function
$\hat a(\rho)=e^{-\frac{\rho}{L}} a(\rho)$,
which remains finite as $\rho \to \infty$.
Similar to the treatment of the non-supersymmetric case,
a new radial compactified
radial coordinate $x$
is introduced,  
with  $0\leq x\leq 1$
and
 $\rho=\frac{x}{1-x}$.
The resulting differential equation for $\hat a(x) $
is written as a set of six first-order equations, which are solved with 
the following boundary conditions 
\begin{eqnarray}
&&
\nonumber
\hat a \big |_{x=0} =0,~~
\frac{d\hat a}{dx}\bigg |_{x=0}=2\sqrt{\frac{2}{11}},~~
\frac{d^2\hat a}{dx^2}\bigg |_{x=0}=0,~
\frac{d^4\hat a}{dx^4}\bigg |_{x=0}=16\sqrt{\frac{2}{11}}+8\frac{d^3\hat a}{dx^3}\bigg |_{x=0}+24\xi,~
\\
&&
 \frac{d^5\hat a}{dx^5}\bigg |=
168\sqrt{\frac{2}{11}}+\sqrt{\frac{11}{8}}\left(\frac{d^3\hat a}{dx^3}\right)^2\bigg |+
56\frac{d^3\hat a}{dx^3}\bigg |+360\xi ~~({\rm with}~~\xi \equiv \alpha_4),
\end{eqnarray} 
and
\begin{eqnarray}
\frac{d\hat a}{dx}\bigg |_{x=1}=0~.
\end{eqnarray} 
The first five algebraic relations above result  directly from  (\ref{sola0n}),
while the condition at infinity is compatible with the asymptotics (\ref{inf-rho}).
Also note that 
the parameter
$\alpha_3=\frac{d^3\hat a}{dx^3}\big |_{x=0}$ is free,
resulting from the numerical approach.
Hence the boundary conditions present a single free parameter $\alpha_4 =\xi$, 
which we use as a control parameter to generate non-trivial solutions.

As an initial solution, we take $\hat a=1$ 
and change the value of  $\xi$ in small steps. 
This setting works well, and the numerical iteration converges quickly
(note that we have used the same solver \cite{COLSYS} 
as in the generic case). 
The solutions have around $2000$ points
in the mesh, with a numerical error of  
$10^{-8}$ or lower for $a(x)$ and its derivatives. 

\begin{figure}[t]
\centering
\subfigure[]{\includegraphics[scale=0.28,angle=-90]{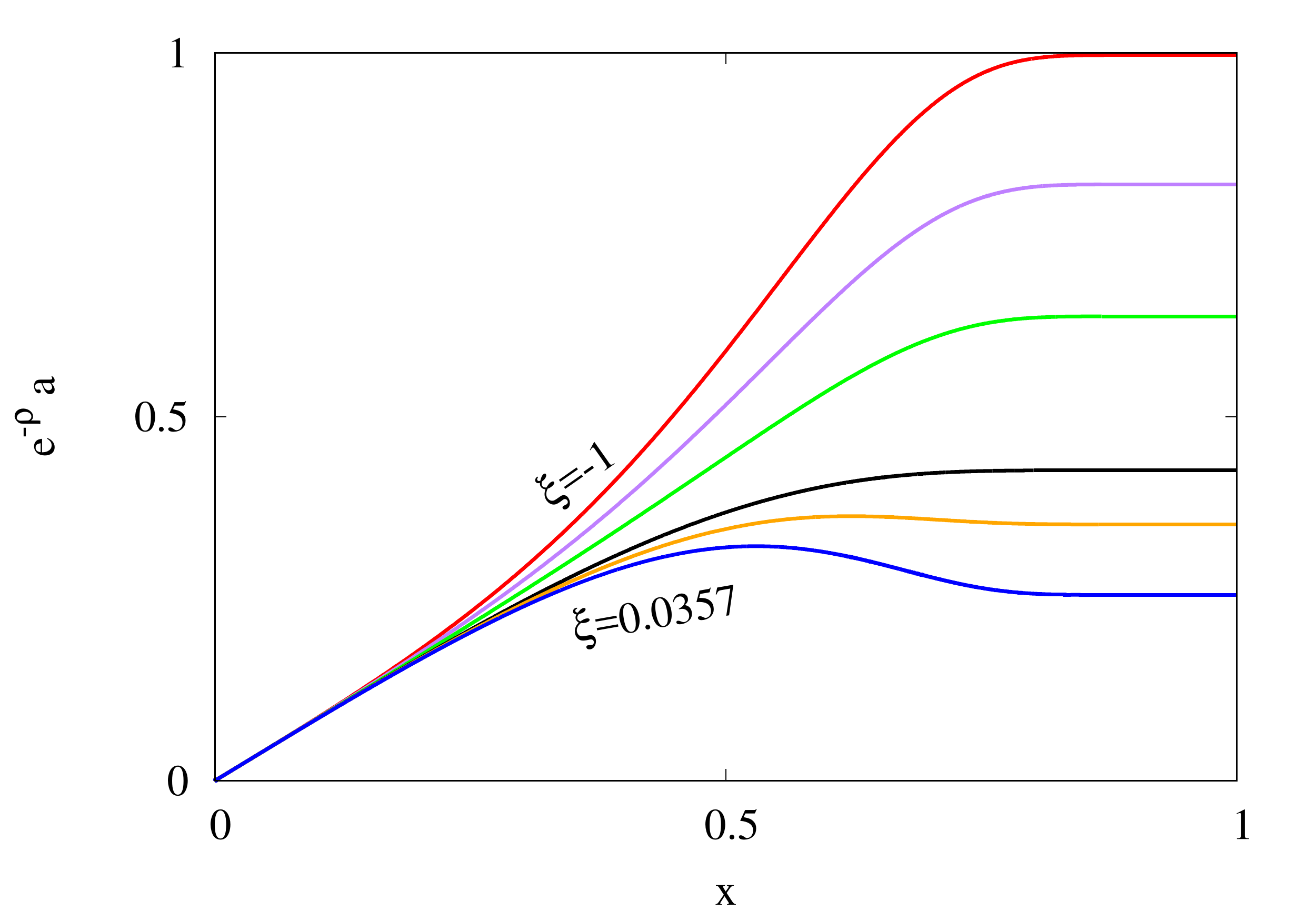}
\label{plot_profile_a}}
\subfigure[]{\includegraphics[scale=0.28,angle=-90]{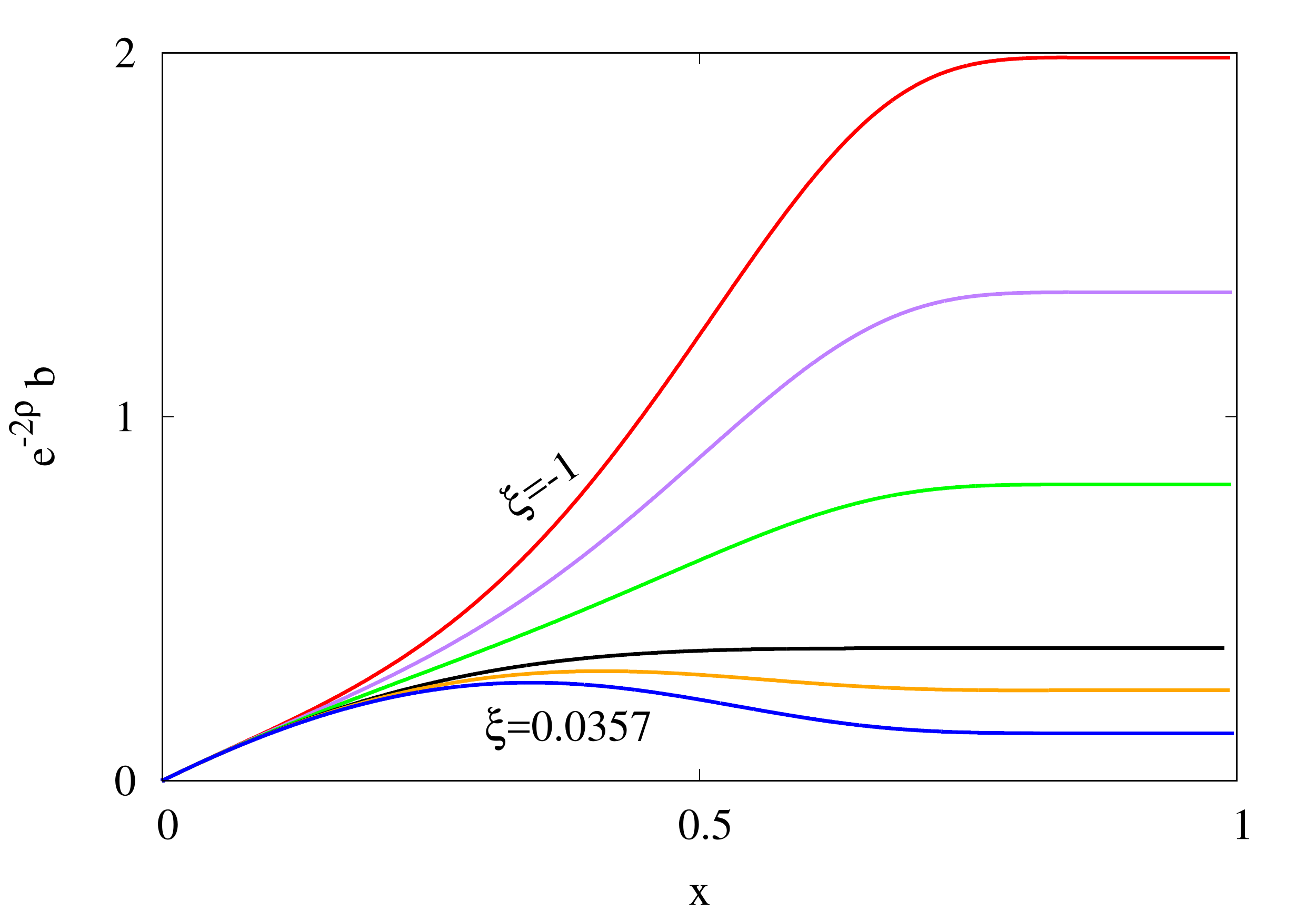}
\label{plot_profile_b}}
\caption{(a) Profiles for the function $a$ (rescaled by a factor $e^{-\rho}$), as a function of the compactified radial coordinate 
$x=\rho/(\rho+1)$. From top to bottom we show the profile for the values $\xi=-1, -0.5, -0.18, 0, 0.025$ and $0.0357$.
(b) Similar profiles for the rescaled metric function $b$. 
Here and in Figures \ref{plot_profile_2}-\ref{plot_profile_Psi}
the results are shown for 
the same values of the near-horizon parameter $\xi \equiv \alpha_4$.
}
\label{plot_profile_1}
\end{figure}
\begin{figure}[t]
\centering
\subfigure[]{\includegraphics[scale=0.28,angle=-90]{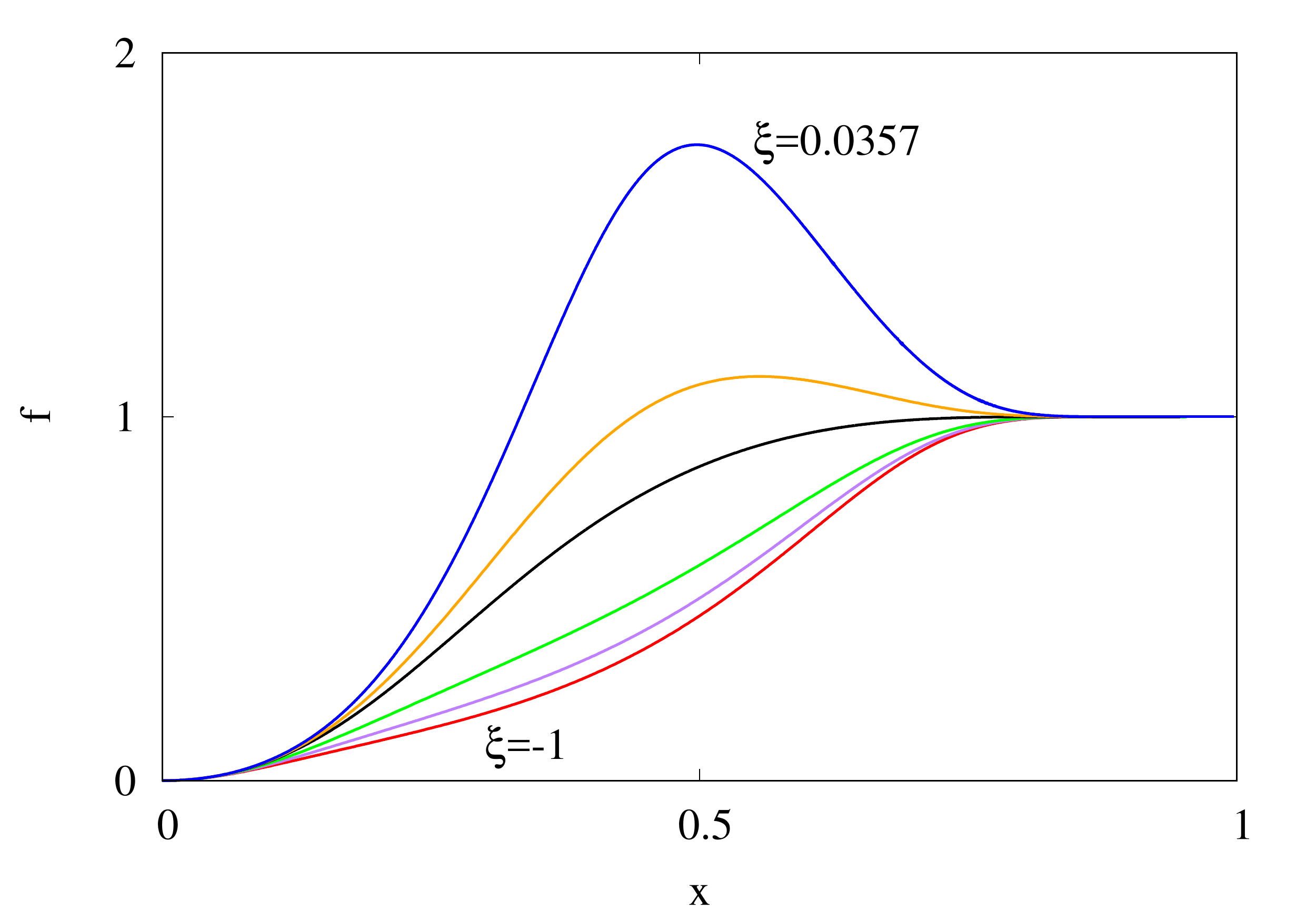}
\label{plot_profile_f}}
\subfigure[]{\includegraphics[scale=0.28,angle=-90]{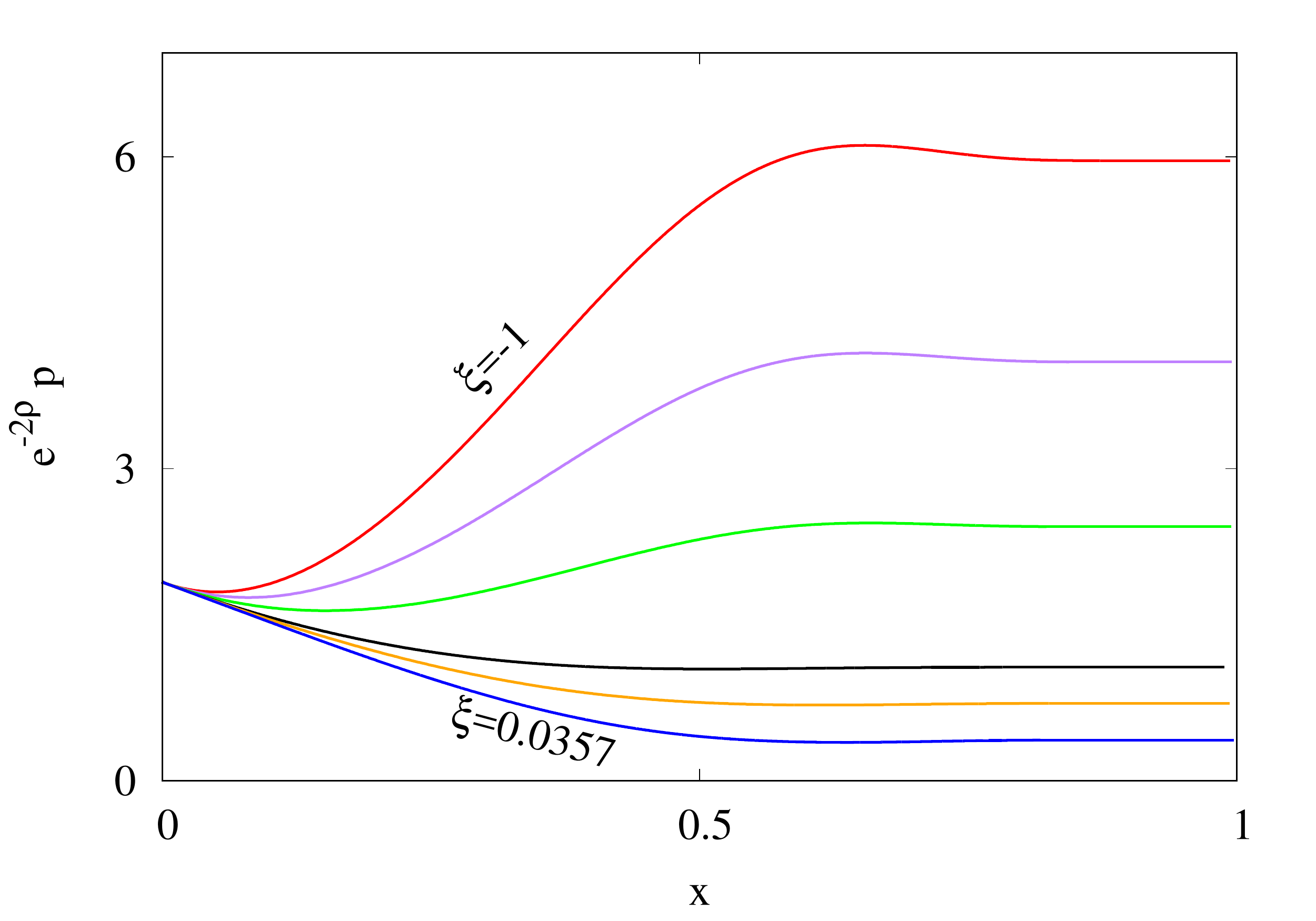}
\label{plot_profile_p}}
\caption{ (a) Similar profiles for the function $f$  and  (b) the rescaled function $p$
 in terms of the compactified radial coordinate $x=\rho/(\rho+1)$.}
\label{plot_profile_2}
\end{figure}
\begin{figure}[t]
\centering
\subfigure[]{\includegraphics[scale=0.28,angle=-90]{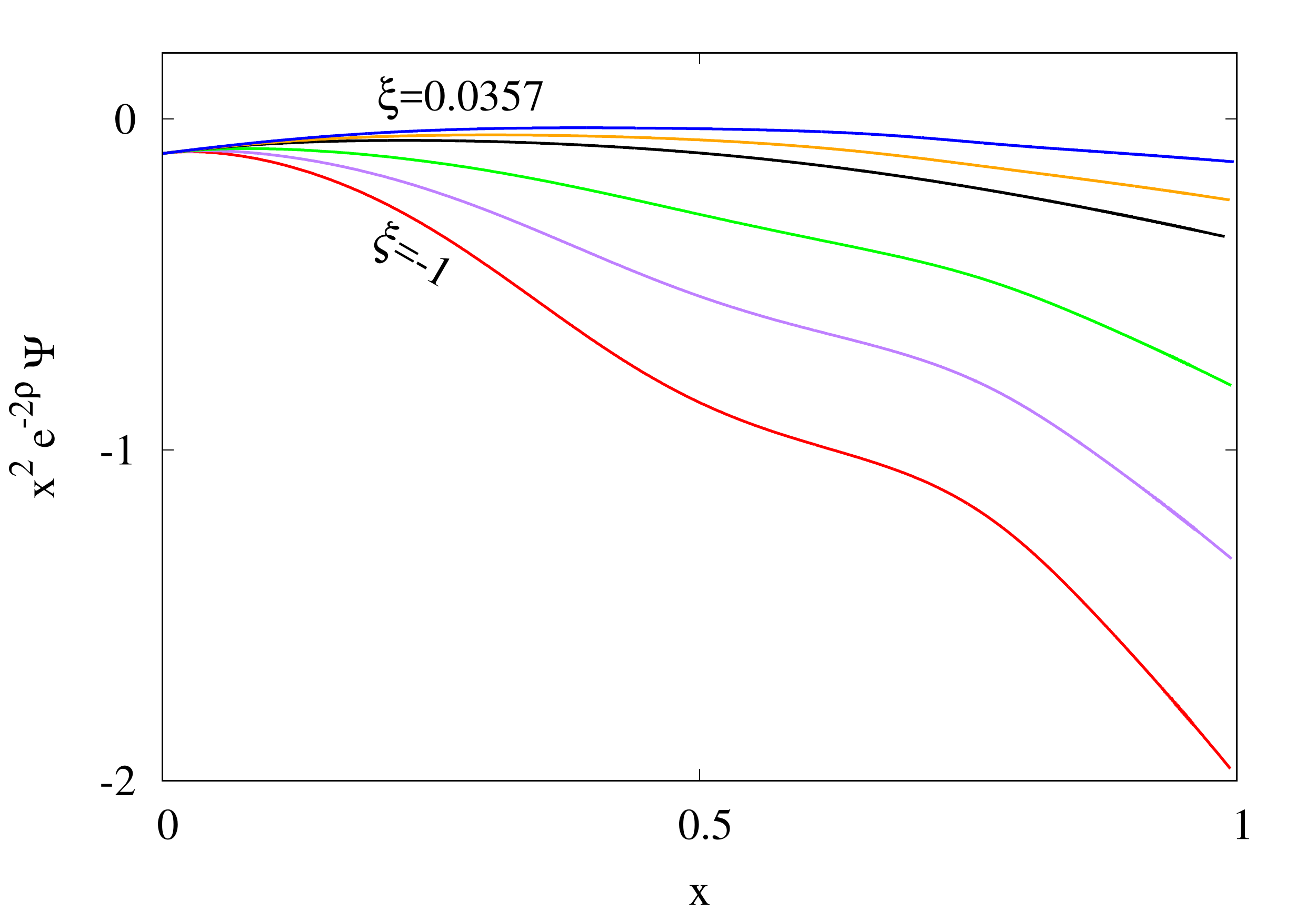}
\label{plot_profile_Psi}}
\subfigure[]{\includegraphics[scale=0.28,angle=-90]{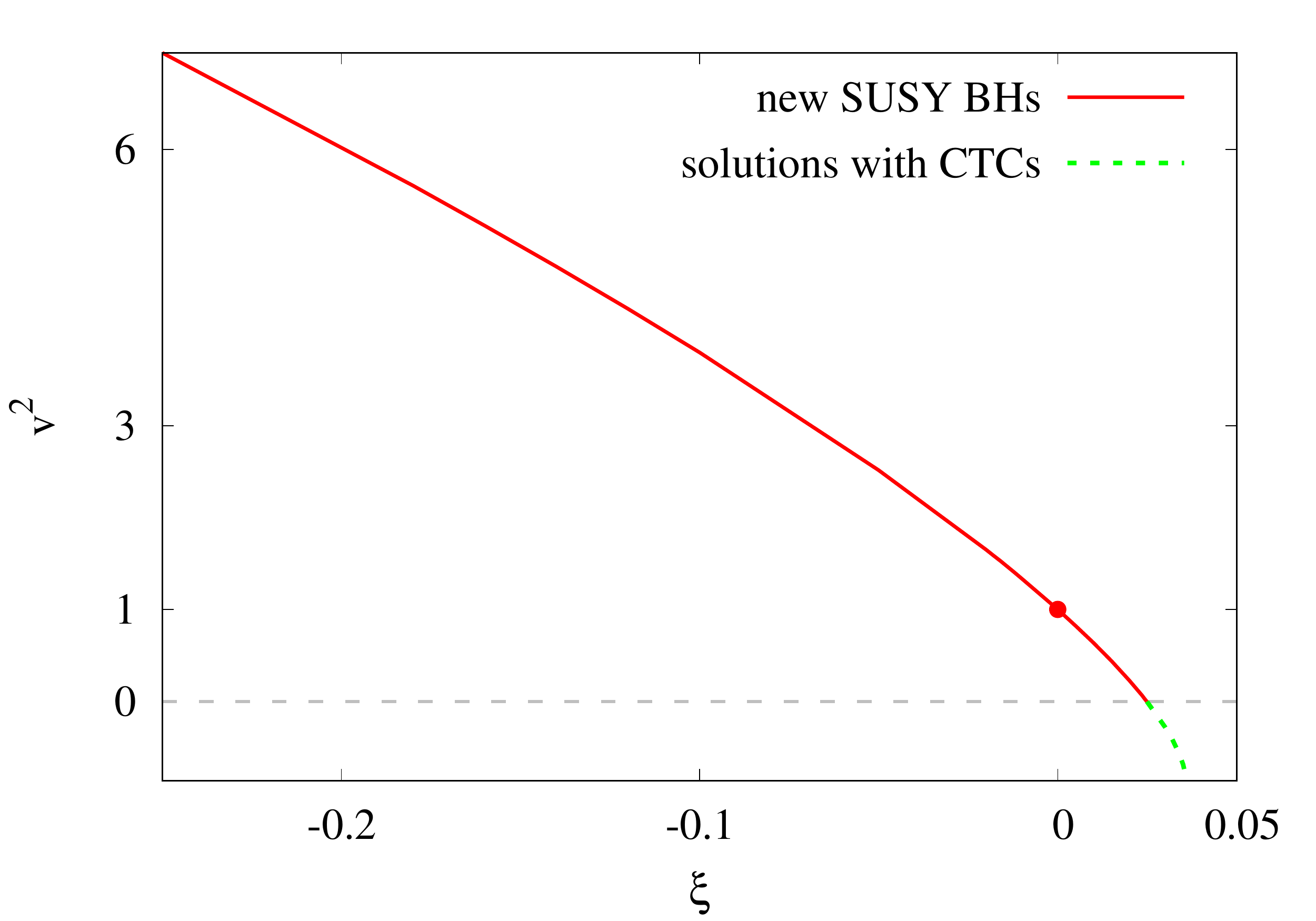}
\label{plot_xi_v2}}
\caption{(a) Similar profiles for the function $\Psi$ vs. the compactified radial coordinate $x=\rho/(\rho+1)$. 
(b) The squashing parameter $v^2$ as a function of the near-horizon parameter $\xi \equiv \alpha_4$.
The red point marks the critical Gutowski-Reall solution from which the new susy black holes emerge.
 Note the existence a set of regular solutions with negative $v^2$ and closed timelike curves (dashed green line).}
\label{plot_profile_3}
\end{figure}
\begin{figure}[t]
\centering
\subfigure[]{\includegraphics[scale=0.28,angle=-90]{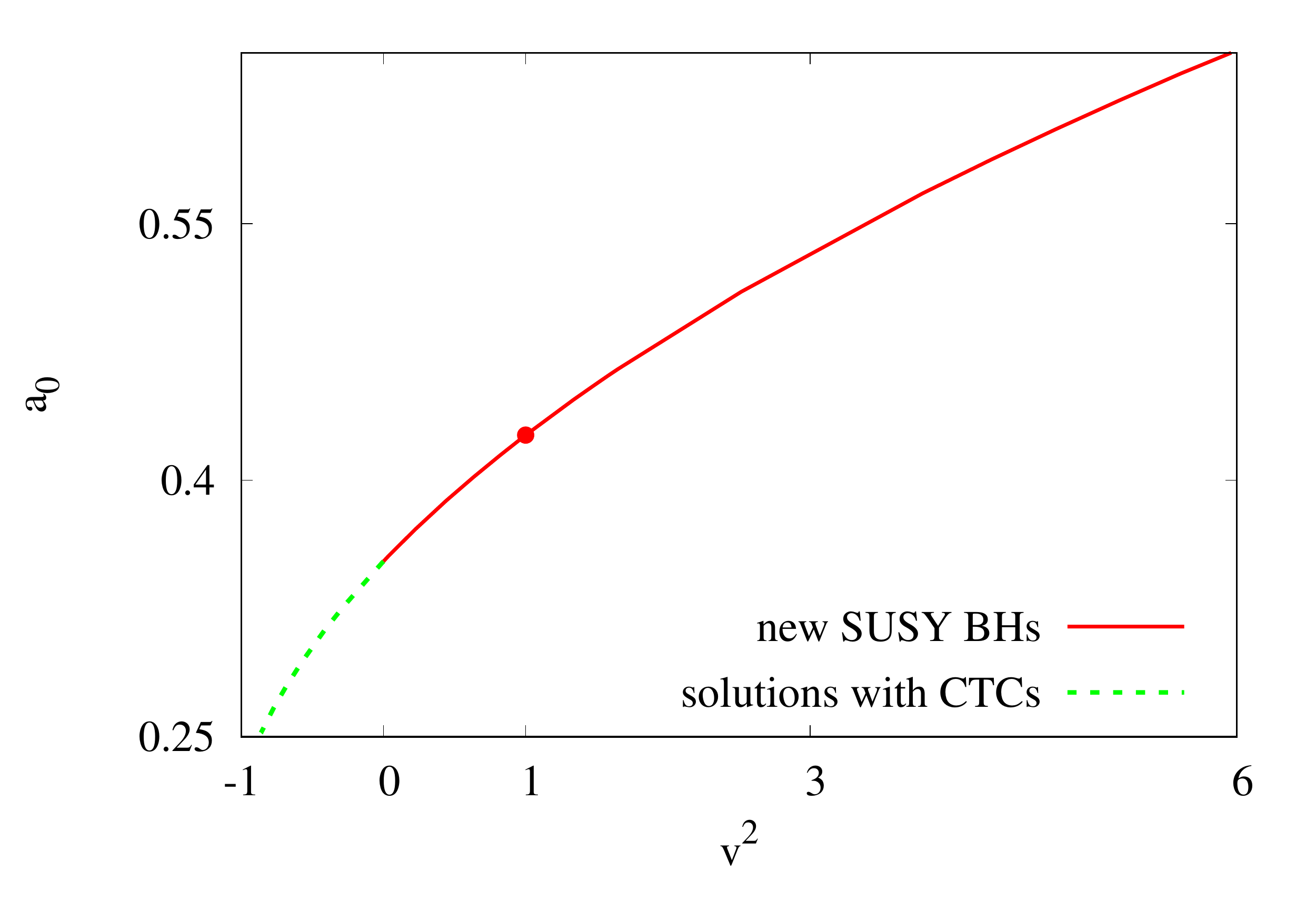}
\label{plot_v2_a0}}
\subfigure[]{\includegraphics[scale=0.28,angle=-90]{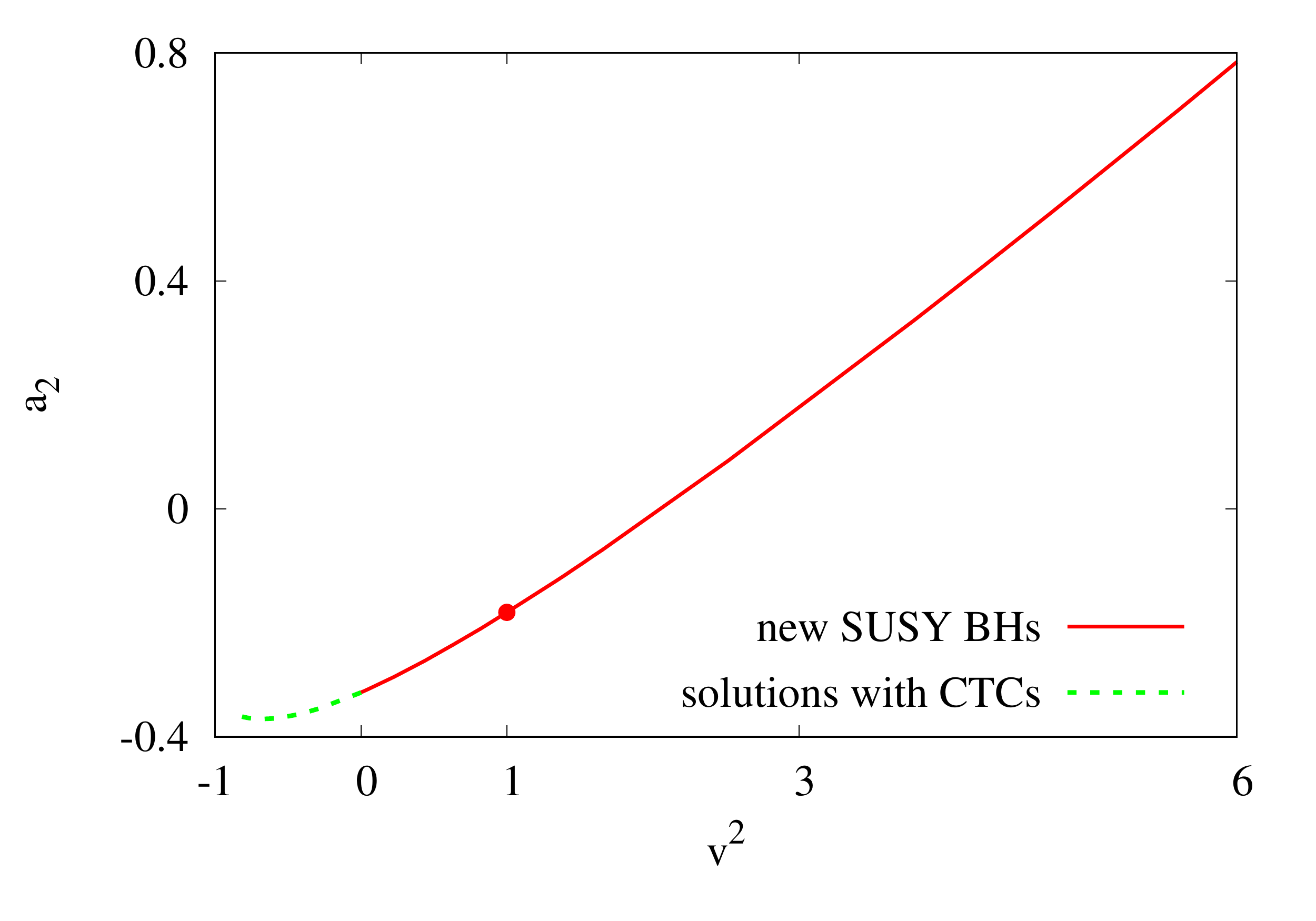}
\label{plot_v2_a2}}
\caption{(a) The large-$\rho$ expansion parameter $a_0$ as a function of the squashing $v$. (b) The same for the parameter $a_2$.
Here and in Figures \ref{plot_v2_2}
the red point marks the Gutowski-Reall solution from where the new susy black holes emerge, and the dashed green line the subset of solutions with  closed timelike curves}
\label{plot_v2_1}
\end{figure}
\begin{figure}[t]
\centering
\subfigure[]{\includegraphics[scale=0.28,angle=-90]{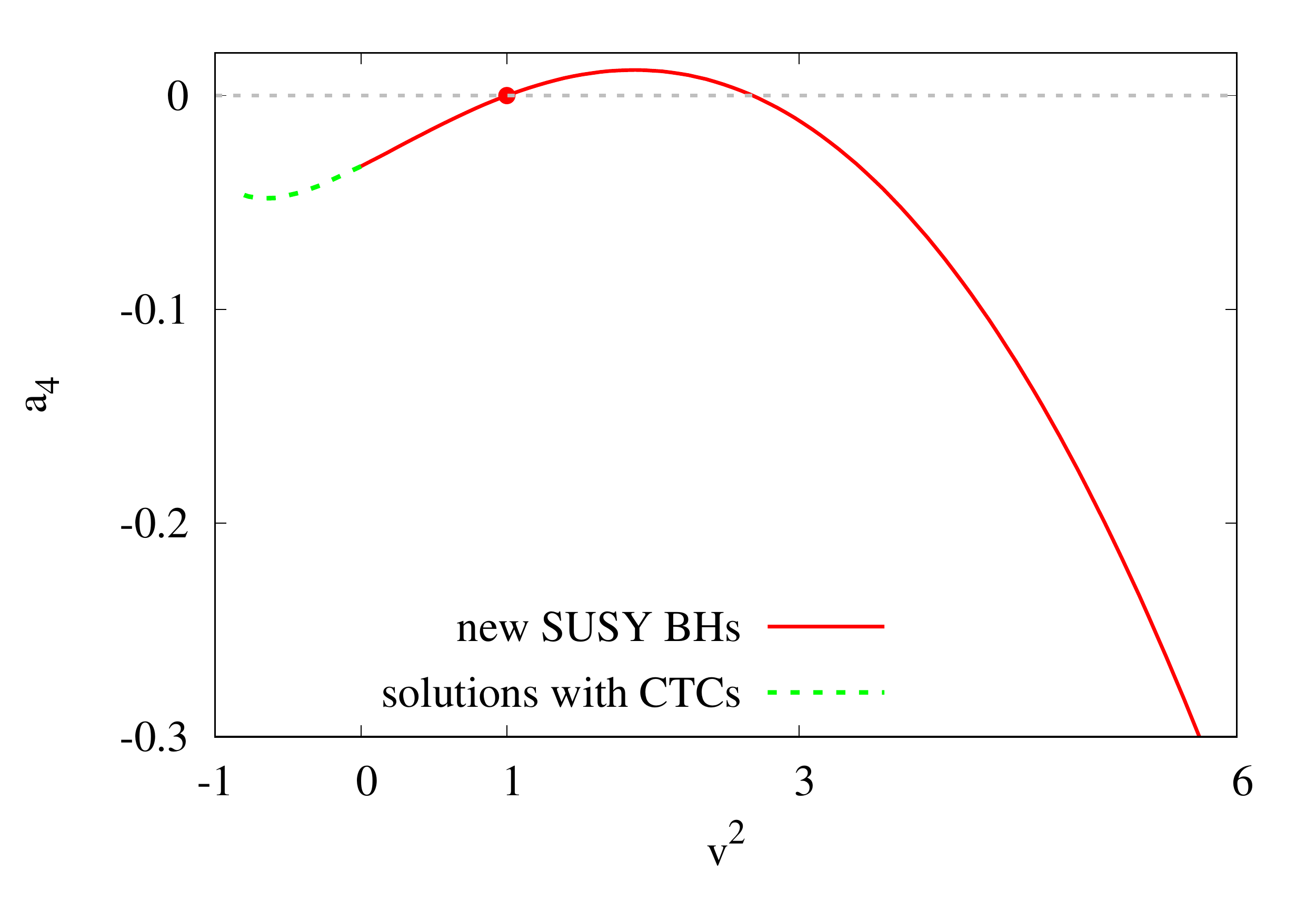}
\label{plot_v2_a4}}
\subfigure[]{\includegraphics[scale=0.28,angle=-90]{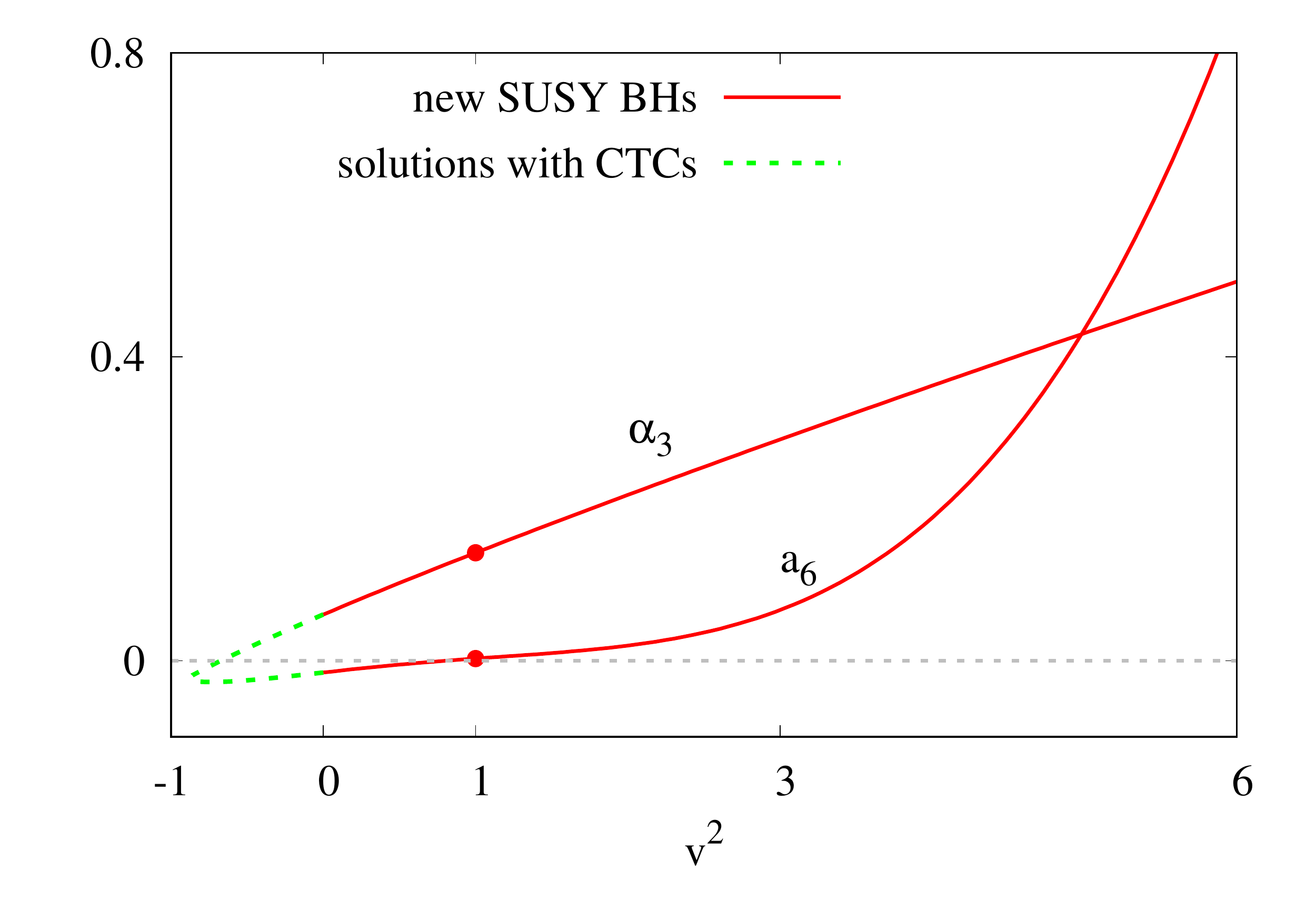}
\label{plot_v2_a3_horizon_a6}}
\caption{(a) The large-$\rho$ expansion parameter $a_4$ as a function of the squashing $v$. 
(b) A similar figure for the parameters $a_6$ and ${\alpha}_3$. 
}
\label{plot_v2_2}
\end{figure}

Once the profile of $a(\rho)$
is known, the full solution is reconstructed from (\ref{set})
together with (\ref{transf2}), (\ref{limit}).
The coefficients $\{ a_0,c,a_2,a_4,a_6 \}$
in the far field expansion  (\ref{inf-rho}) 
are extracted from the numerical output.

In Fig. \ref{plot_profile_a} we present several profiles 
of the function $\hat a (x)$.
The corresponding profiles of the functions of the susy Ansatz 
(multiplied with suitable factors) 
 are shown in Figure \ref{plot_profile_b}, \ref{plot_profile_2} and \ref{plot_profile_Psi}.
 The results in these plots are found for several  values of 
$\xi$.

The relation between the squashing parameter $v^2$ of the far-field expansion and the parameter $\xi$ of the near-horizon behaviour is presented in Figure \ref{plot_xi_v2}.  Note that the solutions are not invariant to changes in the
sign of $\xi$.
For example, the squashing $v^2$ grows for negative $\xi$, 
while for positive values it decreases, with 
$v\to 0$ for  $\xi \simeq 0.03$.
For larger values of $\xi$, a small branch of solutions with CTCs is found (dashed green line). Also note that the solution with $\xi=0$ (red point) is a particular Gutowski-Reall black hole.

The relation of the remaining coefficients $\{ a_0,a_2,a_4,a_6 \}$
in the far field expansion  (\ref{inf-rho}) with the squashing parameter $v$ 
is presented in Figures \ref{plot_v2_1} and \ref{plot_v2_2}, 
where we mark again the Gutowski-Real solution with a red point, and the sector with CTCs in green.

Finally, we also show the dependence of the parameter ${\alpha}_3$ as a function of the squashing $v^2$ in Fig. \ref{plot_v2_a3_horizon_a6}. We mark again the Gutowski-Real solution with a red point, and the sector with CTCs in green.

\subsubsection{General properties} 

The computation of all quantities of interest of the solution is 
a direct application of the general formalism in Section 2.
This results in the following expression
for the mass, angular momentum and electric charge:
\begin{eqnarray}
&&
\nonumber
M=\pi L^2
\left(
\frac{7913}{34848}+\frac{33280}{35937v^2}-\frac{7}{36}v^2+\frac{89}{864}v^4
\right),
 \\
&&
\label{MJQ-susy-BH}
J=-\pi L^3
\left(
 \frac{16640}{35937}
-\frac{2795}{8712}v^2
+\frac{1}{9} v^4
-\frac{1}{27}v^6
\right),
\\
&&
\nonumber
Q= -\pi \sqrt{3} L^2
\frac{1}{13068}
 \left( 6449-1936\,{v}^{2} + 968\,{v}^{4}\right) ,
\end{eqnarray}
which depend on the squashing parameter $v$, only.

The supersymmetric BHs possess a nonzero horizon area
\begin{eqnarray}
A_{H}=7 \pi^2 L^3 \frac{ \sqrt {455}}{121},
\end{eqnarray}
which does not depend on the parameter $v$.
This is precisely what can be seen  in the Figure \ref{plot_Ah_bh3d}, 
where we fixed the conserved charges as in relation (\ref{nh22}), 
and the magnetic parameter $c_m$ such that equation (\ref{cancel}) is satisfied. 
Then the extremal limit yields in fact the susy BHs, 
and all the horizon quantities converge to a single point (red), 
although the global charges are different (see Fig. \ref{plot_M_bh3d}, red line).
Their horizon angular momentum, electrostatic potential,
angular velocity,
horizon mass
and horizon deformation are
\begin{eqnarray}
J_{H}={\frac {9555\,\pi {L}^{3} }{170368 }},
~~
\Phi_{H}=\frac{\sqrt{3}}{2},
~~
\Omega_{H}=\frac{2}{L v^2},
~~
\epsilon_{H}=\sqrt{\frac{65}{44}},
~~
M_{H}=\frac{28665 \pi L^2}{85184 v^2}.
\end{eqnarray}

Now we can return to 
  Figure \ref{MJQ}, where we display a  {\it (mass-angular momentum-electric charge) } 
diagram summarizing the picture for three different classes of supersymmetric 
solutions: $(i)$ the Gutowski-Reall BHs,
$(ii)$  the supersymmetric solitons in \cite{Cassani:2014zwa}
and $(iii)$  the new BHs in this work.
One can notice that the curves for the last two types of solutions do never intersect.
However, one can see that the susy squashed magnetized BHs bifurcate
from a critical Gutowski-Reall solution.
That is where the
two  BH  curves
meet at a critical configuration with
\begin{eqnarray}
&&
M^{(c)}= \frac{49213 L^2 \pi}{42592},~~
J^{(c)}=-\frac{2303 L^3 \pi}{10648},~~ 
Q^{(c)}= -\frac{203\sqrt{3} L^2 \pi}{484}.
\end{eqnarray}
The critical Gutowski-Reall BH has a control parameter (see (\ref{nGR2}))
\begin{eqnarray}
\alpha =2\sqrt{\frac{2}{11}} , 
\end{eqnarray}
while
on the squashed BHs side, this corresponds to the limit\footnote{The supersymmetric solitons in 
\cite{Cassani:2014zwa}
approach instead the globally AdS background as $v\to 1$.}
$v\to 1$ 
($i.e.$ a round $S^3$ sphere at infinity and no boundary magnetic field).

Also, note the solutions with
  $v\simeq 3.61691$ 
possess a vanishing total angular momentum, 
although they still rotate in the bulk ($T_\psi ^t\neq 0$ and $\Omega_H\neq 0$),
while their mass and electric charge are positive.

The limit $v= 0$ ($i.e.$ $c=1/4$ in the far field expansion (\ref{inf-rho}))
is special.
While the horizon geometry does not change,
the asymptotics are different in this case, 
the conformal boundary structure being lost.
One finds $e.g.$
\begin{eqnarray}
&&
F_0=q_0\left(\frac{r}{ L}\right)^2+\dots,~~
F_1(r)=1-\frac{1+8a_2+2\log (\frac{r}{ L})}{6a_0^2}\left(\frac{L}{ r}\right)^2+\dots,~~
\\
\nonumber
&&
F_2(r)=4 a_0^2 r^2+\frac{2}{3}L^2\left(4a_2-1+\log \left(\frac{r}{ L}\right)\right)+\dots,~~
\\
\nonumber
&&
F_3(r)=q_3\left(\frac{L}{ r}\right)^2+\dots,~~
W(r)=q_w \left(\frac{r}{ L}\right)^4+\dots,~~
\end{eqnarray}
with $q_0$, $q_3$ and $q_w$ possessing a complicated dependence on $a_0,a_2,a_4$ and $a_6$.
Thus,
similar to the extremal case in Section 3, the $v\to 0$ limit does not result in a black string configuration.
Moreover, this limiting solution is not asymptotically (locally) AdS 
(despite the absence of (obvious) pathologies). 
We hope to return elsewhere with a detailed study of this interesting limiting solution.

The infinite squashing limit $v\to \infty$
is taken again together with the rescaling (\ref{scale1i}), (\ref{scale2}).
This results in 
a BH solution with a different topology
($e.g.$ the horizon geometry is of the form 
(\ref{h-twist1})),
whose  spacetime asymptotics are again non-standard,
the AdS conformal boundary structure being lost. 
This limit is described by an exact solution,
which is discussed
in Appendix B.

Finally, let us mention that we have found numerical evidence 
for the existence of solutions with $v^2<0$.
However, such configurations 
possess closed timelike curves (in the bulk and on the boundary) and are less interesting.
Non-supersymmetric solutions with this behaviour exist as well. 

\section{Further remarks. Conclusions}

The solutions of the $D=5$ gauged supergravity models
play a central role in the AdS/CFT correspondence, 
providing a dual description of strongly-coupled  
CFT on
the four-dimensional AdS boundary. 
The main purpose of this work was to report the existence
of a new class of BH  solutions 
of the minimal gauged supergravity model
and to provide a discussion of their basic properties 
{(see also \cite{Blazquez-Salcedo:2017kig})}
.
They are built within the same general framework 
as the well-known  Cveti\v c-L\"u-Pope BHs,
 sharing some of their basic properties;
 for example, the horizon has a spherical topology and the two 
angular momenta have equal magnitude.
However, the conformal boundary of the new BHs in this work possesses a squashed sphere,
such that the solutions could become in certain limits black strings or 
black branes.
Moreover, new features occur as one allows for a nonvanishing value of the magnetic field at infinity.
For example, as discussed in Section 3.3, this supports the existence of smooth particle-like solitons,
which satisfy a universal relation between angular momentum and electric charge.
Moreover,  
a particular set of extremal configurations corresponds to a
new one-parameter family of supersymmetric black holes,
which were reported in Section 4.
They satisfy 
 a certain relation between the squashing parameter and the magnetic parameter and
bifurcate from a critical Gutowski-Reall configuration.

\medskip

There are many open questions and avenues for future investigation.
For example, 
the general framework in this paper provides
a ground for further study of the properties of these solutions,
such as a systematic investigation of their domain of existence,
thermodynamics and stability.
Moreover, 
various limits of the solutions briefly
mentioned in Section 3 certainly deserve a systematic study.
Also, our results in the generic non-susy case were found for 
the supersymmetric value of the CS coupling, $\lambda=1$.
However, it would be interesting to consider other values as well.
Here we remark that 
the results in 
\cite{Blazquez-Salcedo:2016rkj}
(valid EMCS BHs with $v=1$ and no boundary magnetic field) 
show the existence of new qualitative features
($e.g.$ the existence of excited solutions)
once the 
CS coupling constant exceeds a critical value.
As yet another possible direction,
we note that 
it is straightforward to extend the framework introduced in Section 2
to other (odd) spacetime dimensions $D>5$
and the same matter content.
Therefore
we predict the existence of AlAdS solutions
also in that case, which would share some basic properties 
of the solutions in this work.

Another interesting question is the relevance of such configurations in an AdS/CFT context.
Here one remarks that 
the supersymmetric solitons have been interpreted in \cite{Cassani:2014zwa}
as providing the
 gravity supersymmetric dual of an ${\cal N}=1$
supersymmetric gauge theory
on a squashed Einstein universe background,
with a nontrivial
background gauge field coupling to the R-symmetry current.
We expect that the 
interpretation of the 
supersymmetric BHs
in this work would be similar,
 describing different phases of the same $D=4$ model.

\medskip
 
Finally, 
let us remark that
the presence of a nontrivial magnetic field on the boundary can be viewed 
in a larger context as a consequence of the `{\it box-type}` behaviour of the
AdS spacetime.
As realized in  
\cite{Herdeiro:2015vaa},
\cite{Herdeiro:2016xnp},
\cite{Herdeiro:2016plq},
for the $D=4$ case,
the AdS asymptotics  
supports the existence of everywhere regular Maxwell-field multipole solutions,
which survive when including the backreaction.
That is, the U(1) field mode, which in the (asymptotically) flat case is divergent at infinity,
gets regularized,  leading to new families of Einstein-Maxwell solutions 
(which include both BHs and solitons).
Although further study is necessary,
this feature appears to be universal,
some partial results 
being reported in 
\cite{Blazquez-Salcedo:2016vwa}
for $D=2k+1\geq 5$ dimensions
(see also
\cite{Chrusciel:2016cvr},
\cite{Chrusciel:2017emq}).

This observation 
leads us to predict the existence of a variety of other 
A(l)AdS
solutions of the $D=5$
minimal gauged supergravity model.
First, we remark that for the solutions in this work, 
the asymptotics of the electric potential  are standard,
while the  magnetic part can be
interpreted as an AdS dipolar field. 
However, this is just the simplest type 
of non-standard boundary conditions
for the U(1) field
 (supported by AdS asymptotics),
which has the advantage to lead to a codimension-1 numerical problem.
More general solutions should exist as well.
For example, 
our preliminary 
numerical results indicate the existence 
of generalizations
of the $D=5$ Reissner-Nordstr\"om BHs with a vanishing
magnetic field
and an asymptotic value  of the electric potential
$a_0=c_e \cos (2\theta)$
(with $c_e$ a control parameter). 
These configurations are static and not spherically symmetric, 
without being 
possible to factorize the $\theta$-dependence 
on both metric and gauge sectors.
Therefore the numerical treatment of this problem is much more complicated,
the solutions being found by solving partial differential equations.
In the absence of an electric charge, they 
possess a nontrivial solitonic limit
and
can be interpreted as AdS electric dipoles.
More general solutions of the $D=5$
minimal gauged supergravity model possessing higher-order multipoles 
for both the electric and magnetic potentials should also exist.
We hope to return elsewhere with a systematic discussion of these aspects.

\bigskip
{\bf Acknowledgement}

We gratefully acknowledge support by
the DFG Research Training Group 1620 ``Models of Gravity''.
 E. R. acknowledges funding from the FCT-IF programme.
This work was also partially supported 
by  the  H2020-MSCA-RISE-2015 Grant No.  StronGrHEP-690904, 
and by the CIDMA project UID/MAT/04106/2013. F. N.-L. acknowledges funding from Complutense University - Santander under project PR26/16-20312.  J. L. B.-S. and J. K. acknowledge support from FP7, Marie Curie Actions, People,
International Research Staff Exchange Scheme (IRSES-
606096).

\appendix

\section{A squashed $S^3$ boundary: limiting behaviour of nutty instantons in $AdS_4$}

A simple model, which  helps to understand the 
limiting
 behaviour of the solutions in this work in terms of the squashing parameter $v$
is provided by the $AdS_4$ nut-charged instantons.
These solutions solve the vacuum Einstein equations on the Euclidean section 
and
can be written in a form
resembling (\ref{metric}), with
\begin{eqnarray}
\label{TN0}
ds^2=F_1(r) dr^2+\frac{1}{4}F_2(r)(\sigma_1^2+\sigma_2^2)+\frac{1}{4}F_3(r)\sigma_3^2,
\end{eqnarray}
($\sigma_i$ being the one-forms on $S^3$ as 
given by (\ref{sigma})),
where
\begin{eqnarray}
\label{FTN0}
\frac{1}{ F_1(r)}= \frac{r^2+n^2}{r^2-n^2}+\frac{-2Mr +\frac{1}{L^2}(r^4-6n^2r^2-3n^4)}{r^2-n^2},
~
F_2(r)=4(r^2-n^2),~F_3(r)=16n^2\frac{1} {F_1(r)}.~{~~~~~}
\end{eqnarray}
This line element can be put into the standard form 
given 
in the literature
via the coordinate transformation
\begin{eqnarray}
\label{psiFTN0}
\psi=-(\phi+\frac{\hat \psi}{2n})~,
\end{eqnarray}
which results in
\begin{eqnarray}
\label{TN}
ds^2=\frac{dr^2}{V(r)}+(r^2-n^2)(d\theta^2+\sin^2 \theta d\phi^2)+V(r)\left(d\hat \psi+4n \sin^2(\frac{\theta}{2})d\phi \right)^2,
~{\rm with}~V(r)=\frac{1}{ F_1(r)}.~{~~~~}
\end{eqnarray}
The  nutty instantons possess two constants:
  $M$, which is a  mass parameter,  and $n$--the nut parameter.
Also,  $r$ is a radial coordinate,
while $\hat \psi$ parameterizes a circle $S^1$, which is fibred over the two
sphere $S^2$, with coordinates $\theta$ and $\phi$.
As a result, 
 the metric
(\ref{TN})
 is only  locally  asymptotically AdS,
while its boundary is a squashed
three sphere as $r \to \infty$ .
As discussed in \cite{Chamblin:1998pz},
this becomes a round $S^3$ for $M=0$, $n=L/2$,
such that (\ref{TN}) corresponds  to  $AdS_4$.
In the generic case, the absence of a conical singularity imposes some constraints
on the value  of $M$, the solutions possessing a variety of interesting features.
However, these aspects are of no interest in the context of this work
(for a detailed analysis,  we refer the reader to Refs.
 \cite{Chamblin:1998pz}, 
\cite{Emparan:1999pm},
\cite{Clarkson:2002uj}, 
\cite{Astefanesei:2004kn}, 
\cite{Astefanesei:2004ji},
\cite{Yonge:2006tn}).
Instead we shall simply only consider their small/large $n$ limits.
 
For $n= 0$,  one recovers the Schwarzschild-AdS Euclideanized solution, 
\begin{eqnarray}
\label{TN1}
ds^2=\frac{dr^2}{1-\frac{2M}{r}+\frac{r^2}{L^2}}+r^2(d\theta^2+\sin^2 \theta d\phi^2)+(1-\frac{2M}{r}+\frac{r^2}{L^2}) d\hat \psi^2~,
\end{eqnarray}
with a $S^2\times S^1$ topology of an $r=const.$ surface. 
Note that,
for the parametrization (\ref{TN0}),
 the limit $n\to 0$
should be taken 
with a rescaled $\psi$-coordinate, which, however has a period fixed by $M,L$.

Another case of interest is $n\to \infty$, in which a different solution is recovered.
To understand this limit, 
one starts again with the line element  (\ref{TN})
and defines the scaled coordinates
\begin{eqnarray}
r=\lambda \bar r,~~\theta= \frac{\Theta}{\lambda},~~\hat\psi= \frac{\Psi}{\lambda},
\end{eqnarray}
together with
\begin{eqnarray}
n=\lambda N,~~M=\bar M\lambda^3.
\end{eqnarray}
Let us now consider the limit $\lambda\to \infty$.
Then the line element (\ref{TN})
becomes
\begin{eqnarray}
\label{mn}
ds^2=f_1(\bar r)d\bar r^2+f_2(\bar r)(d\Theta^2+ \Theta^2 d\phi^2)+f_0(\bar r)\left(d\Psi+  4 N  \left(\frac{\Theta}{2}\right)^2 d\phi \right)^2,
\end{eqnarray}
where
\begin{eqnarray}
 f_0(\bar r)=\frac{1}{f_1(\bar r)}= \frac{-2\bar M \bar r +\frac{1}{L^2}(\bar r^4-6N^2\bar r^2-3N^4)}{\bar r^2-N^2},~~
f_2(\bar r)=\bar r^2-N^2,
\end{eqnarray}
which is the planar version of the Taub-NUT-AdS spacetime \cite{Chamblin:1998pz}.
This limit possesses a number of interesting properties; 
here we mention only 
that an $r=const.$ surface has an $R^3$ topology,
with a warped product $R\times R^2$.
Also, they possess no Misner string singularity
($i.e.$ the periodicity of $\Psi$
is arbitrary) 
with a breakdown of
the entropy/area relationship
\cite{Astefanesei:2004kn}, 
\cite{Astefanesei:2004ji}.

\section{The black branes} 
\subsection{The generic case} 
%
The 
$D=5$
Schwarzschild-AdS BH possesses a
well-known
 generalization with a
planar horizon topology,
which approaches a Poincar\'e patch at infinity.
Adding extra charges 
(and also a boundary magnetic field)
results in generalizations of the 
Schwarzschild black brane, which possess 
a variety of interesting properties
(see $e.g.$
\cite{D'Hoker:2009bc},
\cite{D'Hoker:2010ij}).

We have found that the
infinite squashing limit
of the BHs in this work
leads to a new set of solutions which can be viewed 
as a generalization of the configurations in  
\cite{D'Hoker:2009bc},
\cite{D'Hoker:2010ij}.
Their horizon (and the spacelike part of the boundary metric)
is $R^3$, being
topologically a direct product of a flat direction and the $R^2$-plane.
However, this product is 'twisted' (or warped), and the induced geometry
is not flat. 

These solutions have a line element
\begin{eqnarray}
\label{metric-BB}
&&ds^2 =
-f(r)\frac{r^2}{L^2}dt^2+
\frac{1}{f(r)}
\bigg[
m(r)L^2\frac{dr^2}{r^2}
  + \frac{1}{4} r^2 
	\bigg(
	m(r) ( d\Theta^2+\Theta^2 d\phi^2)
	\\
\nonumber
&&{~~~~~~~~~~~~~~~~~~~~~~~~~~~~~~~~~~~~~~~~~}	
	+ n(r) \big(d\Psi+2v  \left(\frac{\Theta}{2}\right)^2d\phi -\frac{2\omega(r)}{r} dt \big)^2
	\bigg)
	\bigg]
,
\end{eqnarray} 
their gauge potential being
\begin{eqnarray}
A=a_0(r)dt + \frac{1}{2} a_k(r) \left(d\Psi+2v \left(\frac{\Theta}{2}\right)^2 d\phi\right)~.
\end{eqnarray} 

Again, we assume that the configurations are AlAdS, with a conformal
boundary metric given by (\ref{twist2}),
while $a_k \to c_m$  as $r\to \infty$.
Then a far field solution can be constructed in a systematic way,
the leading order terms in the asymptotics being
\begin{eqnarray}
\nonumber
&&
f(r)=1-\frac{4 v^2}{9} \left( \frac{L}{r}\right)^2
+\left[
\frac{\hat \alpha}{L^4}-\frac{4v^2}{15} \left(\frac{9c_m^2}{L^2}-4v^2) \log \left(\frac{r}{L}\right) \right)
\right] \left( \frac{L}{r}\right)^4+\dots,
\\
\nonumber
&&
m(r)=1+\frac{v^2}{9} \left( \frac{L}{r}\right)^2
+\left[
\frac{\hat \beta}{L^4}+\frac{4v^2}{15} \left(\frac{3c_m^2}{L^2}+2v^2\right) \log \left(\frac{r}{L}\right)  
\right]\left( \frac{L}{r}\right)^4+\dots,
\\
&&
\label{BB-inf}
n(r)= 
1-\frac{17v^2}{9}\left( \frac{L}{r}\right)^2
+\bigg[
\frac{3(\hat \alpha-\hat \beta)}{L^4}+\frac{4c_m^2 v^2}{15L^2}
+\frac{497v^4}{405}  
\\
\nonumber
&&
{~~~~~~~~~~~~~~~~~~~~~~~~~~~~~~~~~~~~~~~~~~~~~~~~}
+\frac{8v^2}{5}\left(\frac{3c_m^2}{L^2}- 3v^2\right) \log \left(\frac{r}{L}\right) 
\bigg]\left( \frac{L}{r}\right)^4
       +\dots,
\\
&&
\nonumber
w(r)=\hat j\frac{1}{r^3}+\dots,
~~~
a_0(r)= -\frac{q}{r^2}+\dots,
~~~
a_k(r)=c_m+\left[\mu-2c_m L^2v^2 \log \left(\frac{r}{L}\right)\right]\frac{1}{r^2}+\dots .
\end{eqnarray}
Restricting to the non-extremal case,
the black branes possess a horizon at $r=r_H>0$,
with approximate solution  
very similar to (\ref{eh-expansion}),
with
$f(r)=f_2(r-r_H)+\dots,$
$m(r)=f_2(r-r_H)+\dots,$
$n(r)=f_2(r-r_H)+\dots,$
and nonvanishing 
$\omega$, $a_0$ and $a_k$.
This leads to an induced metric on the horizon 
with a form 
very similar to 
(\ref{h-twist1}).
The horizon area density and Hawking temperature are
\begin{eqnarray}
A_H=\pi  \Delta \Psi r_H^3\frac{m_2 }{32 f_2 }\sqrt{\frac{n_2 }{f_2}},~~
T_H=\frac{1}{2\pi}\frac{r_H^2}{L^2}\frac{f_2}{\sqrt{m_2}}~,
\end{eqnarray}
 with $\Delta \Psi$ the (arbitrary) periodicity of the $\Psi$-coordinate.

The global charges for the black branes 
are computed by using the approach described in Section 2.
A major difference in this case is that one deals with
densities of relevant charges, since $\Theta$ has an infinite range.
One finds 
\begin{eqnarray}
M=\frac{1 }{8}
\left(
-\frac{(3\hat \alpha+\hat \beta)}{8L^2}
+\frac{ c_m^2 v^2 }{30}
+\frac{1883L^2v^4}{12960} 
\right)\Delta \Psi,
~
J=-\frac{\hat j  v^3}{64 \pi} \Delta \Psi~,~
Q=-\pi \left(q v-\frac{4\lambda}{3\sqrt{3}}c_m^2 \right) \Delta \Psi.~~{~~}
\end{eqnarray}
 
\begin{figure}[t]
\centering
\includegraphics[scale=0.4,angle=-90]{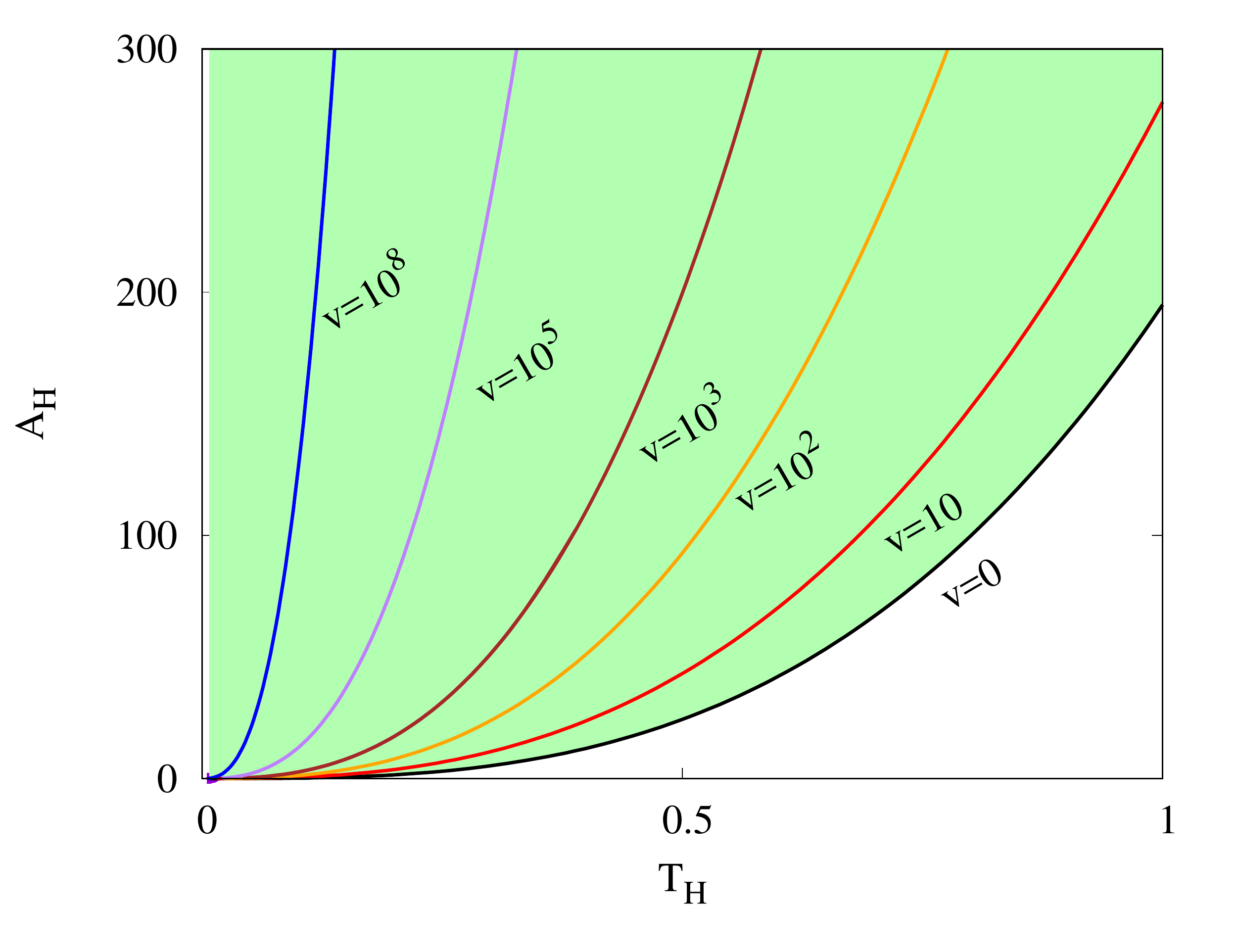}
\caption{Area vs. temperature of static twisted black branes without electric charge. The lines represent families of black branes with constant twisting parameter $v$.}
\label{plot_static_vac_Ah_vs_Th}
\end{figure}

The study of these configurations can be performed in a similar way 
to the (spherical horizon topology)
 BHs in the paper. 
So far the only case we have investigated more systematically corresponds to 
vacuum, static black branes.
The horizon area-temperature diagram of these solutions is shown in 
 Figure \ref{plot_static_vac_Ah_vs_Th}.
One can see that these twisted branes present similar properties to the standard
$v=0$ Schwarzschild black brane 
(represented with a black line in this figure). 
We have also found numerical evidence for the existence of regular configurations in the more general case of twisted black branes in EMCS theory, with (in principle) arbitrary values of the $J,Q,c_m$ and $v$. 
However, a systematic  study of their properties is beyond the purposes of this work.
 
\subsection{A supersymmetric solution} 
An interesting question here concerns the possible existence of a supersymmetric 
limit of the black brane solutions discussed above.
To address it, we use a slightly modified version of the framework 
employed for black holes with a spherical horizon topology.
These solutions are a particular member of the timelike case
in 
\cite{Gauntlett:2003fk} 
(see also \cite{Behrndt:2004pn} 
for a related study).
Then  
we consider the usual framework with
\begin{eqnarray}
\label{metrick01}
ds^2=-f^2(\rho)  ( dy+w)^2+\frac{1}{f}ds_B^2,~~
A=\frac{\sqrt{3}}{2}\left( f(dy+w)+\frac{L}{3}P  \right),
\end{eqnarray}
with a K\"ahler metric $ds^2_{B}$ on a four-dimensional base space
(characterized by a one-form $X^1$, while 
 $P$ is the Ricci one-form potential)
and $w$ a transverse one-form.
The solutions of interest are found for the following choice 
of the  base space 
\begin{eqnarray}
\label{metrickb01}
ds_B^2 =d\rho^2+a^2(\rho)( { \sigma}_1^2+  {\sigma}_2^2)+ b^2(\rho)  {\sigma}_3^2 ,
\end{eqnarray}
with the one forms
\begin{eqnarray}
\nonumber
  {\sigma}_1 = dx,
~~
  {\sigma}_2 = dy ,
~~
 {\sigma}_3 = dz+\frac{1}{2}(xdy-ydx),  
\end{eqnarray}
$x,y,z$ possessing the usual range.

Then a similar reasoning as in the case above of a Bergmann manifold as the base space,
leads to a
formulation of the problem in terms of a single function $a(\rho)$,
with
\begin{eqnarray} 
w=\Psi(\rho)\sigma_3,~~P=p(\rho)\sigma_3,~~X=-a^2(\rho) \sigma_1\wedge \sigma_2+b(\rho) \sigma_2 \wedge d\rho~,
 \end{eqnarray}
and the gauge potential
\begin{eqnarray} 
\label{A-susy0}
A= \frac{\sqrt{3}}{2} \left[ f(\rho) \diff y + \left(f(\rho)\Psi(\rho)+\frac{L}{3}p(\rho) \right)  {\sigma}_3  \right ].
\end{eqnarray}
The function $a(\rho)$ satisfies again a sixth-order equation
which can
formally be written in the compact form  (\ref{eqa}).
However, the expressions for
$p$,  
$f$
and
$g$
are (slightly) different, with some  terms which are absent here:
\begin{eqnarray} 
\label{set-k0}
p=4a'^2+2aa'',
~
f^{-1} \ = \ \frac{L^2}{12 a^2 a'}[4 (a')^3 + 7 a\, a' a''  + a^2 a'''],
~
 g =-\frac{a'''}{a'} - 3 \frac{a''}{a} + 4 \frac{(a')^2}{a^2},~~{~~~}
\end{eqnarray}
while $b$, $\Psi$ and the operator $\nabla^2$ are the same.
We would like to emphasize that the explicit $a$-equation 
here
does not coincide with the one resulting from (\ref{eqa}).
Despite that, 
\begin{eqnarray} 
\label{exk0}
a(\rho)=\alpha L \sinh\left(\frac{\rho}{L}\right) , 
\end{eqnarray}
is still a solution.
After defining
\begin{eqnarray} 
\label{texk0}
x=\Theta \cos \phi,~~y=\Theta \sin \phi ,
\end{eqnarray}
together with a new  radial coordinate 
\begin{eqnarray} 
\label{trk0}
\rho=\frac{L}{2}{\rm ArcCosh}(\frac{1}{3}(2r^2+1)) ,
\end{eqnarray}
this supersymmetric 
configuration can be written  in a compact form,
with
\begin{eqnarray} 
\label{s1k0}
&&ds^2=
-\left(1-\frac{L^2}{r^2}\right)^2 \left[dy+\frac{2\alpha^2 L}{3}
                           \left(
													         \frac{r^2}{L^2}+2+\frac{3}{2(\frac{r^2}{L^2}-1)} 
                              \right)
\left(d\psi+\frac{\Theta^2}{2}d\phi\right) \right]^2
\\
\nonumber
&&{~~~~~~~~~~}
+
\frac{dr^2}{(\frac{L^2}{r^2}-1)^2(\frac{r^2}{L^2}+2)}
+\frac{\alpha^2 r^2}{3}
\left[
d\Theta^2+\Theta^2 d\phi^2
+
\frac{4}{3}\alpha^2\left(2+\frac{r^2}{L^2}\right)\left(d\psi+\frac{\Theta^2}{2} d\phi\right)^2
\right],
\\
&&
A= \frac{\sqrt{3}}{2}\left(\frac{L^2}{r^2}-1\right)dt-\frac{\alpha^2 L^3}{2\sqrt{3}r^2}\left(d\psi+\frac{\Theta^2}{2}d\phi\right)~.
\end{eqnarray}
This corresponds, in fact, to the solution  obtained by Gutowski
and Reall in Ref. \cite{Gutowski:2004ez},
as the large size limit of the 
configuration (\ref{nGR1}).
It describes a black brane, 
the above line element possessing a horizon at $r=L$,
whose properties are functions of the parameter $\alpha$
($e.g.$ the event horizon area density
is
$A_H=\alpha^4L^3/(3\sqrt{3})$).
However, as $r\to \infty$, a non-asymptotically AdS 
geometry is approached, which corresponds to a plane-fronted wave
with a vanishing magnetic field  \cite{Gutowski:2004ez}.

Based on the results in Section 4, 
one may expect the existence of other solutions
of the sixth-order $a$-equation,
different from (\ref{trk0}).
However, we have failed to find any, the 
freedom
 in the choice of the boundary conditions
at $\rho=0$ 
(as implied
by 
(\ref{nh-susy-r2}))
being absent in this case.
Thus one finds a single possible form of the solutions at $\rho\to 0$,
corresponding to (\ref{sola0}).
Then the numerical integration of the $a$-equation leads always to the solution (\ref{exk0}).

 \begin{small}

 \end{small}

\end{document}